\documentclass[1p]{elsarticle}

\usepackage{hyperref}
\usepackage{subfig}
\usepackage{amsmath}
\usepackage{booktabs}
\usepackage{graphicx}
\usepackage{epstopdf}
\usepackage{algorithm}
\usepackage{algorithmic}
\usepackage{morefloats}
\usepackage{url}
\usepackage{color}

\graphicspath{{.}{images/}}
\DeclareGraphicsExtensions{.pdf}

\journal{Journal of Parallel and Distributed Computing - Elsevier}

\begin{document}

\begin{frontmatter}

\title{Highly intensive data dissemination in complex networks\footnotemark[0]}

\footnotetext[0]{The publisher version of this paper is available at \url{http://dx.doi.org/10.1016/j.jpdc.2016.08.004}.
\textbf{{\color{red}Please cite this paper as: ``Gabriele D'Angelo, Stefano Ferretti. Highly intensive data dissemination in complex networks. Journal of Parallel and Distributed Computing, Elsevier, vol. 99 (January 2017)''.}}}

\author{Gabriele D'Angelo\corref{cor1}}
\ead{g.dangelo@unibo.it}

\author{Stefano Ferretti}
\ead{s.ferretti@unibo.it}

\address{Department of Computer Science and Engineering. University of Bologna, Italy.}

\cortext[cor1]{Corresponding Author. Address: Department of Computer Science and Engineering. University of Bologna. Mura Anteo Zamboni 7. I-40127, Bologna. Italy. Phone +39 051 2094511, Fax +39 051 2094510} 

\begin{abstract}
This paper presents a study on data dissemination in unstructured Peer-to-Peer (P2P) network overlays. The absence of a structure in unstructured overlays eases the network management, at the cost of non-optimal mechanisms to spread messages in the network. Thus, dissemination schemes must be employed that allow covering a large portion of the network with a high probability (e.g.~gossip based approaches). 
We identify principal metrics, provide a theoretical model and perform the assessment evaluation using a high performance simulator that is based on a parallel and distributed architecture. A main point of this study is that our simulation model considers implementation technical details, such as the use of caching and Time To Live (TTL) in message dissemination, that are usually neglected in simulations, due to the additional overhead they cause. Outcomes confirm that these technical details have an important influence on the performance of dissemination schemes and that the studied schemes are quite effective to spread information in P2P overlay networks, whatever their topology. Moreover, the practical usage of such dissemination mechanisms requires a fine tuning of many parameters, the choice between different network topologies and the assessment of behaviors such as free riding. All this can be done only using efficient simulation tools to support both the network design phase and, in some cases, at runtime.
\end{abstract}

\begin{keyword}
Data dissemination \sep Simulation \sep Complex Networks \sep Performance Evaluation
\end{keyword}

\end{frontmatter}

\section{Introduction}\label{sec:introduction}

Unstructured Peer-to-Peer (P2P) systems have been recognized as a good practice to build effective distributed applications. 
This is particularly evident when peers composing the network are dynamic, with frequent arrivals and departures. 
In fact, in this case, the use of agile attachment strategies to create an overlay network (i.e.~the network composed of links representing an interaction/connection between nodes), plus the use of a simple dissemination protocol to let nodes interact, offer an easy way to manage interaction substrate, on top of which it is possible executing distributed applications.
Nodes create links based on an attachment process that does not depend on the ``type'' of involved nodes. Thus, for instance, if we are dealing with a content management system, links are not created based on the contents owned by peers; rather, links are established based on other criteria, e.g.~arbitrarily.

As concerns the spread of information, an interesting solution is based on gossip. 
This epidemic dissemination strategy uses randomized communication that distributes 
contents without a specific, content based, routing scheme.
Gossip has been recognized as a robust and scalable communication paradigm to be employed in large-scale distributed environments~\cite{gda-simutools-09,KincaidA05,dobrescu}. In fact, although it has communication costs usually higher than other, optimized solutions, e.g. tree-based protocols, a gossip-based dissemination scheme is intrinsically fault tolerant.

There is a vast literature on gossip. Related studies are mainly theoretical, since their aim is to prove that large-scale networks can reliably and effectively employ these strategies to disseminate information \cite{Ferretti:2012,Kempe:2004,Alistarh:2010:EG:1880999.1881012,Fernandess:2007,Georgiou:2013}. 
For instance, in \cite{Alistarh:2010:EG:1880999.1881012} it is shown that, depending on the system model, certain gossip-based protocols can achieve a message complexity around $O(n \log^3 n)$, or even $O(n)$, with high probability.
It is worth mentioning that in the proposed models the behavior of nodes is usually simplified, and several practical issues are not considered, that instead should be took into consideration when building a distributed system.
Some other works exploit simulation to evaluate epidemic strategies \cite{Ferretti:2012,Jelasity:2005,KincaidA05,dobrescu}. Also in these works, nodes have a very simple behavior.
The rationale behind this choice is twofold, usually. Firstly, often a simple behavior of nodes/agents allows verifying quite easily if an interesting emergent behavior occurs at the whole system/network level.
Secondly, these simplifications allow having a lightweight simulation model that enables the simulator to scale up to large networks.
However, also in this case, while a general main result is obtained, there is a lack of technical details, which are instead important during the real deployment of these strategies.

Today, the use of parallel and distributed simulation and the advent of multi/many-core processors make possible adding more details on the behavior of simulated entities.
Adding such complexity in simulation corresponds to give more emphasis on the impact of some important algorithmic details and expedients that can affect the dissemination performance in a P2P overlay network. 
In this work, we assess the performance of different dissemination protocols on different P2P overlays, and study the impact of caching and of the Time To Live (TTL) to distribute messages. In our assessment, we employ parallel and distributed simulation.
The metrics employed during the assessment are the coverage of the network and the delay for disseminating messages.
Not only, a theoretical model is provided that, given a gossip protocol and the topology of the underlying network overlay, allows estimating the threshold values corresponding to a phase transition between the ability of a given gossip protocol to spread a message to a significant set of nodes in the net, and a local dissemination that reaches a limited neighborhood of a node, only.

The contributions of this work are the following.
\begin{itemize}
 \item We study a \emph{degree dependent} dissemination algorithm, that relays messages to nodes based on their degree. We employ different degree dependent functions in our simulations.
 \item We provide a theoretical model that based on the dissemination protocol and the degree distribution of nodes composing the network, is able to determine the threshold values for the parameters of the dissemination algorithm. Such a threshold identifies a phase transition: below the threshold a disseminated message reaches a small, local fraction of network nodes, while above the threshold the message reaches a giant component of the network, i.e.~a set of nodes of the order of the network size.
 \item We study the impact of cache and TTL on data dissemination.
 \item We perform a parallel and distributed simulation of dissemination algorithms in large scale networks; different network topologies are employed (i.e.~random graphs, scale-free networks, Watts-Strogatz small-world networks, k-regular). To the best of our knowledge, this is the first contribution that shows results of large scale simulations over different topologies, where the dissemination is so highly intensive and nodes behavior considers cache and TTL management. Moreover, contrary to other typical simulation studies, where a single node acts as the source generating messages, in our simulations all nodes generate messages, concurrently. From a simulation point of view, this a more complex problem that mimics network intensive networking applications, e.g.~P2P online gaming and distributed virtual environments \cite{gda-disio-11-2}, P2P file sharing or wireless sensor networks.
 \item A preliminary evaluation of the impact of free riding~\cite{hughes2005free} on data dissemination with different gossip protocols and network topologies is reported.
 \item To assess the performance of the considered dissemination protocols, we use a metric termed ``overhead ratio'', that measures the total number of delivered messages (for a given protocol) over the minimum number of messages needed to obtain a complete coverage (the lower bound), given the considered graph. The rationale behind this metric is to quantify the overhead, in terms of sent messages, for a communication protocol. It allows to compare the behavior of different dissemination strategies over different network topologies.
 \item Given a set of nodes in a P2P overlay, and the need to create a given overlay, an issue is how to set the network in order to guarantee certain communication properties. Our work permits to understand, during the design phase of a P2P overlay, how to set network parameters so as to obtain a certain overhead, that would guarantee a certain network coverage and delay. In most cases, this can be done only exploring the space of parameters of the available dissemination protocols. In practice, this requires the execution of a high number of simulations runs. This confirms, the need of scalable and efficient simulation tools.
\end{itemize}

The reminder of this paper is structured as follows. 
Section~\ref{sec:related} presents some related works available in the literature.
In Section~\ref{sec:background}, we discuss some background needed to understand the gossip algorithms and the performance assessment.
Section~\ref{sec:gossip} presents the considered dissemination algorithms.
Section \ref{sec:model} presents the theoretical model and employs it to study the ability of the considered dissemination strategies to spread messages over an overlay network, given its degree distribution. 
In Section~\ref{sec:simulation}, the simulation testbed is described.
Section \ref{sec:performance} reports on simulation experiments we carried out. 
The main results from the performance evaluation are discussed more fully in Section~\ref{sec:discussion}.
Finally, concluding remarks are reported in Section~\ref{sec:conclusions}.

\section{Related Work}\label{sec:related}

In this section, we review some works concerned with data dissemination in unstructured P2P networks.
Since the considered schemes are based on gossip-style epidemic protocols, we focus on gossip-related approaches.
Moreover, a main rationale for this choice is that gossiping poses some challenges when we try to guarantee a high (possibly full) network coverage and low delay, in scalable and highly intensive scenarios.
Gossip is a simple, yet effective strategy to disseminate information. 
Its main feature is the use of randomization to propagate data. It has been proved that in certain contexts this provides better reliability and scalability than deterministic approaches \cite{Eugster:2007}. 

Gossip-based communication can use either push, pull or push-pull schemes.
According to a push based dissemination, it is the sender that decides which nodes will receive a message it is relaying through the network. 
Pull-based approaches let receivers trigger a communication with another node that will send some data. 
Finally, in push-pull protocols both nodes gossiping with each other share their owned data.

There is a vast literature on the use of dissemination and gossip-based approaches in distributed systems. Many works propose its utilization in several application domains. Examples are information spreading in mobile ad-hoc environments \cite{Anagnostopoulos:2012:AEI:2365374.2365840,conf/nca/GarbinatoRT07,PitreyS13}, multicast \cite{Riviere:2007}, multiplayer online games and distributed virtual environments \cite{gda-disio-11-2}, multiresolution data representations for sensor networks \cite{Sarkar:2011}, opportunistic networks \cite{Chaintreau:2008,Ferretti2013481}, publish-subscribe systems \cite{Baldoni:2007,Ferretti:2012,VoulgarisRKS06}, query processing over XML data \cite{Slavov:2014}, resource discovery \cite{Haeupler:2012,Ferretti20131631,Kempe:2004,khatibi,shah}, resource management in cloud computing and distributed systems \cite{mbp-2011,Sharifkhani}, social networks \cite{Asthana:2013}, community detection \cite{Bae:2015}, etc. 

Moreover, examples of real systems exist that employ gossip strategies. It is well known that Amazon S3 uses a gossip protocol to quickly spread server state information throughout the system \cite{amazon}. Not only, Amazon's Dynamo storage system employs a gossip based distributed failure detection and membership protocol \cite{DeCandia:2007}.
The Facebook team developed Cassandra \cite{Lakshman:2010}, a distributed storage system employing a gossip strategy called Scuttlebutt for membership management and for disseminating system control state messages \cite{vanRenesse:2008}.
This is an anti-entropy gossip based mechanism, exploited to guarantee an
efficient utilization of CPUs and communication channels. 
Tribler is an anonymous open source P2P BitTorrent client, that adds keyword search ability to the BitTorrent file download protocol using gossip.

These solutions are employed into cloud or within the internal mechanisms of distributed systems, with a limited amount of nodes involved. Instead, we are interested here in large scale systems composed of thousands of nodes.
In this sense, many studies have been presented on epidemic algorithms to disseminate contents in communication networks \cite{Ferretti:2012,Kempe:2004,sf_complenet,Alistarh:2010:EG:1880999.1881012,Fernandess:2007,Georgiou:2013}. 
Due to the need to scale up to very large numbers of nodes, these studies were mainly theoretical, modeling gossip as a particular instance of general percolation problems and epidemic spread of viruses. Indeed, gossiping a message means that a node sends it, with a certain probability function, to nodes it is connected to (i.e.~its neighbors). From a modeling point of view, this is equivalent to imagine a node that passes a virus (with a given probability) to another node it interacts with.
While elegant these mathematical approaches do not take into consideration several practical issues that should be considered when building a distributed system. 
For instance, an assumption can be that nodes have a full knowledge of the interaction history, or conversely, that they are completely unaware of previous nodes communications. 
In this sense, a gossip strategy can be represented as a Susceptible, Infective, Recovered (SIR) scheme \cite{wkermack27,Newman:2010:NI:1809753}, according to which a (susceptible) node $n$ becomes infected upon reception of a message and spreads it to its neighbors. Hereinafter, $n$ will not relay the message anymore (in case some neighbor would resend the message to it), i.e.~$n$ is in a recovered state.
On the other hand, the Susceptible, Infective, Susceptible (SIS) scheme allows to represent the situation when a node $n$ retransmits always a received message, even if it has done it already \cite{wkermack27,Newman:2010:NI:1809753}. In fact, when $n$ receives a message (i.e.~$n$ becomes infected) it sends the message to its neighbors (via gossip) and then it returns to the susceptible state.
In other words, $n$ does not maintain any log of messages it processed already.

As a matter of facts, while it is impractical to assume that nodes can hold forever information about relayed messages, it is a common practice to employ a caching system, that stores a limited amount of managed messages (e.g.~newer ones). 
Indeed, in many scenarios it is reasonable to assume the need to spread a given item for a limited amount of time.
When a node receives a message, if it has the message id in its cache already, this is means that the message has been already processed. Thus, the message can be discarded. 
In some sense, this resembles the Susceptible, Infective, Recovered, Susceptible (SIRS) model, which is an extension of the SIR model where a recovered node comes back to a susceptible state after a while (this resembles a node which removes a message entry from its cache).
We will show that the caching policy and the cache size are two key factors for the gossip performance.

The use of theoretical, simplified mathematical models is also due to the difficulty to resort to viable and effective simulators, able to run a large amount of node instances in a simulated network \cite{gda-simutools-09,KincaidA05,dobrescu}. 
Today, the use of parallel and distributed simulation, coupled with the use of actual multi/many-core processors, makes this possible.
Some previous works have demonstrated via simulation that gossip strategies can be
proficiently employed to disseminate data in P2P overlay
networks~\cite{gda-simutools-09,gda-disio-11-2,XgaetaX,Jelasity:2005}.
In particular, \cite{Jelasity:2005} presents a study of a gossip-based protocol for computing aggregate values over network components in a fully decentralized fashion. Simulation results were obtained using a discrete time stepped simulator.
In \cite{gda-simutools-09}, the authors of this paper made a study on the performance of gossip dissemination protocols over scale-free networks. 
While the main focus of that work was on tools to simulate scale-free networks, the ``overhead'' metric was used to assess the performance of a gossip dissemination scheme. A discussion on this metric is reported in Section \ref{sec:metrics} of this paper.
In this work, we extend such results using a larger set of simulation algorithms, different network topologies, providing a theoretical model for the considered schemes, using a wider simulation assessment, and by resorting to an improved parallel and distributed simulation tool~\cite{gda-mospas-11}.

\section{Background}\label{sec:background}

In this section, we provide some background on main issues concerned with gossip in unstructured networks.

\subsection{Network Topologies}

In this work, we deal with unstructured networks. In an unstructured network, links among nodes are established arbitrarily. 
Nodes locally manage their connections with others, and these links do not depend on the contents being disseminated \cite{EberspacherS05a}.
The specific attachment protocol, that determines how nodes connect each other, allows building specific general topologies. 
These solutions are particularly simple to build and manage, with little maintenance costs, yet at the price of a non-optimal organization of the overlay. 

Different network topologies are considered in this paper.
In the following, we describe the general characteristics of such topologies, together with the methods employed to build them. 

\subsubsection{Random Graph Networks}
A random graph corresponds to a general network where links among nodes are randomly generated. In this case, the Erd\"{o}s-R\'{e}nyi (ER) model is employed to build such random graphs \cite{ER}.
According to it, a graph is constructed by connecting nodes randomly. Each edge is included in the graph with a probability $p$, independently from every other edge. 
Thus, $p$ determines how much the graph is connected.

\subsubsection{Scale-Free Networks}
A scale-free network possesses the distinctive feature of having nodes with a degree distribution that can be well approximated by a power law function. Hence, the majority of nodes have a relatively low number of neighbors, while a non-negligible percentage of nodes (``hubs'') exists with higher degrees \cite{gda-simutools-09}. 
The presence of hubs has an important impact on the connectivity of the net. 
In fact, the peculiarity of these networks is that they possess a very small diameter, thus allowing to propagate information in a low number of hops. 

To build such a kind of networks, we employed the Barab\'asi-Albert (BA) model \cite{RevModPhys.74.47}.
The BA model is based on a preferential attachment that shows the well known “rich-get-richer” effect. In substance, upon arrival of a novel node in the network, it creates novel links with other nodes. In this case, a link is most likely to attach to nodes with higher degrees. 
The network begins with an initial network of $m$ nodes.
Then, other nodes are added to the network one at a time. Each new node is connected to $m$ existing nodes with a probability that is proportional to the number of links that the existing nodes already have.

\subsubsection{Watts-Strogatz Small World Networks}
The Watts and Strogatz (WS) model is a random graph generation model that produces graphs with small-world properties \cite{watts1998cds}. 
It does so by interpolating between an ER random graph and a regular ring lattice.
An initial $d$-dimensional lattice structure is used to generate a WS model. Each node in the network is initially linked to its $2d$ closest neighbors (for each dimension, a node has $2$ neighbors i.e.~its predecessor and its successor along that dimension). 
Then, a $p$ parameter is employed as the rewiring probability. Each edge has a probability $p$ that it will be rewired to the graph as a random edge.

\subsubsection{k-Regular Networks}
k-regular networks are those where all nodes start with the same degree $k$. 
These networks are quite common in several (P2P) systems, where the software running on peers is configured to have a given number of links in the overlay. 
This is usually accomplished for load balancing purposes \cite{TDGsIMC07}.

\subsection{Implementing a Dissemination Protocol}

A dissemination protocol over unstructured networks implies that messages are transmitted through links among nodes in the overlay. Thus, each node receiving a message analyzes it (passing this message to the application module, if necessary) and relays the message to its neighbors, via a dissemination algorithm. 
Since networks are usually graphs, it is possible that a node receives the same message multiple times. 
To avoid that such a message is relayed forever and to avoid that the communication among nodes is congested by redundant transmissions, two practical implementation expedients are employed, i.e.~i) caching identifiers of already processed messages and ii) assigning a deadline to the life of a message in the network.

\subsubsection{Caching}
Caching is a common practice in all computing related problems. 
In this context, caching is exploited by nodes to maintain the last identifiers of messages already processed. This way, if a node receives a message whose identifier is in the node's cache, such a message can be dropped to avoid redundant message processing.
A key factor is the cache size. In fact, the higher the cache size the easier to avoid redundant retransmissions, but the larger the memory requirements.
On the other hand, when dealing with large scale networks, with all nodes that concurrently generate novel messages that need to be disseminated, it is likely that nodes are required to handle a high amount of messages in a short time interval, thus overwhelming caches. 
We study the impact of the cache size on the performance of the considered dissemination protocols.

\subsubsection{Time-To-Live}
Messages are associated to a Time-To-Live (TTL) value. The TTL avoids that messages are forwarded forever in the net. Each time a message is relayed from a node, its TTL is decreased. When this value becomes $0$, the receiving node does not relay that message to its neighbors.
The tuning of the TTL is an important issue for the performance of the dissemination protocol.
It should be sufficiently large to guarantee that the message can be spread through the whole network.
However, usually this is an aspect which is not considered in mathematical works. 
Finding the ``optimal'' value for the TTL is very important and very different values are employed in different works.
For instance, in \cite{gda-simutools-09,gda-disio-11-2,Ozkasap:2009,sacha,Tanta-ngai} TTL values range from $6$ up to $100$.
When a flooding protocol is employed, then the TTL can be set equal to the network diameter. 
(Then one needs an estimation of this value \cite{Ferretti20131631}.)
However, when we consider the diameter of a network, we are focusing on the shortest paths among nodes, and in general this does not mean that the average path length among nodes scales with the diameter value.
When the dissemination protocol is based on gossip, it is not guaranteed that a message from a source will reach a given node through their shortest path. In this work, we analyze the impact of the TTL on the network coverage.

\subsection{Message Dissemination Evaluation Metrics}\label{sec:metrics}
Once a dissemination protocol is employed to spread information, it is important to define the metrics to evaluate it. 
The first considered measure is a common one, and it is strictly related to the ability of a communication protocol to disseminate a message to a ``sufficiently large'' subset of the population. 
This measure comes from percolation theory, which describes the behavior of connected clusters in graphs and studies the formation of long-range connectivity. 
A dissemination protocol works well if a message reaches a giant component of the order of the network size. If we imagine to spread a message into an infinite network, an infinite amount of nodes should receive the message.
As described with more details in the next sections, the considered protocols refer to a dissemination probability value that allows deciding if a node sends a message to other nodes, or not.
Based on the overlay topology (i.e.~the nodes' degree distribution), a {\bf threshold value} exists for this dissemination probability that corresponds to a phase transition. Below the threshold, a giant component does not exist (thus, typically a message reaches only a local neighborhood of the node that originated the message), while above it there exists a giant component.
In this paper, we will show that this phase transition point can be measured using a theoretical model that is based on the considered gossip protocol and the degree distribution of the network overlay
(while other considered metrics relate to more technical aspects; hence, they can be measured through simulation).
However, this threshold value should be considered as a sort of lower bound for ensuring data dissemination. In fact, the mentioned implementation parameters, such as nodes' cache size and TTL of messages, that are employed to viably deploy gossip algorithms in unstructured P2P overlays, can strongly influence the performance of a dissemination protocol. We will study their impact through other metrics, measured via simulation.

A desirable property of a dissemination protocol is the extension of the previous one, and is that of being able to reach all nodes, and this should
happen as quickly as possible. Thus, we use a measure called {\bf coverage}, which denotes the fraction of nodes which actually
received the messages. Ideally, we wish to obtain $100\%$ coverage,
meaning that all nodes received all the generated messages. 
It is difficult to estimate such a value theoretically, especially when TTL and caching techniques are considered. We measure this measure through simulation.

The third considered
measure is called {\bf delay}, and represents the average number of
hops that a message traverses before reaching a node (lower is
better). The delay is computed as follows: when a message is received
by a node for the first time, that node records the number of hops the
message traversed from its generation. The delay is computed as the
number of hops, averaged over all nodes which received the message,
and over all messages sent during a simulation run.

It is also important to identify appropriate cost metrics, so that all
dissemination protocols can be compared in the same conditions. 
in this sense, a useful measure is the {\bf overhead ratio} $\rho$ which is measured as follows \cite{gda-disio-11-2}:

\[
\rho = \frac{\textit{Delivered messages}}{\textit{Lower bound}}
\]

\noindent where ``delivered messages" is the total number of messages
that are delivered by a specific dissemination
protocol and the ``lower bound'' is the minimum number of messages (in
each graph) that are necessary to obtain a complete coverage. Thus,
the lower bound represents the number of messages sent by a broadcast
protocol which deliver events along the edges of a spanning tree, and
never sends duplicates. The lower bound depends on the graph and is
independent from the dissemination protocol to be used. For example,
in a graph of $n$ nodes and in which $m$ different events are
generated, the lower bound to the number of delivered messages is
$\Omega(n m)$. Each newly generated message has to traverse at least
$n-1$ links to eventually reach all nodes in the graph. Observe that
$n-1$ is precisely the number of edges on any spanning tree on a graph
with $n$ vertices.

\section{Dissemination Protocols}\label{sec:gossip}

In this section, we describe the considered dissemination protocols; they are basically all push gossip-based approaches.
According to our model, all nodes are able to generate a new message to be 
disseminated in the
network. When the generation procedure is invoked at a given node, a single 
message is created
with a certain probability, as described in Algorithm \ref{alg:generation}. The 
generation of a
message simulates the occurrence of a new event produced at a given node that 
must be propagated.
If the message is created, then it is sent through the net, using a 
\textsc{disseminate()} procedure
(line 6 of the algorithm). The message is also inserted in a cache (line 5).

\begin{algorithm}
\caption{Generation of a Message} \label{alg:generation}
\begin{algorithmic}[1]
\STATE function \textsc{generate}()
\STATE $\mathit{t} \gets$ \textsc{generationThreshold()} %
\IF{\textsc{random()} $< \mathit{t}$}  %
    \STATE \emph{msg} $\gets$ \textsc{createMessage}() %
    \STATE \textsc{cache}(\emph{msg}) %
    \STATE \textsc{disseminate}(\emph{msg}) %
\ENDIF%
\end{algorithmic}
\end{algorithm}

\begin{algorithm}
\caption{Reception of a Message} \label{alg:receive}
\begin{algorithmic}[1]
\STATE function \textsc{receive}(\emph{msg})
\IF{(\textsc{notCached}(\emph{msg}) $\wedge$ $\mathit{msg.ttl} > 0$)}  %
    \STATE \textsc{cache}(\emph{msg}) %
    \STATE $\mathit{msg.ttl \gets msg.ttl -1}$
    \STATE \textsc{disseminate}(\emph{msg}) %
\ENDIF%
\end{algorithmic}
\end{algorithm}

Upon reception of a given message (see Algorithm \ref{alg:receive}), the 
receiving node forwards
the message to its neighbors by calling the 
\textsc{disseminate()} function
(line 5 in the algorithm). This is accomplished only if the message is not 
already in the node's
cache. In fact, if the message is in cache, it has already been 
disseminated; hence, the node
has nothing to do with the message \emph{msg} (line 2). Conversely, \emph{msg} 
is transmitted and
cached (line 3 of Algorithm \ref{alg:receive}).
Needless to say, due to the possible memory constraints of a node, the cache is 
limited in size (\emph{cache.size}).

\subsection{Dissemination \#1: Gossip with Fixed Probability }

According to the first protocol, Algorithm \ref{alg:simple}, the node (say $\mathit{n_i}$)
randomly selects those edges through which the message \emph{msg} must be propagated
\cite{conf/nca/GarbinatoRT07,verma}. Specifically, all $\mathit{n_i}$'s neighbors 
(i.e.~$\Pi_\mathit{i}$) are considered and a threshold value $\gamma \leq 1$ is maintained, which
determines the probability that \emph{msg} is gossiped to the neighbor (when $\gamma = 1$ we
obtain a flooding algorithm). At each step the message is propagated from $\mathit{n_i}$ to $\gamma |\Pi_\mathit{i}|$ other nodes.
Thus, the higher the node degree, the higher its workload.

\begin{algorithm}[th]
\caption{Dissemination: Gossip with Fixed Prob.~of Dissemination (at $\mathit{n_i}$)} \label{alg:simple}
\begin{algorithmic}[1]
\STATE function \textsc{initialization}()
\STATE $\gamma \gets$ \textsc{chooseProbability()} %
\STATE

\STATE function \textsc{disseminate}(\emph{msg})%
\FORALL{$\mathit{n_j} \in \Pi_\mathit{i}$} %
    \IF{\textsc{random()} $< \gamma$}  %
        \STATE \textsc{send}(\emph{msg},$\mathit{n_j}$)%
    \ENDIF%
\ENDFOR
\end{algorithmic}
\end{algorithm}

\subsection{Dissemination \#2: Probabilistic Broadcast}

The second distribution protocol we consider is a probabilistic broadcast scheme (see in Algorithm
\ref{alg:broadcast}). Once the \textsc{disseminate()} procedure is called, if the message has been
locally generated at the node and \emph{msg} still needs to be spread to the network (we assume
this check is performed in \textsc{firstTransmission()}, line 5), \emph{msg} is sent to all node's
neighbors (lines 6-8). Conversely, if \emph{msg} has been received from someone else, the node decides with a certain
probability $\beta$ (defined at the beginning of the protocol) to forward \emph{msg} (line
5). In the positive case, the message is sent to all node's neighbors.

\begin{algorithm}
\caption{Probabilistic Broadcast} \label{alg:broadcast}
\begin{algorithmic}[1]
\STATE function \textsc{initialization}()
\STATE $\beta \gets$ \textsc{probabilityBroadcast()} %
\STATE
\STATE function \textsc{disseminate}(\emph{msg}) %
\IF{(\textsc{random()} $< \beta$ $\vee$ \textsc{firstTransmission()}) }  %
    \FORALL{$\mathit{n_j} \in \Pi_\mathit{i}$} %
        \STATE \textsc{send}(\emph{msg},$\mathit{n_j}$)%
    \ENDFOR
\ENDIF%
\end{algorithmic}
\end{algorithm}

\subsection{Dissemination \#3: Degree Dependent Gossip}
\label{sec:gossip_ddg}

According to this scheme, a node decides to relay the message, based on some specific features of its neighbors.
Thus, a gossip probability is used that is not the same for all nodes, but it depends on the degree of nodes. 
If a node $m$ has degree $i$, it will receive a message from a neighbor with a probability $\gamma(i)$, which is a function of the degree.
The algorithm related to this kind of schemes can be obtained by taking Algorithm \ref{alg:simple} and substituting $\gamma$ with a function $\gamma(i)$, if $i$ is the degree of the considered node (thus, it is not reported here).
The rationale is that nodes with a low degree might ``compensate'' their little amount of links with a higher reception probability (i.e., neighbors increase their dissemination probability), whereas nodes with high degree have a higher probability to receive a message from
one of their neighbors, thus reception probability can be safely lowered. This
last countermeasure is taken to avoid a flood of redundant messages.

There is a vast set of possible functions that might be employed in the algorithm. 
In this work, we experiment with two different $\gamma(i)$ functions, i.e.  
\[\gamma_1(i) = \left\{
  \begin{array}{ll}
    1 			& i = 1, 2\\
    \frac{1}{i^\alpha} & i > 2
  \end{array}
\right.
\]
and 
\[\gamma_2(i) = \left\{
  \begin{array}{ll}
    \frac{1}{\ln(\alpha i)} & i > \text{max}(2, \mathrm{e} / \alpha)\\
    1 			& \text{otherwise}
  \end{array}
\right.
\]
where $\alpha$ is a parameter. 
The rationale behind these functions is that if a node has a single connection or just two connections (it might be a node in a chain), then it floods the message, since stopping the gossip could stop the message dissemination into a network subset.
Moreover, as concerns $\gamma_2(i)$, we avoid that the inverse logarithmic function returns a value not comprised between $0$ and $1$.
In the performance evaluation (Section~\ref{sec:performance}), these protocols will be referred to as Degree Dependent Function 1 (DDF1) and Degree Dependent Function 2 (DDF2), respectively. Many other functions can be tested on the different network topologies but this is out of the scope of this paper, in which we prefer to investigate the general approach instead of studying in deep a specific aspect.

To obtain information on the degree of neighbors, we assume that nodes exchange degree information. In particular, degree information is piggybacked within messages relayed to neighbors when a node $n$ relays a message. This allows nodes exchanging this information without the need for further control messages. This means that, under the communication point of view, the overhead introduced by this mechanism is of a few bytes for each exchanged message. Furthermore, appropriate techniques can be used to further reduce this overhead (e.g.~update the node degree only when it changes). In others words, this means that the overhead introduced by this procedure is in most cases negligible. In the early steps of simulation, neighborhood information is missing, thus nodes are forced to broadcast each message to all neighbors.

\section{Modeling Dissemination Protocols}\label{sec:model}

In this section, we present an analytical model for dissemination strategies. 
The model follows a general framework widely exploited in complex networks theory \cite{Ferretti20131631,newmanHandbook}.
It shows that dissemination strategies can disseminate information data over a network, and that a proper tuning of the protocol parameters, based on the underlying network topology, enables reaching a giant component of the network (i.e., messages percolate through the overlay network).
As mentioned in Section \ref{sec:background}, such a model allows to evaluate if a threshold value exists for the parameters, given a specific dissemination protocol and the topology of the underlying overlay network over which the protocol is run.

In line with typical theoretical models, some simplifications are made on technical details and parameters. 
Thus, as previously described theoretical results should be considered as qualitative outcomes. 
In any case, trends are confirmed via simulation, where we include the mentioned technical details not considered in the theoretical model. 

\subsection{System Model}

We consider push-based communication schemes, in accordance with schemes described in Section \ref{sec:gossip} \cite{gda-simutools-09,DBLP:conf/bioadit/OkuyamaTK06,Portmann20031159}. 
From a modeling point of view, push communication protocols resemble the spread of an epidemic in a contact network. Hence, they can be modeled using the frameworks typical of complex network theory \cite{Ferretti:2012,Ferretti20131631}.

We are interested here in assessing if it is possible to have a significant coverage of the network by tuning properly the dissemination protocol parameters.
Since we are dealing with very large networks, an approach which is typical in complex network theory is to examine infinite networks rather than just large ones. 
This assumption on the network size does not impact the behavior and the modeling of nodes, since peers know only their neighbors and manage contents based on this local knowledge.

Following the presented approach, we assume that links among nodes are randomly generated, based on a given node degree distribution. 
A consequence of the random nature of the attachment process is that, regardless of the node degree distribution, the probability that one of the second neighbors (i.e.~nodes at two hops from the considered node) is also a first neighbor of the same node, goes as $N^{-1}$, being $N$ the number of nodes in the overlay \cite{newmanHandbook}. 
Hence, this situation can be ignored since the number of nodes is high. 

\subsection{General Metrics}

We denote with $p_i$ the probability that a peer $n$ has degree equal to $i$, while 
$q_i$ is the \emph{excess degree distribution}, 
$q_i = \frac{(i+1)p_{i+1}}{\langle p \rangle}$ \cite{Newman:2010:NI:1809753}.
Probabilities $p_i$ and $q_i$ represent two similar concepts i.e.~the number of contacts of a considered peer (its degree), and the number of contacts obtained following a link (its excess degree), respectively. In the following, we introduce measures obtained by considering the degree $p_i$ of a node, and considering the excess degree $q_i$ of a link. In this last case, with a slight abuse of notation we denote all the probabilities/functions related to the excess degree with the same letter used for the degree, with an arrow on top of it, just to recall that the quantity refers to a link.

With $\langle p \rangle$ we denote the average degree, which depends on the degree distribution of the overlay.
Then, $\langle q \rangle$ is the mean value of the excess degree, that is \cite{Newman:2010:NI:1809753}
\begin{eqnarray}\label{eq:mean_q} 
  \langle q \rangle = \sum_i i q_i = \frac{\langle p^2 \rangle - \langle p \rangle}{\langle p \rangle}.
\end{eqnarray}

Given a peer $n$ in charge of relaying a message, the probability that $n$ forwards it to $i$ of its neighbors is denoted with $f_i$. This measure depends on the particular probabilistic communication protocol in use.
The probability that, following a link we arrive to a node that forwards the query to $i$ other nodes is denoted with $\overrightarrow{f_i}$.

To proceed with the reasoning, we need to introduce the generating functions for all these measures. Hence, $G$ is the generating function for $p_i$ coefficients,
$G(x) = \sum_i p_i x^i$, 
$\overrightarrow{G}$ is the generating function for $q_i$ coefficients,
$\overrightarrow{G}(x) = \sum_i q_i x^i$, 
$F$ is the generating function for $f_i$ coefficients,
$F(x) = \sum_i f_i x^i$,
and $\overrightarrow{F}$ is the generating function for $\overrightarrow{f_i}$ coefficients,
$\overrightarrow{F}(x) = \sum_i \overrightarrow{f_i} x^i.$
An interesting observation is that once one has characterized these $f_i, \overrightarrow{f_i}$ values and the related generating functions $F, \overrightarrow{F}$, it is possible to determine the average amount of receivers. This is obtained through a reasoning which is inspired from \cite{newmanHandbook}.

Let denote with $r_i$ the probability that $i$ peers receive a message, starting from a given node. Similarly, denote with $\overrightarrow{r}_i$ the probability that $i$ peers receive the message, starting from a link. In general, $\overrightarrow{r}_i$ can be defined using the following recurrence,
\begin{align*}
  \overrightarrow{r}_0 & =  0,\nonumber \\
  \overrightarrow{r}_{i+1} & =  \sum_{j \geq 0} \overrightarrow{f}_j \sum_{a_1 + a_2 + \ldots + a_j = i} \overrightarrow{r}_{a_1} \overrightarrow{r}_{a_2} \ldots \overrightarrow{r}_{a_j}.
\end{align*}
This equation can be explained as follows. It measures the probability that following a link we disseminate the message to $i+1$ peers. 
(The case $\overrightarrow{r}_0$ is impossible, since at the end of a link there must be a node.) In general, one peer is that reached at the end of the link itself. Then, we consider the probability that the peer has other $j$ links (varying the value of $j$). Each link $k$ allows to disseminate the message to $a_k$ peers, and the sum of all these reached peers equals to $i$.

Similarly, we can calculate $r_i$ as follows
\begin{align*}\label{eq:r_k}
  r_0  &= 0,\nonumber \\
  r_{i+1}  &= \sum_{j \geq 0} f_j \sum_{a_1 + a_2 + \ldots + a_j = i} \overrightarrow{r}_{a_1} \overrightarrow{r}_{a_2} \ldots \overrightarrow{r}_{a_j}.
\end{align*}
In this case, we start from the peer itself, considering it has a degree equal to $j$; and as before, from its $j$ links we can reach $i$ other peers, globally.

The use of generating functions may be of help to handle these two equations \cite{Wilf_1994}. In fact, if we consider the generating functions for $r_i$ and $\overrightarrow{r}_i$,
\begin{eqnarray}\label{eq:R} 
  R(x) = \sum_i r_i x^i, & & \overrightarrow{R}(x) = \sum_i \overrightarrow{r}_i x^i
\end{eqnarray}
then, after some manipulation we arrive to the following result
\begin{eqnarray}\label{eq:R_q_rec} 
  \overrightarrow{R}(x)  =  x \sum_{j \geq 0} \overrightarrow{f}_j [\overrightarrow{R}(x)]^j
     =  x \overrightarrow{F}(\overrightarrow{R}(x))
\end{eqnarray}
and, similarly,
\begin{eqnarray}\label{eq:R_rec} 
  R(x)  =  x \sum_{j \geq 0} f_j [\overrightarrow{R}(x)]^j
     =  x F(\overrightarrow{R}(x)).
\end{eqnarray}
From the generating functions, we might recover the elements $r_i$, $\overrightarrow{r}_i$ composing them. Unfortunately, equations (\ref{eq:R_q_rec}), (\ref{eq:R_rec}) may be difficult to solve, depending on the degree probability distribution $p_i$ which controls the whole introduced measures.
But actually, we are not interested that much in the single values of $r_i$, $\overrightarrow{r}_i$. In fact, it is easier and more useful to measure the average number $\langle r \rangle$ of peers that receive a given message through the dissemination protocol. To this aim, we can employ the typical formula for generating functions
$\langle r \rangle = R'(1)$ \cite{Wilf_1994}.
In fact, taking the first equation of (\ref{eq:R}), differentiating and evaluating the result for $x=1$, and since $r_0 = 0$, we have 
$$R'(x) \Bigl\lvert_{x=1} = \sum_i i r_i,$$
which is the mean value related to the distribution of $r_i$ coefficients.
Coefficients of the introduced generating functions are probabilities, and thus
$F(1)= \sum_i f_i = 1$, and similarly $\overrightarrow{F}(1)=1$, $R(1)=1$, $\overrightarrow{R}(1)=1$. Hence, taking (\ref{eq:R_rec}) and differentiating 
\begin{eqnarray*}\label{eq:r} 
\langle r \rangle & =  & R'(1) = \big[F(\overrightarrow{R}(x)) + x F'(\overrightarrow{R}(x))\overrightarrow{R}'(x)\big]_{x=1}\nonumber \\
 &  = & 1 + F'(1)\overrightarrow{R}'(1).
\end{eqnarray*}
Similarly, from (\ref{eq:R_q_rec}),
\begin{eqnarray*}\label{eq:R_q_der} 
\overrightarrow{R}'(1) & = & \big[\overrightarrow{F}(\overrightarrow{R}(x)) + x \overrightarrow{F}'(\overrightarrow{R}(x))\overrightarrow{R}'(x)\big]_{x=1}\nonumber \\
 & = & 1 + \overrightarrow{F}'(1)\overrightarrow{R}'(1).
\end{eqnarray*}

Thus,
\begin{equation*}\label{eq:R_q_der_f}
\overrightarrow{R}'(1) = \frac{1}{1 - \overrightarrow{F}'(1)}.
\end{equation*}
This last equation allows to find the final formula for $\langle r \rangle$,
\begin{eqnarray}\label{eq:r_final}
\langle r \rangle & = & 1 + \frac{F'(1)}{1 - \overrightarrow{F}'(1)}. 
\end{eqnarray}

Equation (\ref{eq:r_final}) has a divergence when $\overrightarrow{F}'(1) = 1$.
This implies that in an infinite network, the messages are spread through the net, reaching an infinite amount of nodes.
In the next subsection, we identify the values of the $\overrightarrow{f}_i$ coefficients of the $\overrightarrow{F}$ generating function, that depend strictly on the topology of the underlying network overlay and on the particular probabilistic communication protocol in use on top of it.

\subsection{Phase transitions for dissemination strategies}

We consider now the protocols described in the previous section, and we identify conditions under which we have a phase transition for message dissemination throughout the network. 
Above the phase transition a vast majority of nodes in the network will receive each message, i.e.~a high coverage can be obtained. 

\subsubsection{Fixed probability}

This scheme is based on a gossip probability $\gamma$, which is independent from any feature of connected nodes. 
Based on this scheme, the probability $f_i$ that a node $n$ decide to relay a message to $i$ of its neighbors is
$$
  f_i = \gamma^i \sum_{j \geq i} p_j \binom{j}{i} (1-\gamma)^{j-i}.
$$
This equation accounts for all probabilities of $n$ having a degree higher than $i$, which decides to independently relay the message to $i$ nodes (with probability $\gamma$), chosen among its neighboring set; moreover, other neighbors do not receive the message (with probability $1 -\gamma$).

A similar reasoning can be made to measure the probability that, following a link we arrive to a node that forwards the query to $i$ other nodes. This probability is readily obtained by substituting, in the equation above, $p_j$ with $q_j$, i.e.
$\overrightarrow{f}_i = \gamma^i \sum_{j \geq i} q_j \binom{j}{i} (1 - \gamma)^{j-i}.$

If we consider the generating function $F$ of the $f_i$ coefficients, we have 
$$F(x) = \sum_i f_i x^i 
     =  \sum_i \gamma^i x^i 
              \sum_{j \geq i} p_j \binom{j}{i} (1 - \gamma)^{j-i}
     =  \sum_j p_j (\gamma x + 1 - \gamma)^j
   =  G\big(\gamma x + 1 - \gamma\big)$$

Now, it is also possible to evaluate the average of the values $f_i$, by calculating the derivative of $F$ measured at $x=1$, since $F'(1) = \sum_i i f_i$ \cite{Wilf_1994}.
We have
$$F'(x)\Bigl\lvert_{x=1}\ \ = \ \ \frac{dG}{dx}\big(\gamma x + 1 - \gamma\big)\Bigl\lvert_{x=1}\ \
                        = \ \ \gamma G'(1)\ \ = \ \ \gamma \langle p \rangle,$$
where $\langle p \rangle = G'(1)$ is the mean node degree \cite{Ferretti20131631,newmanHandbook}.
From a similar reasoning,
$\overrightarrow{F}'(x)\Bigl\lvert_{x=1}
  = \gamma \overrightarrow{G}'(1) = \gamma \langle q \rangle.$


In this case, the condition  $\overrightarrow{F}'(1) = 1$ for having that the message percolates through the network is satisfied
when 
$$\gamma = 1 / \langle q \rangle$$
i.e.,~the gossip probability is equal to the inverse of the mean value of the excess degree.
Actually, this is a result which is well known, since this is the critical transmission of the spread of an epidemic in a network \cite{PhysRevLett.85.4626}.
This formula allows measuring the phase transition, for the fixed probability gossip scheme, in whatever topology of interest; one just needs a measure (or numerical estimation) of the average excess degree. 

\begin{figure}[ht]
\centering
\subfloat[Random graph\label{fig:fixed_rg}]{\includegraphics[width=6.5cm]{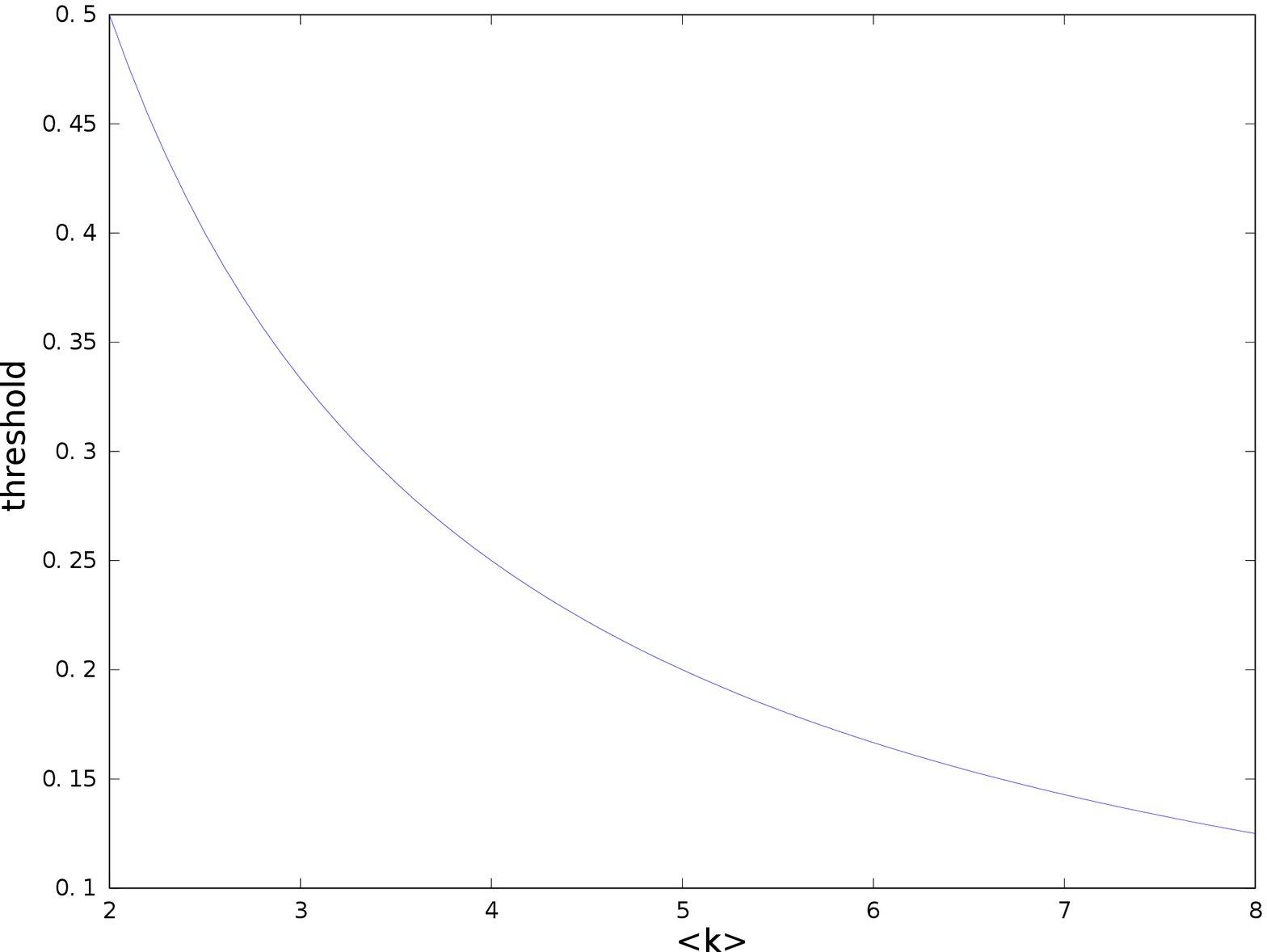}}
\subfloat[k-regular graph\label{fig:fixed_k_reg}]{\includegraphics[width=6.5cm]{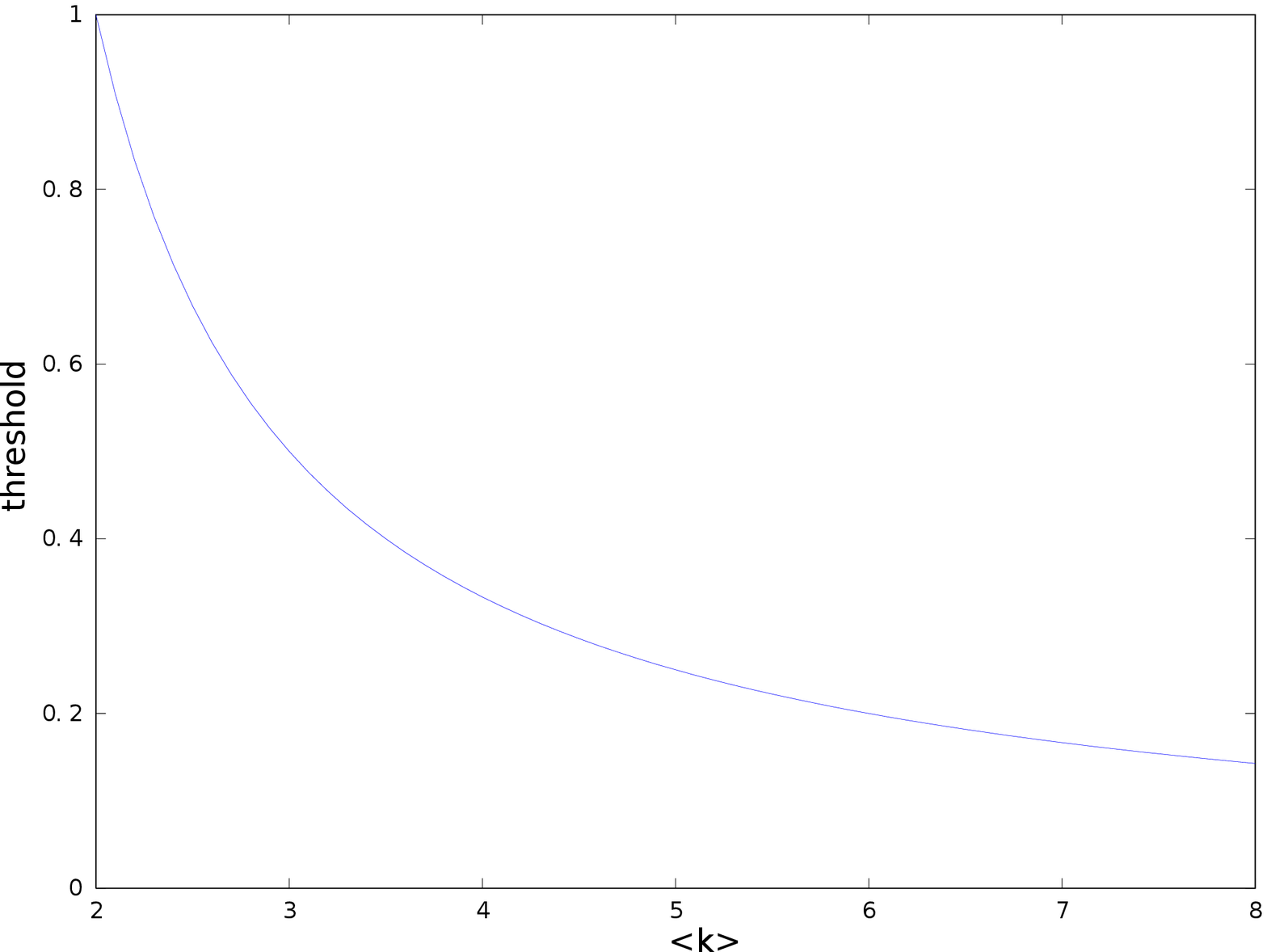}}
\caption{Fixed probability gossip (and probabilistic broadcast) -- phase transitions of the $\gamma$ (and $\beta$) parameter.}
\label{fig:fixed_model}
\end{figure}

Now, $\langle q \rangle$ has been measured for certain network topologies \cite{newmanHandbook}. 
We can thus easily provide the phase transition thresholds for entire classes of networks.
For instance, in random graphs with a Poisson degree distribution with mean $\langle k \rangle$, the excess degree generating function is $\overrightarrow{G}(x) = e^{\langle k \rangle (x-1)}$. Hence, 
$\langle q \rangle_{\text{rg}} = \overrightarrow{G}'(x) |_{x=1} = \langle k \rangle e^{\langle k \rangle (x-1)} |_{x=1} = \langle k \rangle.$
An thus, we have a phase transition when $\gamma_{\text{rg}} = 1 / \langle k \rangle$ (see Figure \ref{fig:fixed_rg}).
That is, we have a high coverage in a random graph when it is quite likely that each node will relay a message to at least one among its ($\langle k \rangle$, on average) neighbors. 
As mentioned, this formula does not account how fast the message would percolate, and does not consider the fact that a message might have a TTL associated, that can stop the message spread.
Indeed, if one sets the TTL equal to (an estimation of) the network diameter, a message can reach an amount of nodes that is of the order of the network size.
However, in general the threshold value, identified by this theoretical model, is a sort of lower bound for having a data dissemination.

Similarly, in a $k$-regular graph, where nodes have all the same degree $k$ ($p_k =1, p_i = 0 (i \neq k)$), the average excess degree is $\langle q \rangle_{k\text{-reg}} = k-1$ (see eq. \ref{eq:mean_q}) (see Figure \ref{fig:fixed_k_reg}). 
This is evident, since in such a network an edge will lead to a node that (by construction) has $k-1$ other edges.
An thus, we have a phase transition when $\gamma_{k\text{-reg}} = \frac{1}{(k-1)}$.

\subsubsection{Probabilistic broadcast}

In this case, the node decides
whether to forward the received message with a certain probability $\beta$. 
However, if the message has been locally generated at the node, it is sent to all node's neighbors.
If the message is forwarded, it is always sent to all neighbors.
Since we are considering a large scale network, we can neglect the first relay operation that creates the message and sends it to all its neighbors. Then, for all other nodes, the probability $f_i$ that a node $n$ relays the message to $i$ neighbors is
$f_i = \beta p_{i+1}$. In fact, with probability $\beta$, node $n$ performs a broadcast to all its neighbors (apart from the node from which it has received the message); hence, $i$ nodes will receive such message when $n$ has $(i+1)$ neighbors (with probability $p_{i+1}$).

Similarly, $\overrightarrow{f}_i = \beta q_i$, i.e., we select an edge and we arrive to a node that has an excess degree $q_i$. Thus,
$$
  \overrightarrow{F}(x) = \sum_i \beta q_i x^i = \beta \overrightarrow{G}(x).
$$

Then, $\overrightarrow{F}'(1) = \beta \overrightarrow{G}'(1) = \beta \langle q \rangle$. Surprisingly (or probably not), we obtain a formula which is identical to that obtained for the fixed probability. 
In the former case, we used a gossip probability $\gamma$, here we have a broadcast probability $\beta$, but the average amount of receivers depends on the specific value of a probabilistic parameter and on the average excess degree, which depends on the topology of the overlay. 
Summing up, according to the mathematical machinery utilized in this model, the two approaches behave in the same way.

\subsubsection{Degree Dependent Gossip}

In this case, the gossip probability depends on the degree of the nodes. Hence, if a node $m$ has degree $i$, it will receive a message from a neighbor with a probability $\gamma(i)$, which is a function of the degree.
In order to determine if $n$ relays the message to a neighbor $m$, we should sum all the possible cases that $m$ has a certain excess degree, considering the probability to relay the message to a node with such a degree, i.e.~$\Theta = \sum_j q_j \gamma(j)$. 
With this in view, the probability $f_i$ to forward a message to $i$ neighbors becomes
$$
  f_i = \Theta^i \sum_{k \geq i} p_k \binom{k}{i} (1- \Theta)^{k-i},
$$
which is obtained by considering all the possible cases of $n$, having a degree higher than $i$, which forwards the query to $i$ neighbors. Moreover, $n$ does not gossip the query to its remaining $k-i$ neighbors.
Following the same reasoning utilized for other gossip strategies, 
$$  \overrightarrow{f_i} = \Theta^i \sum_{k \geq i} q_k \binom{k}{i} (1- \Theta)^{k-i}.$$
Thus, its generating function is
\begin{eqnarray*}\label{eq:F_calcolo} 
  \overrightarrow{F}(x) & = & \sum_i \overrightarrow{f_i} x^i 
     \ = \ \sum_i \Theta^i x^i 
              \sum_{k \geq i} q_k \binom{k}{i} (1 - \Theta)^{k-i}\nonumber \\
  & = & \sum_k q_k \sum_{i =0}^k \binom{k}{i}  \Theta^i x^i (1 - \Theta)^{k-i}
   \ = \ \sum_k q_k (\Theta x + 1 - \Theta)^k\nonumber \\
  & = & \overrightarrow{G}\big(\Theta  x + 1 - \Theta \big). 
\end{eqnarray*}

Thus, $\overrightarrow{F}'(1) = \Theta \overrightarrow{G}'(1) = \Theta \langle q \rangle$, where in this case, $\Theta$ is not a simple parameter, but a function depending on the degree of nodes.

\begin{figure}[ht]
\centering
\includegraphics[width=6.5cm]{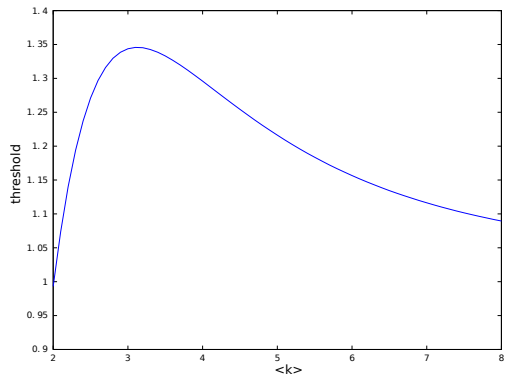}
\caption{Degree dependent gossip, function 1 -- phase transition with random graph networks of the $\alpha$ parameter.}
\label{fig:model}
\end{figure}

As an example, Figure \ref{fig:model} shows when a phase transition occurs if we consider a degree dependent gossip exploiting function $\gamma_1(i)$ of Section \ref{sec:gossip_ddg}, over random graphs. In particular, we report on the x-axis the average degree, while on the y-axis reports the $\alpha$ value employed on the formula of the gossip strategy. For that value of $\alpha$, a related degree dependent gossip function is obtained.

\section{Simulation Testbed}\label{sec:simulation}

This section presents the details of the simulator we employed to evaluate the performance metrics and assess the dissemination protocols over different network topologies.

LUNES (Large Unstructured NEtwork Simulator) is an easy-to-use tool for the simulation of complex protocols on top of large graphs of whatever topology~\cite{gda-mospas-11}. It is modular and separates into different software components the tasks of: i) network creation, ii) implementation of the protocols and iii) analysis of the results. The use of a modular approach has the advantage of permitting the (re-)use and integration of existing software tools, and facilitates the update and extensibility of the tool. The flow of data processing is linear, i.e.~a network is created by the network creation topology module; then a communication protocol is executed on top of such a network by the protocol simulator; its results are analyzed by the trace analysis module. It is worth mentioning that all such tools have been designed and implemented to work in parallel and therefore they are able to exploit all the computational resources provided by parallel (multi-processor or multi-core) or distributed (e.g.~clusters of PCs) architectures. In other words, while, for instance, a network (generated by the network creation topology module) is used by the protocol simulator, the network creation topology module may be active for the generation of another network. Similarly, while the protocol simulator is running, its outcomes from previous executions can be analyzed by the trace analysis module. Outcomes from a given module are exploited by the other one via simple template files (such as the graphviz dot language). These modules are described in isolation in the rest of the section.

Many other software tools have been used to investigate complex networks, the most popular of them is PeerSim~\cite{p2p09-peersim}. While PeerSim has demonstrated a good scalability and a comparison between LUNES and PeerSim is out of the scope of this paper, it is worth noting that the main goal of LUNES is to simulate environments that are much more data intensive with respect to the models that have been studied up to now (e.g.~for the total number of messages that must be delivered in the network). In other words, we aim to deal with simulation models that can not be addressed using a centralized (Java-based) approach such as in PeerSim. All cases in which the usage of multi-core CPUs, clusters of PCs or High Performance Computing architectures are necessary.

\subsection{Network Topology Creation}
LUNES is able to import the graph topologies generated by other tools. In the current version of LUNES, we employ \textit{igraph}, a well-known tool for creating and manipulating undirected and directed graphs~\cite{igraph}. It includes algorithms for network analysis methods and graph theory and allows handling graphs with millions of vertices and edges. The graphs generated by igraph (or other tools) can be directly used for protocol simulation or much more often are stored in ``corpuses'', that are collections of homogeneous graphs. Each corpus can be seen as a testbed environment in which the simulator compares the behavior and outcomes of protocols under exactly the same conditions. For example, given a specific dissemination protocol, different execution parameters (e.g.~dissemination probability) can be tested to find what is the best configuration with respect to the desired dissemination properties. For obtaining results that are statistically correct, the evaluation of each metric (i.e.~coverage, delay, overhead ratio) requires the execution of multiple independent runs. This means that, in LUNES, each graph that is part of a corpus is the configuration for an independent run. Under the computation complexity viewpoint, this means that the size of the problem is enormously increased with respect to a single run that tests a specific configuration of a given dissemination protocol.

It is worth pointing out that the overlay creation and management does not consider issues concerned with the underlying physical network and proximity of nodes. This is a common practice during the evaluation of a P2P protocol over an unstructured overlay \cite{gda-simutools-09,KincaidA05,Ferretti:2012,Kempe:2004,Jelasity:2005,gda-disio-11-2,Baldoni:2007,mbp-2011,p2p09-peersim,Boyd:2006}.
Indeed, adding variables related to physical networks, such as network proximity and variable delay in message transmission, would increase the complexity of the model, hence making more difficult to extract and compare some general results related to the performance of a dissemination protocol in an overlay. 
Thus, it is simpler and more effective counting the amount of hops, as a measure of the delay, rather than measuring network latencies that vary depending on the physical mapping of the overlay in a geographical network.

\subsection{Protocol Simulation} 
\label{sec:protocol_simulation}
In LUNES, the simulation services are demanded to the ART\`IS middleware and the GAIA framework~\cite{gda-simpat-2014,pads}.
This means that, in the implementation of new protocols to be simulated in LUNES, there is no need for dealing with low-level simulation details. The only Application Programming Interface (API) used in LUNES is quite high level and is provided by GAIA. Furthermore, for the implementation of new protocols LUNES already offers a set of primitives and functions that can be used and modified without the need of starting from scratch. For example, in the current version all the most common features of dissemination protocols are already implemented and adding new variants or more complex protocols is straightforward.

\subsection{Trace Analysis}
Under the performance and scalability viewpoint, the most demanding points are the protocol simulation and the traces analysis. As to the traces analysis, it has been excluded from the simulation tasks and some specific software tools have been implemented. The simulation of a network with a few hundred nodes for the time necessary for studying some common properties can generate a huge amount of simulation traces that have to be stored, parsed and analyzed (in the order of few gigabytes and, for the performance evaluation shown in this paper, up to 300 million of delivered messages per run). This means that, very simple metrics used for performance evaluation of the simulated protocol can require a lot of effort. In the current version of LUNES, this task is implemented using a set of shell scripts and some specific tools that have been implemented in C language for efficiency. This mix is both quite efficient and easy to extend and personalize. We have intentionally avoided to build a monolithic application to provide users with an easily customizable tool.

\subsection{Time Evolution}
LUNES exploits a time-stepped approach to perform simulations. A reason is that this choice simplifies the deployment of the simulation over parallel and distributed simulation architectures. Moreover, it allows exploiting the load balancing techniques and simulation entities migration approaches offered by the ART\`IS middleware and the GAIA framework. This means that, in order to simulate asynchronous scenarios using time-stepped simulations, a reasonable time granularity must be identified, so that the size of each timeslot allows properly handling successive events that occurs in time, thus guaranteeing a correct event ordering for subsequent events.

As concerns P2P systems and complex networks, it is quite common to simulate them using a time-stepped approach \cite{gda-simutools-09,KincaidA05,Ferretti:2012,Jelasity:2005,gda-disio-11-2,Baldoni:2007,mbp-2011,p2p09-peersim,Boyd:2006}.

\section{Performance Evaluation}\label{sec:performance}

In this section, the different dissemination protocols are evaluated on top of the graph topologies previously introduced in Section~\ref{sec:background}. Each different topology is studied by means of a corpus that is composed of 10 graphs generated using the igraph generators reported in Table~\ref{tab:igraph_generators}. For all the graphs in this section, each point is obtained averaging the results on the graphs of the specific corpus. Confidence intervals are very narrow and are left out from the figures for better readability.

\begin{table}[ht]
\begin{center}
\begin{tabular}{ |l|l|c|c|c| }
\hline
\multicolumn{5}{|c|}{igraph generators} \\
\hline
& method & nodes & edges per node & edges \\
\hline
\hline
 Random & igraph\_erdos\_renyi\_game & 500 & 2, 3, 4 & 1000, 1500, 2000 \\
 Scale-free & igraph\_barabasi\_game & 500 & 2, 3, 4 & 1000, 1500, 2000 \\
 Small-world & igraph\_watts\_strogatz\_game & 500 & 2, 3, 4 & 997, 1494, 1990 \\
 k-regular & igraph\_k\_regular\_game & 500 & 4, 6, 8 & 1000, 1500, 2000 \\
\hline
\end{tabular}
\end{center}
\caption{igraph generators used for building the graph corpuses, main parameters.}
\label{tab:igraph_generators}
\end{table}

Each simulation run is 1000 timesteps long, an amount of steps that is necessary to avoid transient effects (e.g.~the cache efficiency). Each node in the network can generate new messages during the whole simulation lifespan except for the last $TTL$ timesteps. If nodes are allowed to generate new messages in the very last timesteps then these messages would not have the possibility to reach their destination before the simulation ends. Clearly, this must be avoided because it would affect the measured coverage. The time between successive messages is generated according to a typical exponential distribution (mean value 10 timesteps in this performance evaluation). As already mentioned, each node implements a cache structure with the aim of reducing the number of duplicate messages in the network. This cache is managed using the Least Recently Used (LRU) replacement algorithm; the cache size is one of the parameters that will be studied in the following. The study of more specific (and efficient) cache replacement algorithms is left as future work.

Another main model parameter is the lifetime of each message in the network (Time-To-Live, TTL). We study how this parameter influences the protocols performance but, as a rule of thumb, when not differently stated it is equal to the 130\% of the max diameter of all the graphs in a given corpus. It is clear that a TTL that is lower than the graph diameter does not permit a full coverage and conversely, a too  large TTL could lead to some extra overhead. Given its main importance, this last point will be addressed in the first part of the following performance evaluation.\\

To foster the reproducibility of our experiments, all the source code used in this performance evaluation, and the raw data obtained in the experiments execution, are freely available on the research group website~\cite{pads}.

\subsection{Random Graph Networks}

\begin{figure}[ht]
\centering
\subfloat[\label{fig:random_1000edges-coverage_FP-CB}]{\includegraphics[angle=270,width=6.5cm]{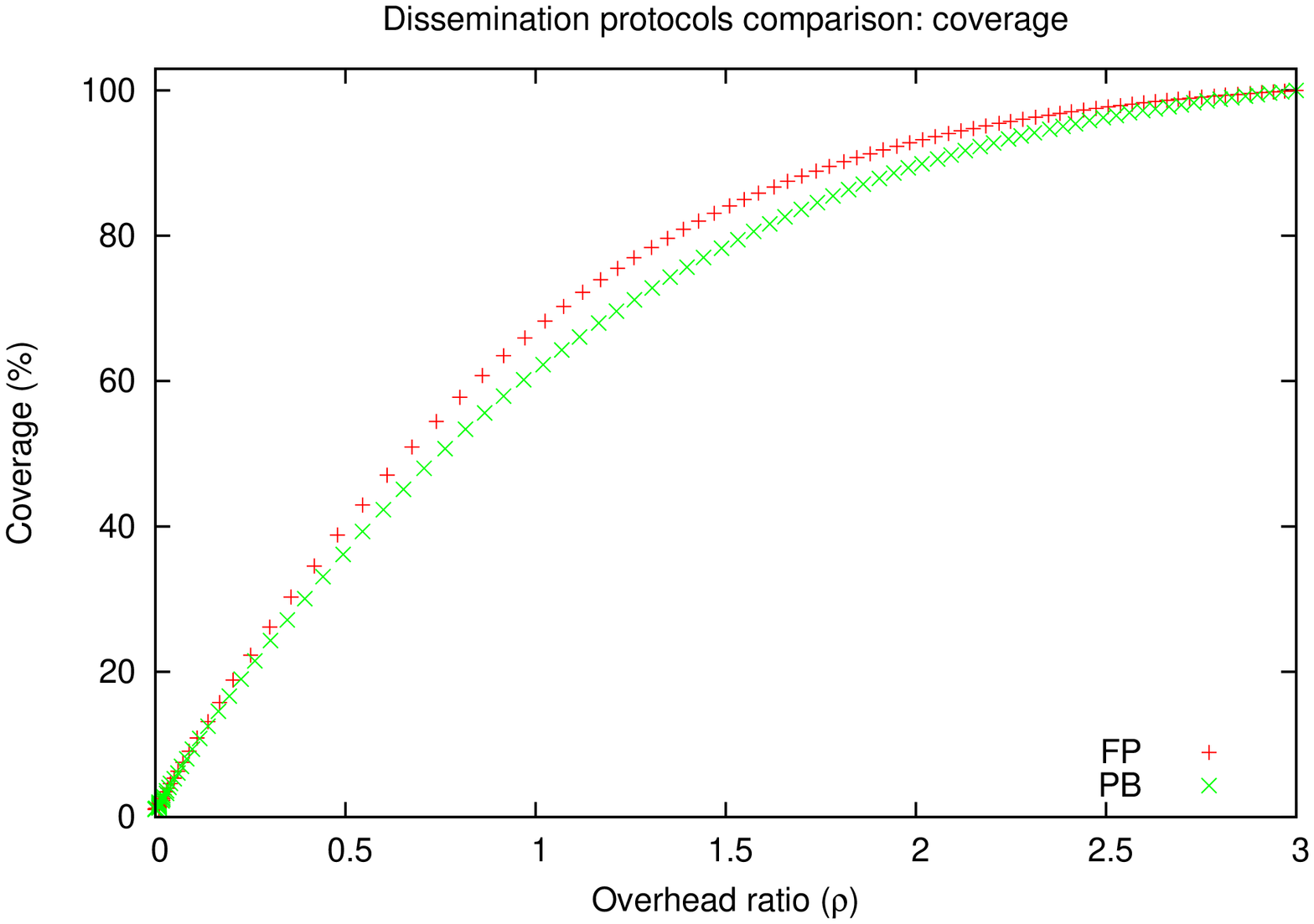}}
\subfloat[\label{fig:random_1000edges-delay_FP-CB}]{\includegraphics[angle=270,width=6.5cm]{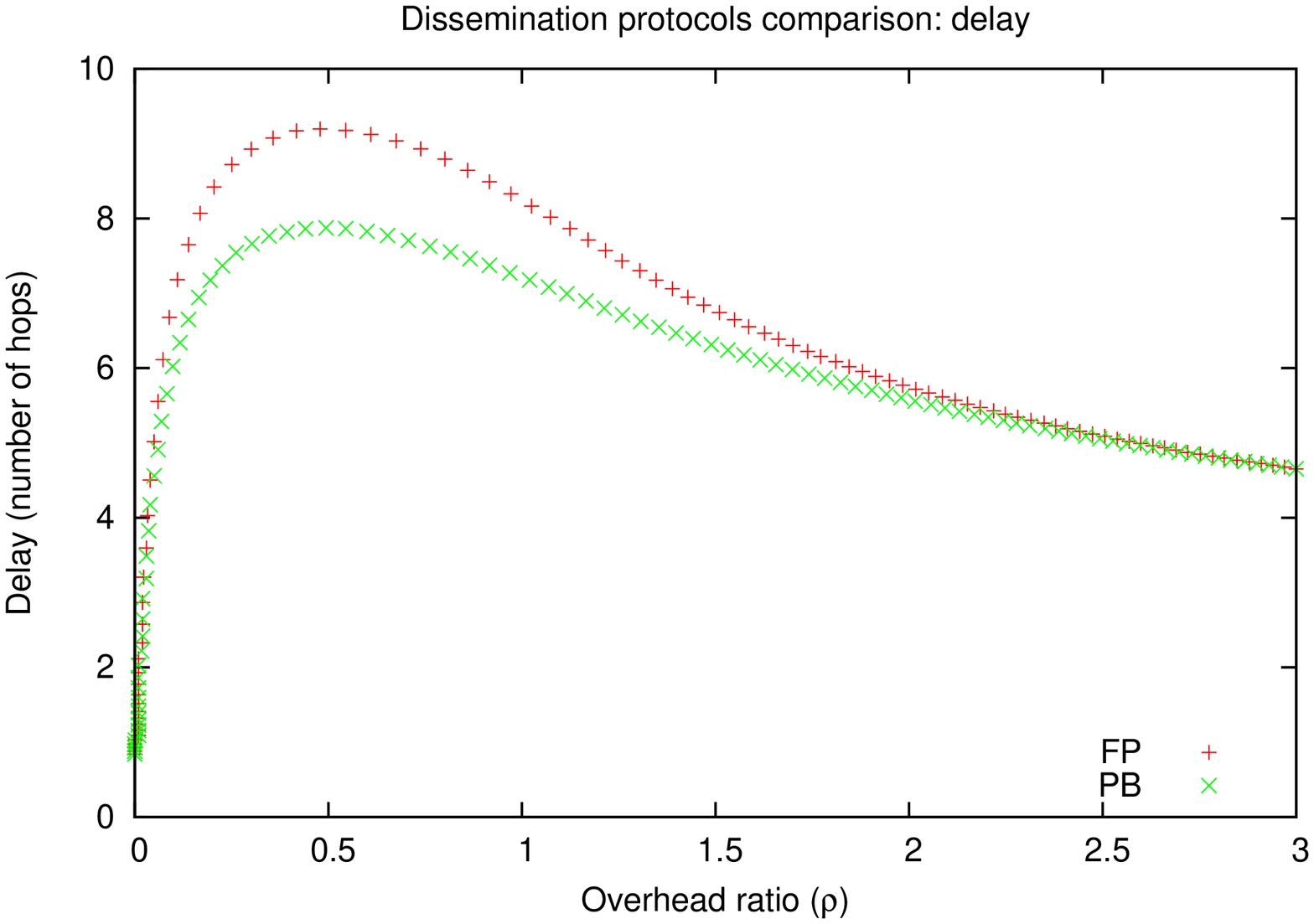}}
\caption{Random graph networks, 500 nodes, 1000 edges, max diameter=10, TTL=16, cache=256.}
\label{fig:random_1000edges}
\subfloat[\label{fig:random_1500edges-coverage_FP-CB}]{\includegraphics[angle=270,width=6.5cm]{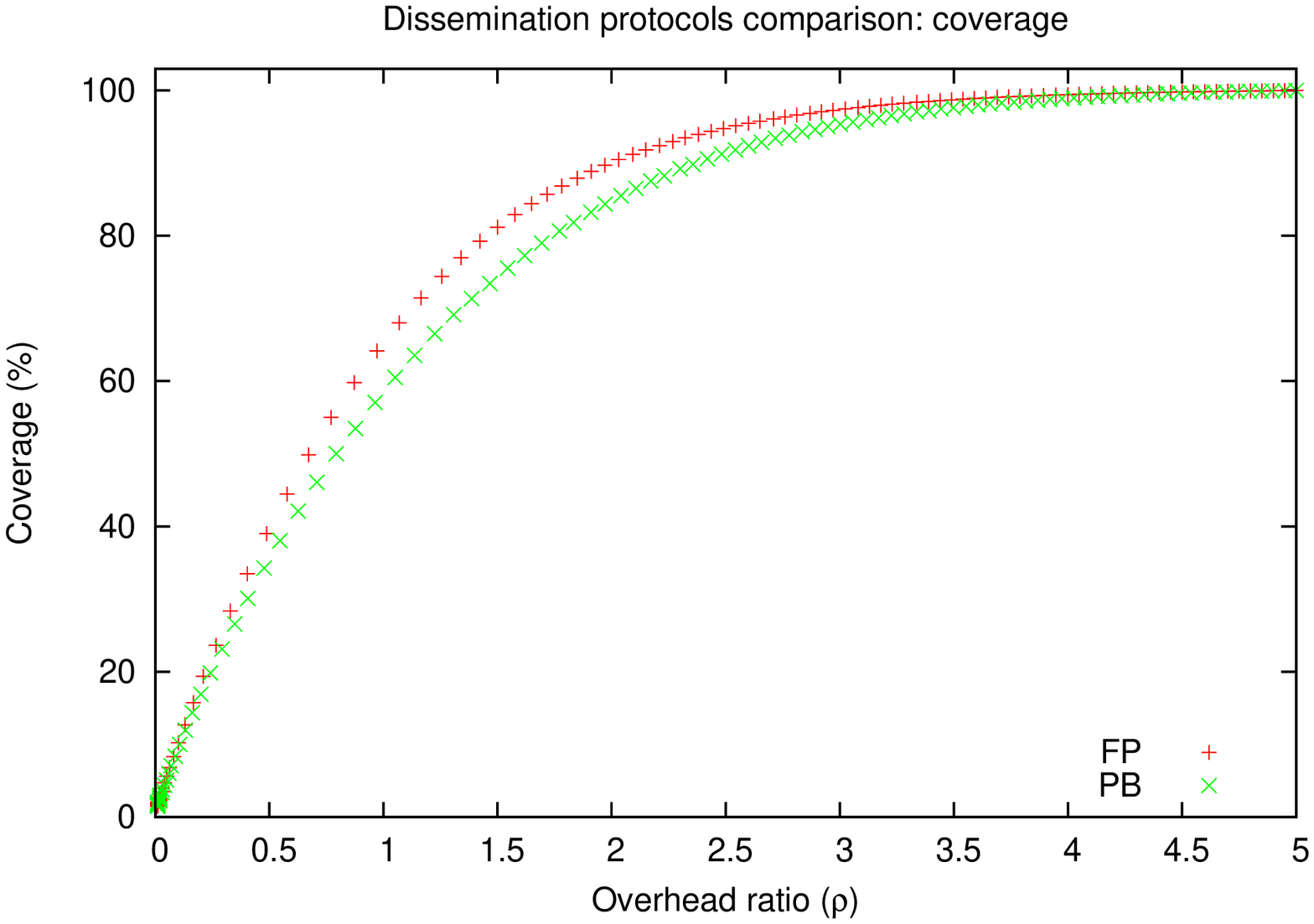}}
\subfloat[\label{fig:random_1500edges-delay_FP-CB}]{\includegraphics[angle=270,width=6.5cm]{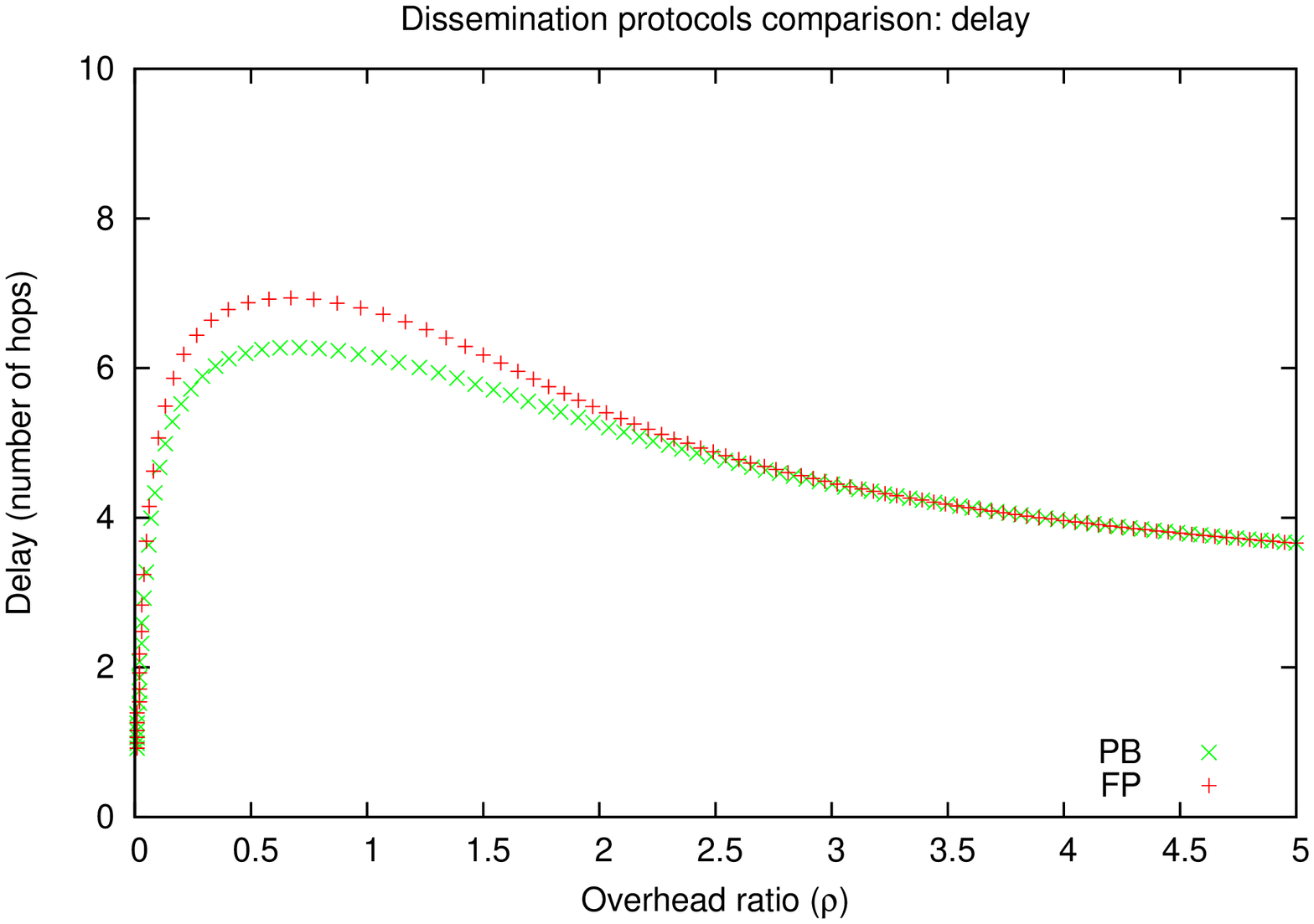}}
\caption{Random graph networks, 500 nodes, 1500 edges, max diameter=7, TTL=10, cache=256.}
\label{fig:random_1500edges}
\subfloat[\label{fig:random_2000edges-coverage_FP-CB}]{\includegraphics[angle=270,width=6.5cm]{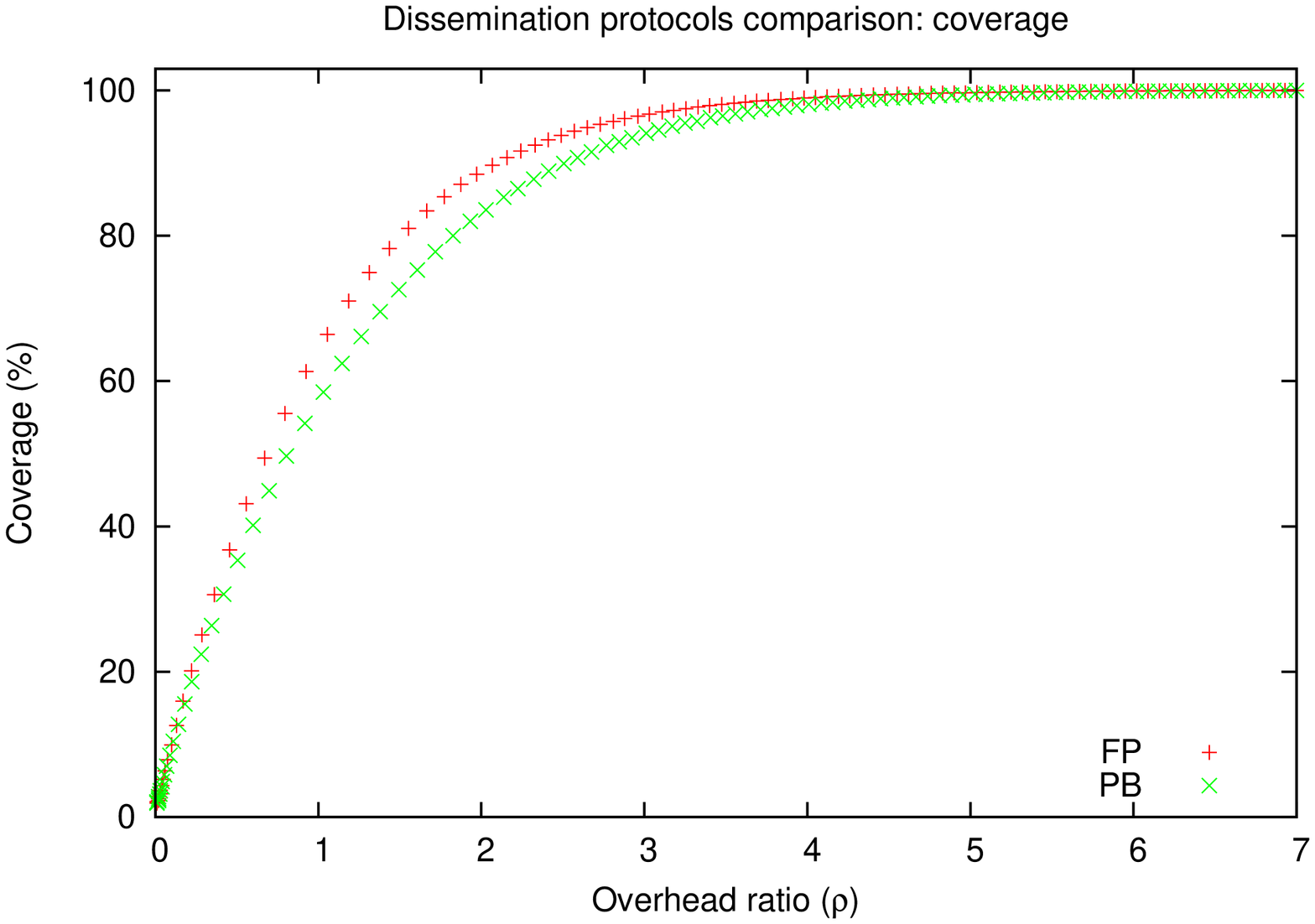}}
\subfloat[\label{fig:random_2000edges-delay_FP-CB}]{\includegraphics[angle=270,width=6.5cm]{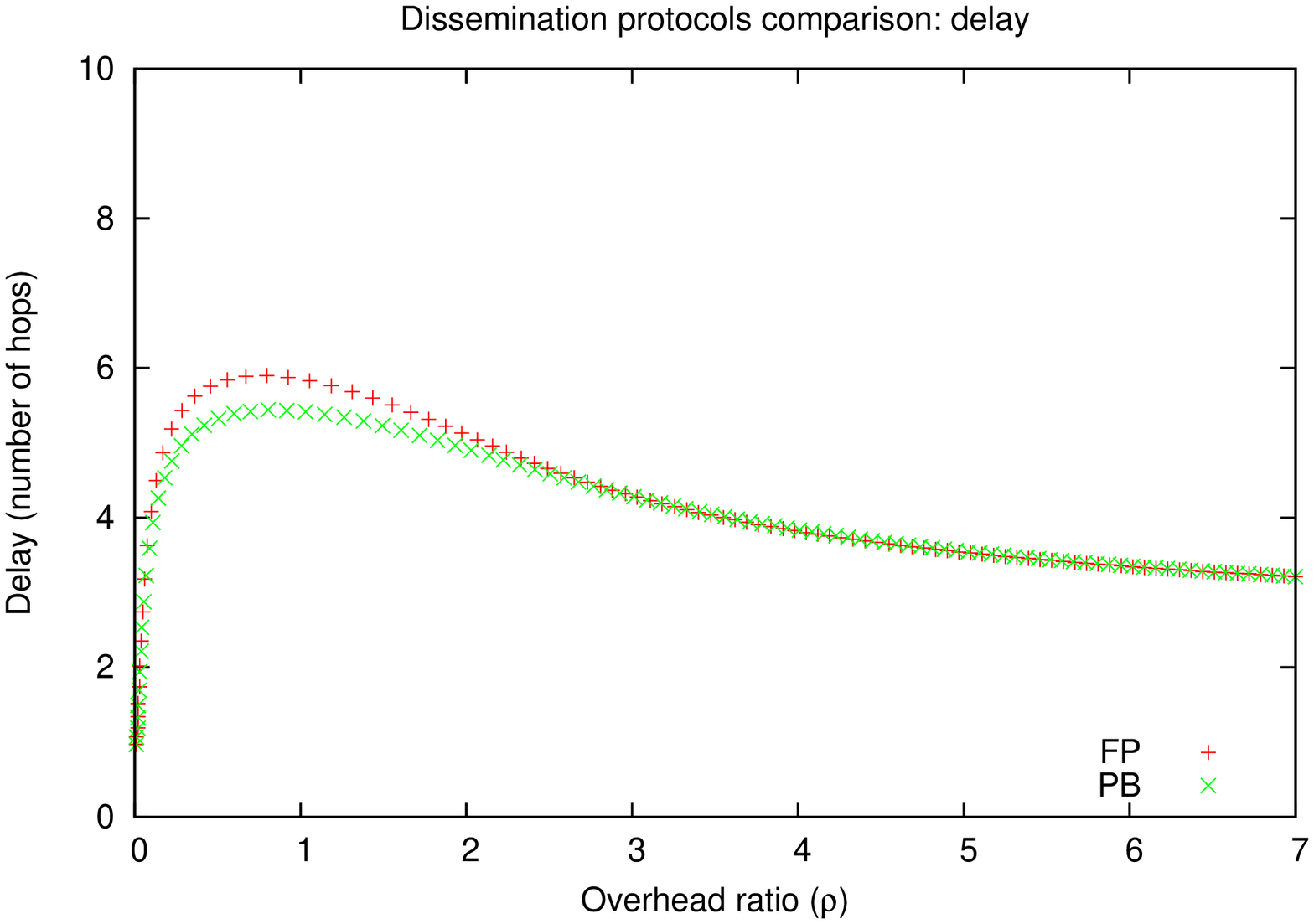}}
\caption{Random graph networks, 500 nodes, 2000 edges, max diameter=6, TTL=8, cache=256.}
\label{fig:random_2000edges}
\end{figure}

Figures~\ref{fig:random_1000edges},~\ref{fig:random_1500edges} and \ref{fig:random_2000edges} show the behavior of the Fixed Probability (FP) and Probabilistic Broadcast (PB) protocols on random graphs with 1000, 1500 and 2000 edges. On the left side of each figure, we show the coverage obtained with a given overhead ratio, while on the right side we show the corresponding delay. In both cases, all the results are obtained using an indirect method. In fact, FP and PB are executed varying the dissemination probability and collecting the resulting overhead ratio, coverage and delay. This means that for each set of figures (i.e.~(a) and (b)), $1000$ simulation runs have been executed (that is $100$ different dissemination probabilities and for each of them $10$ multiple independent runs). In this first set of experiments, the cache size has been set to 256 positions. 

Both algorithms are able to provide full dissemination (i.e.~100\% coverage) but FP has higher coverage for medium overhead values. Because of this, for applications that do not require a full coverage, FP is a better choice than PB. On the other hand, the downside of FP is the slightly higher delay with respect to PB. 
Indeed, a reduced overhead corresponds to a reduced amount of message copies (for the same data) being transmitted in the overlay and a higher delay. In fact, the higher the amount of copies of the same message spread through the network, the lower the delay (number of hops to reach a given node). 

Increasing the number of edges, the overhead required to obtain full dissemination also increases. 
This is due to the fact that the higher the number of edges in the network, the higher the average node degrees, and thus, the higher the amount of duplicate messages generated by the dissemination. 
Clearly, the caching strategy reduces the retransmission of duplicates (since it avoids retransmission of cached messages), yet without solving the problem entirely. 
However, a message can travel through different paths to reach a node, and increasing the amount of edges in a random graph increases proportionally the probability of this situation.

In Figures~\ref{fig:random_1000edges},~\ref{fig:random_1500edges} and \ref{fig:random_2000edges}, the (b) charts show a rapid increment of the delay (depending on the overhead) and a progressive decrement. In general, this trend is common to all the experiments we made.
An explanation seems to be as follows. With low overheads, only few paths are enabled for sent messages, with local disseminations; this results in short delays for transmitted messages. Indeed, by looking at the (a) charts in these figures, with such overheads on average a message is able to reach less than the $50\%$ of network nodes.
Then, the higher the overhead the higher the probability of dissemination and the higher the path lengths of transmitted messages. Thus, we measure higher delays.
However, if we further increase the overhead, we have a tipping point above which the higher dissemination probability allows messages to travel through alternative paths in the network with the effect of lowering on average the delay path between nodes pairs.

\begin{figure}[ht]
\centering
\subfloat[\label{fig:random_1000edges-coverage_cache}]{\includegraphics[angle=270,width=6.5cm]{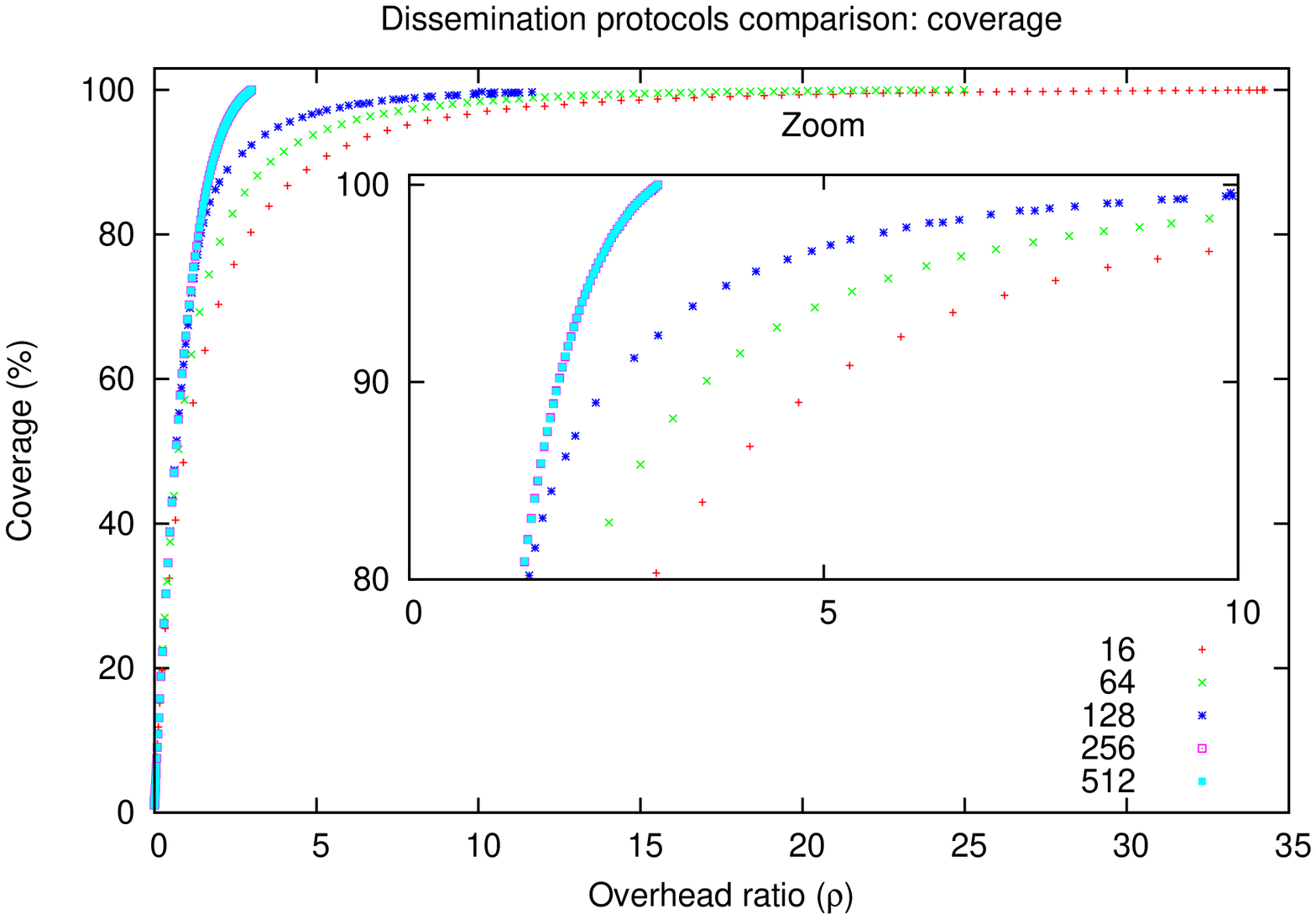}}
\subfloat[\label{fig:random_1000edges-coverage_ttl}]{\includegraphics[angle=270,width=6.5cm]{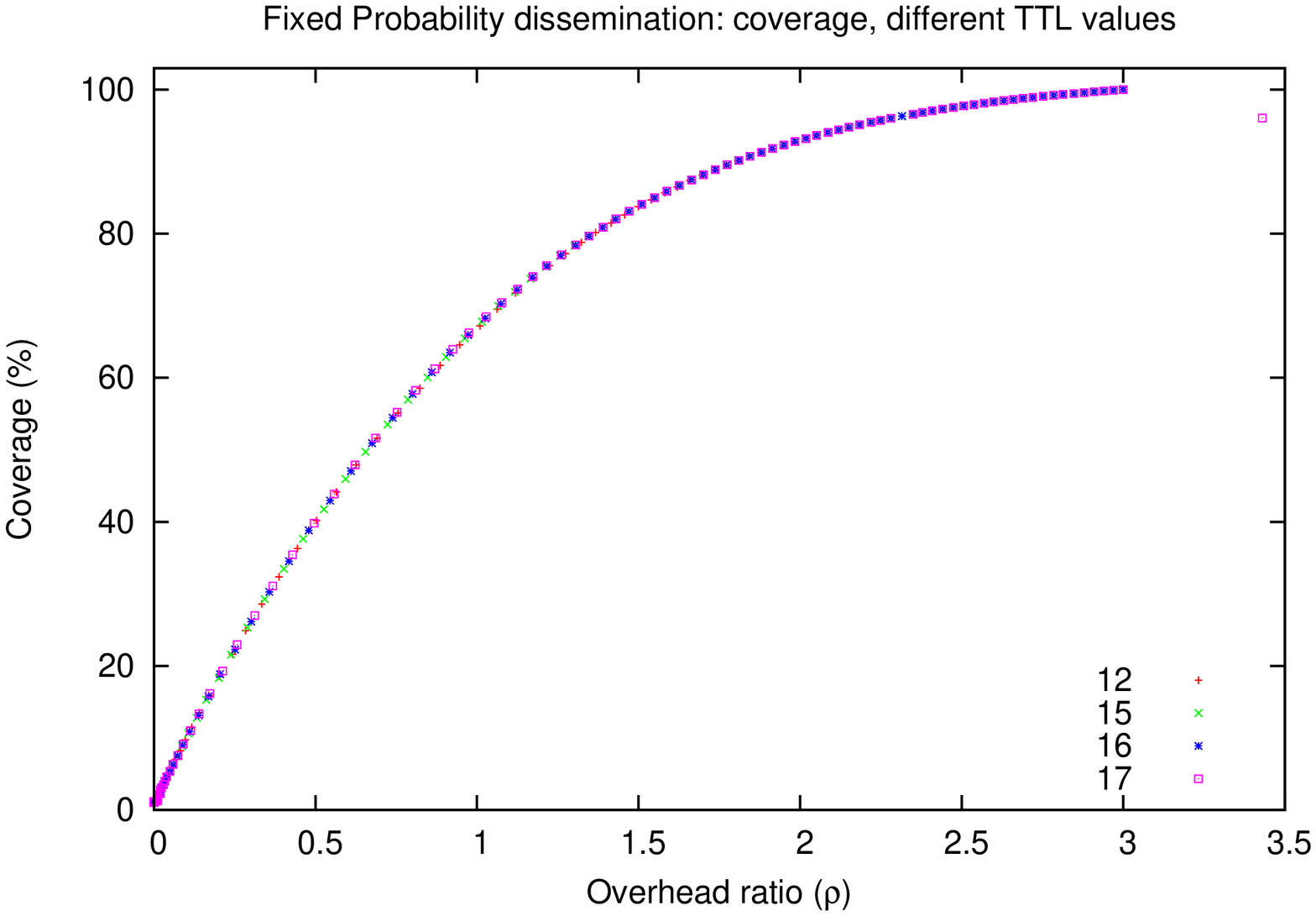}}
\caption{Random graph networks, 500 nodes, 1000 edges, max diameter=10. a) TTL=16, b) cache=256. FP dissemination protocol.}
\end{figure}

Figure~\ref{fig:random_1000edges-coverage_cache} shows the effect of different cache sizes on the FP protocol; the testbed is a random graph network composed of 500 nodes and 1000 edges. It is obvious that the cache efficiency has a big effect on the dissemination overhead. In fact, the role of the cache is to ``absorb'' as many as possible duplicated messages. If a cache is too small with respect to the amount of different messages that are in the network then all these duplicated messages will continue to flow until their TTL becomes $0$; then they are discarded. For example, with a cache of 16 positions the overhead to obtain full coverage is $34.24$. Increasing the cache size has the effect to lower the overhead for the coverage level. Obviously, the cache efficiency has a limit. In fact, the points for cache size $256$ and $512$ are overlapped. This means that this cache size, in this specific scenario, is enough to obtain the best possible caching effect (all the duplicated messages are absorbed by the cache). In the following of this section, all experiments will be executed with $cache=256$ positions. Even if only few bytes are necessary to identify each generated message in a dissemination, we don't think that a feasible approach would be to store (in each node) all the unique messages that have reached the nodes. In our view, this approach would limit the scalability with respect to the number of nodes and generated messages.
 
The TTL dimensioning is another key point in the dissemination protocols evaluation. As already mentioned, a TTL value that is lower than the graph diameter would make impossible to obtain full dissemination; but what is the effect of an excessively large TTL value? The answer it that it depends on the cache efficiency. If the cache is unable to absorb duplicate messages then the larger the TTL the longer the messages will stay in the network (i.e.~increasing the overhead). Figure~\ref{fig:random_1000edges-coverage_ttl} shows that when the cache size is adequate (e.g.~$256$ positions) the effect of TTL on the overhead is negligible. This means that setting the TTL to 130\% of the max diameter of all the graphs in a given corpus is a correct assumption.\\

\begin{figure}[ht]
\centering
\subfloat[\label{fig:random_1000edges-coverage_FP-DDL-DDP}]{\includegraphics[angle=270,width=6.5cm]{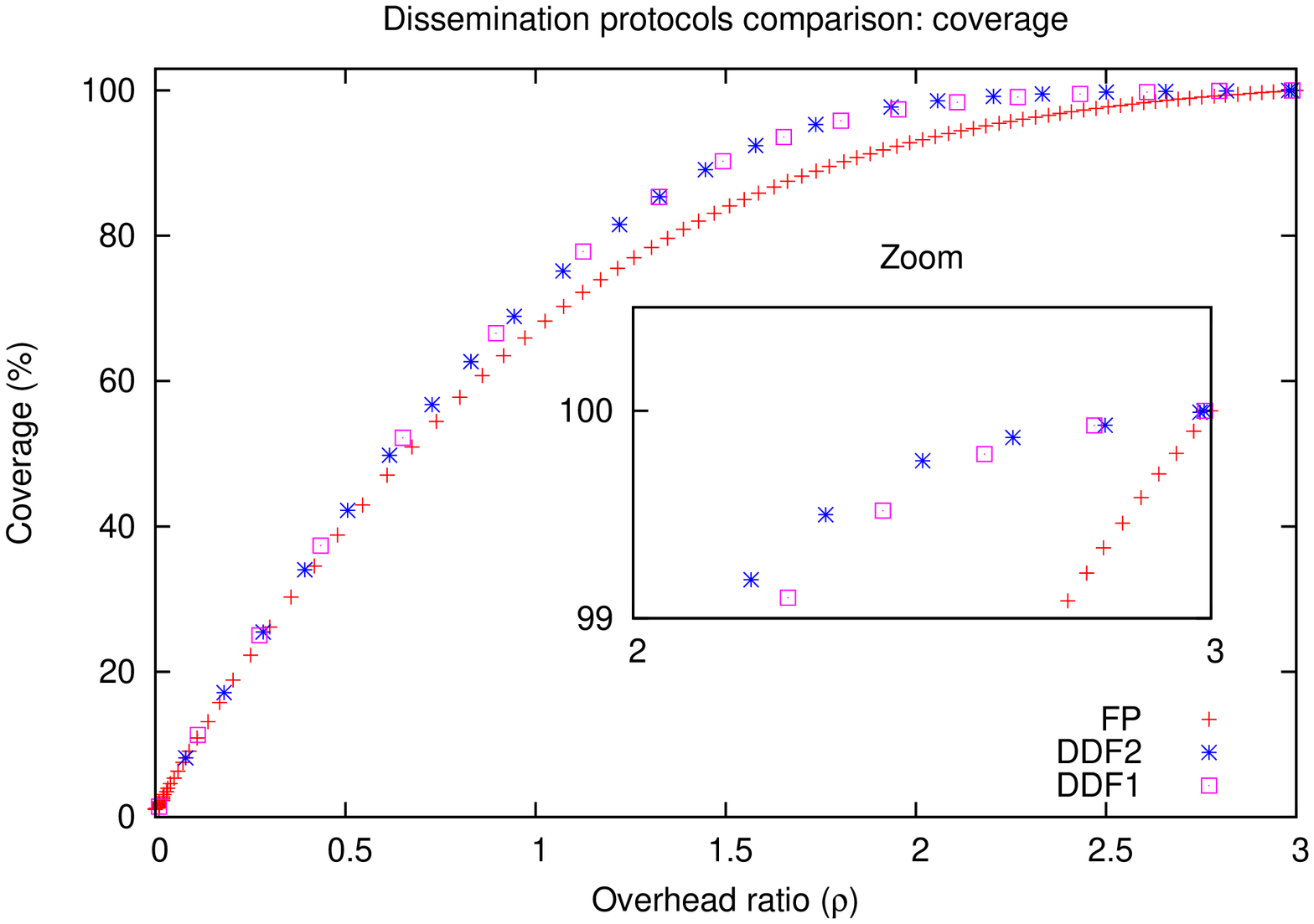}}
\subfloat[\label{fig:random_1000edges-delay_FP-DDL-DDP}]{\includegraphics[angle=270,width=6.5cm]{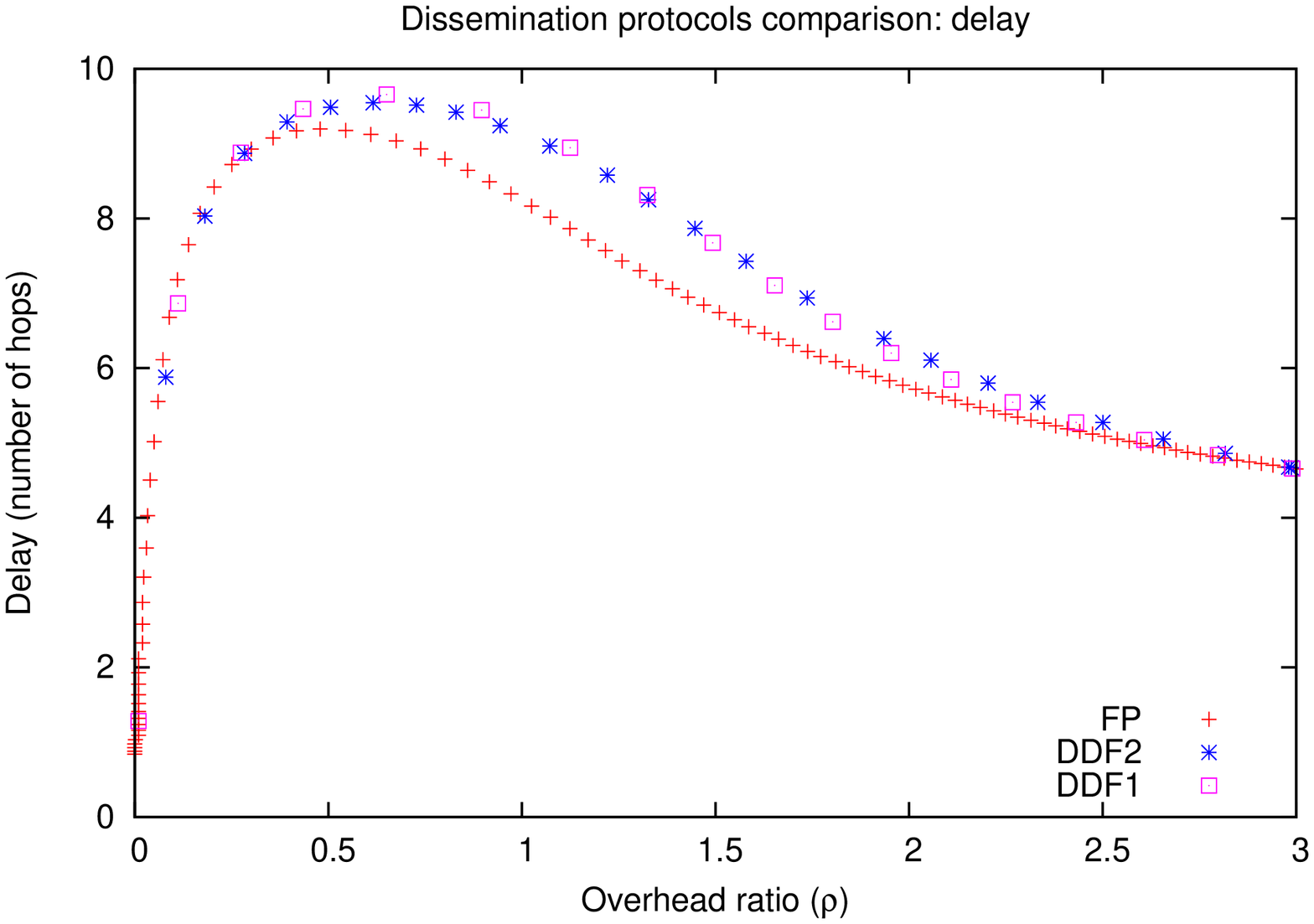}}
\caption{Random graph networks, 500 nodes, 1000 edges, max diameter=10, TTL=16, cache=256.}
\label{fig:random_1000edges-extra}
\subfloat[\label{fig:random_1500edges-coverage_FP-DDL-DDP}]{\includegraphics[angle=270,width=6.5cm]{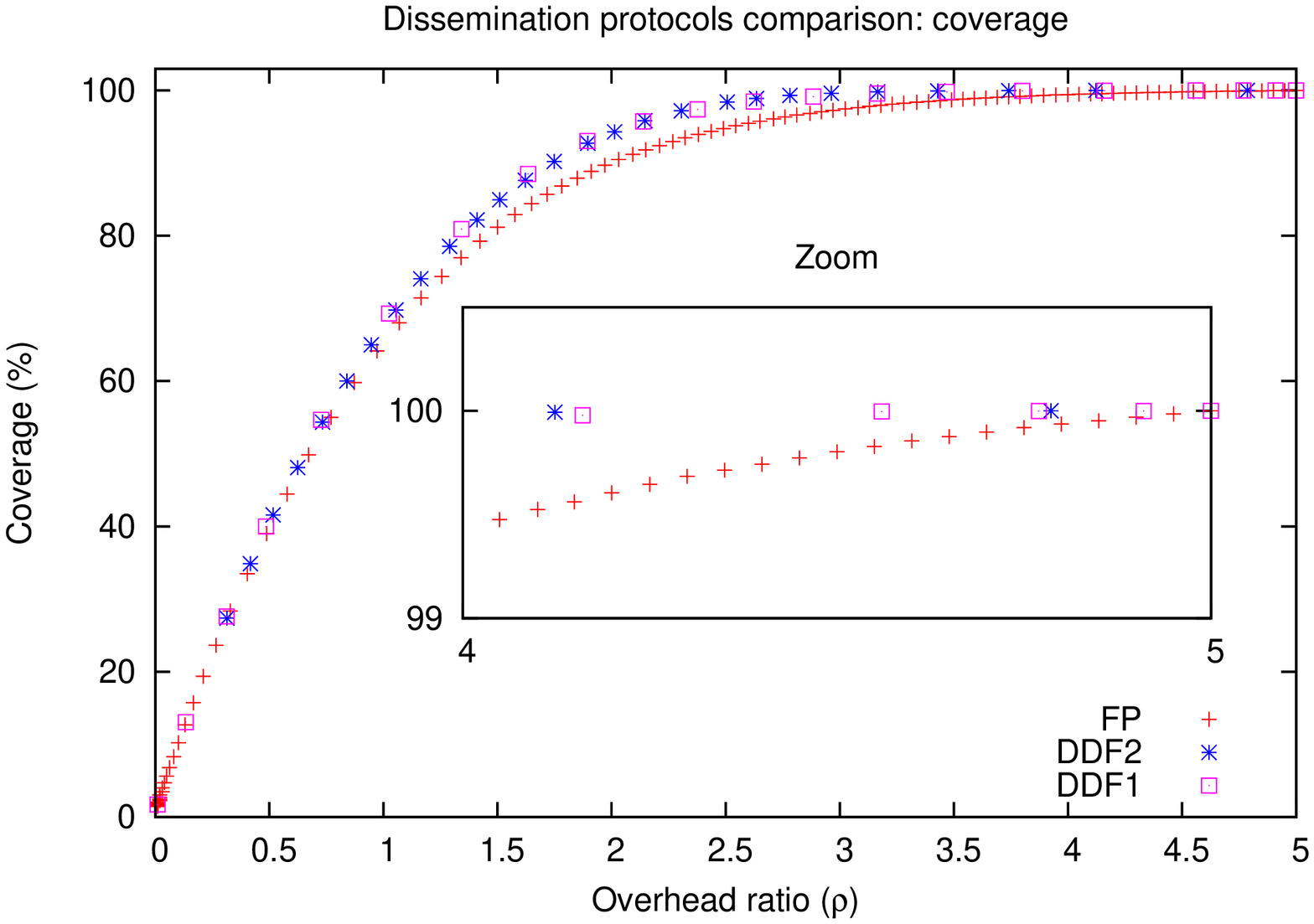}}
\subfloat[\label{fig:random_1500edges-delay_FP-DDL-DDP}]{\includegraphics[angle=270,width=6.5cm]{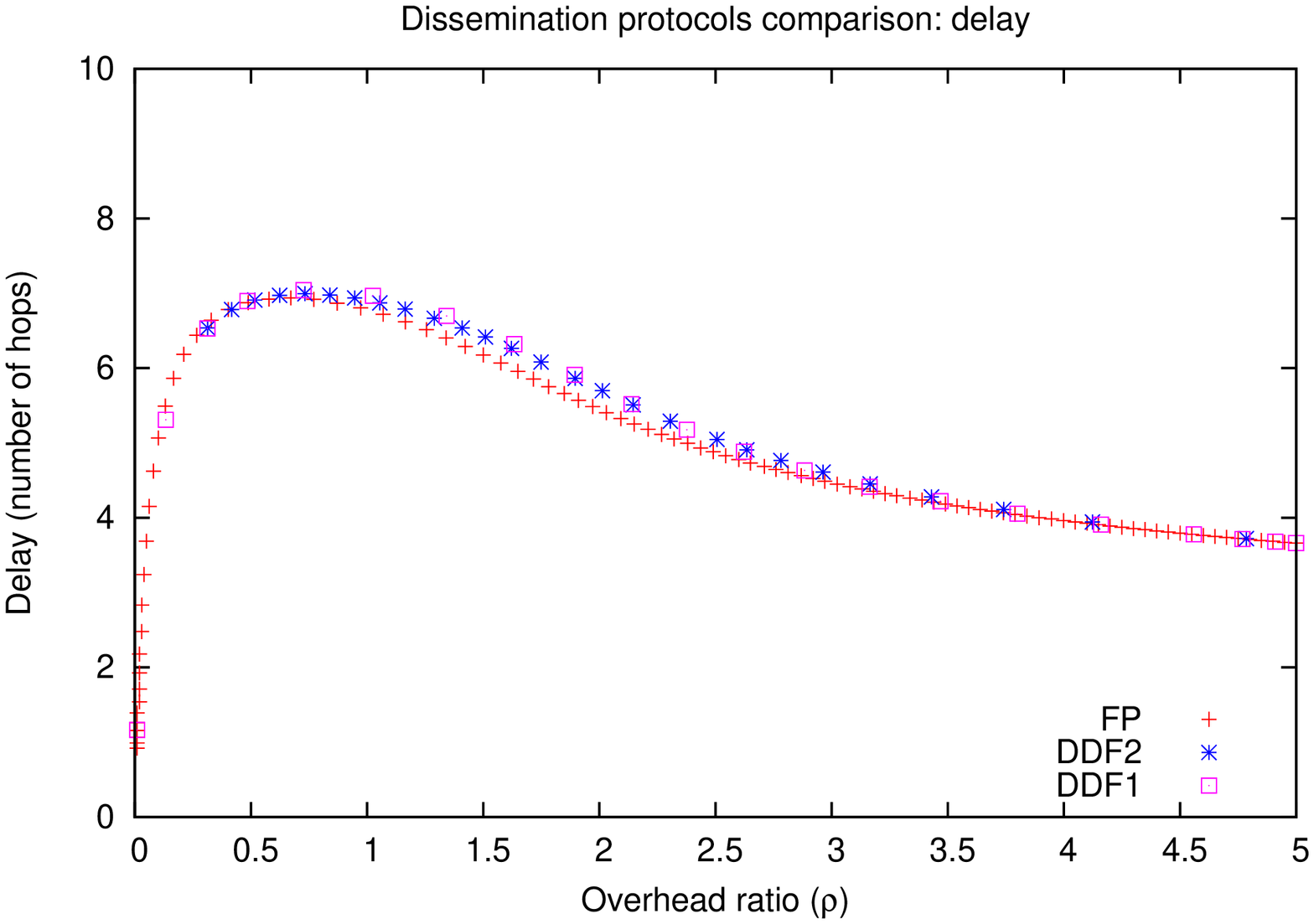}}
\caption{Random graph networks, 500 nodes, 1500 edges, max diameter=7, TTL=10, cache=256.}
\label{fig:random_1500edges-extra}
\subfloat[\label{fig:random_2000edges-coverage_FP-DDL-DDP}]{\includegraphics[angle=270,width=6.5cm]{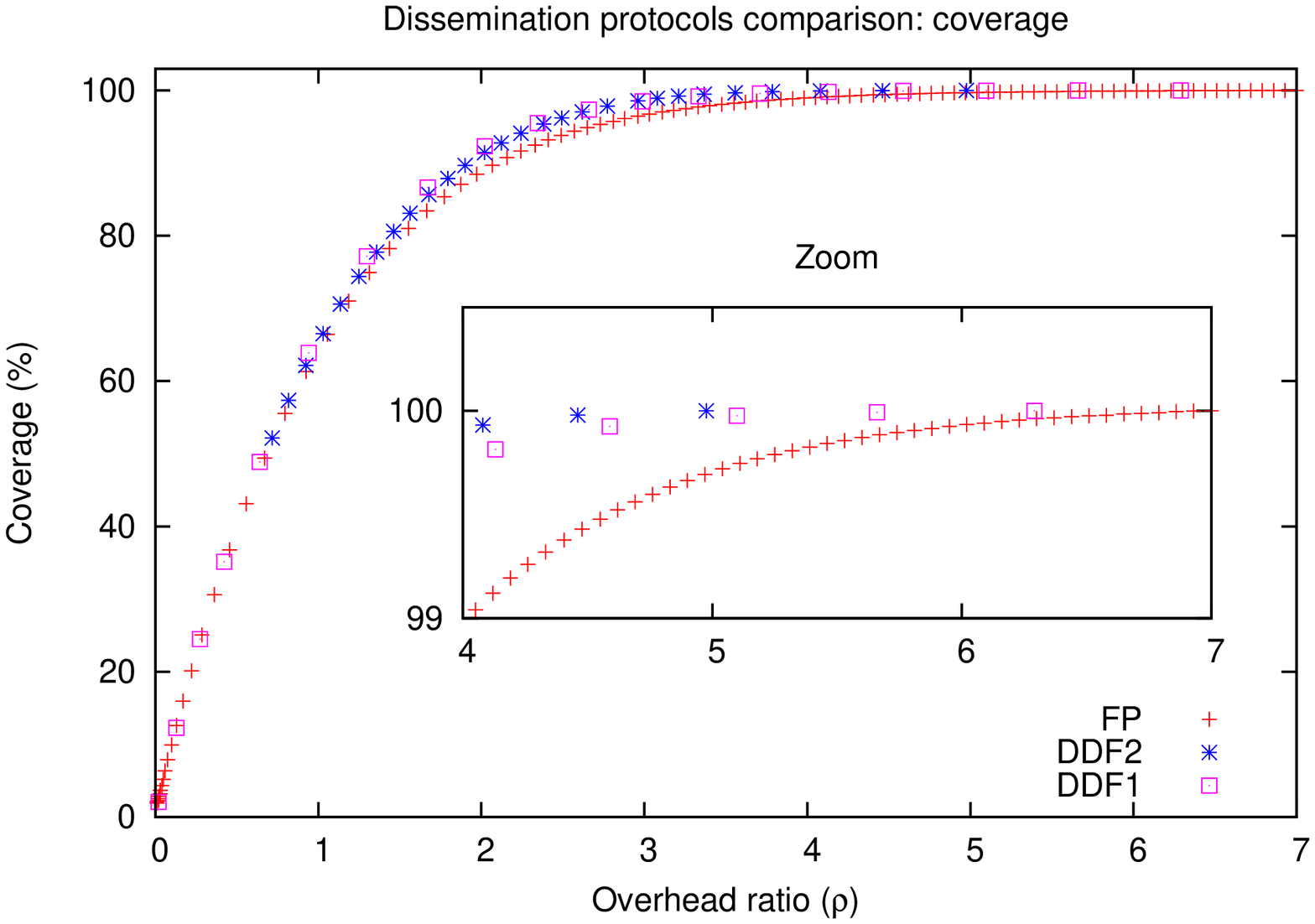}}
\subfloat[\label{fig:random_2000edges-delay_FP-DDL-DDP}]{\includegraphics[angle=270,width=6.5cm]{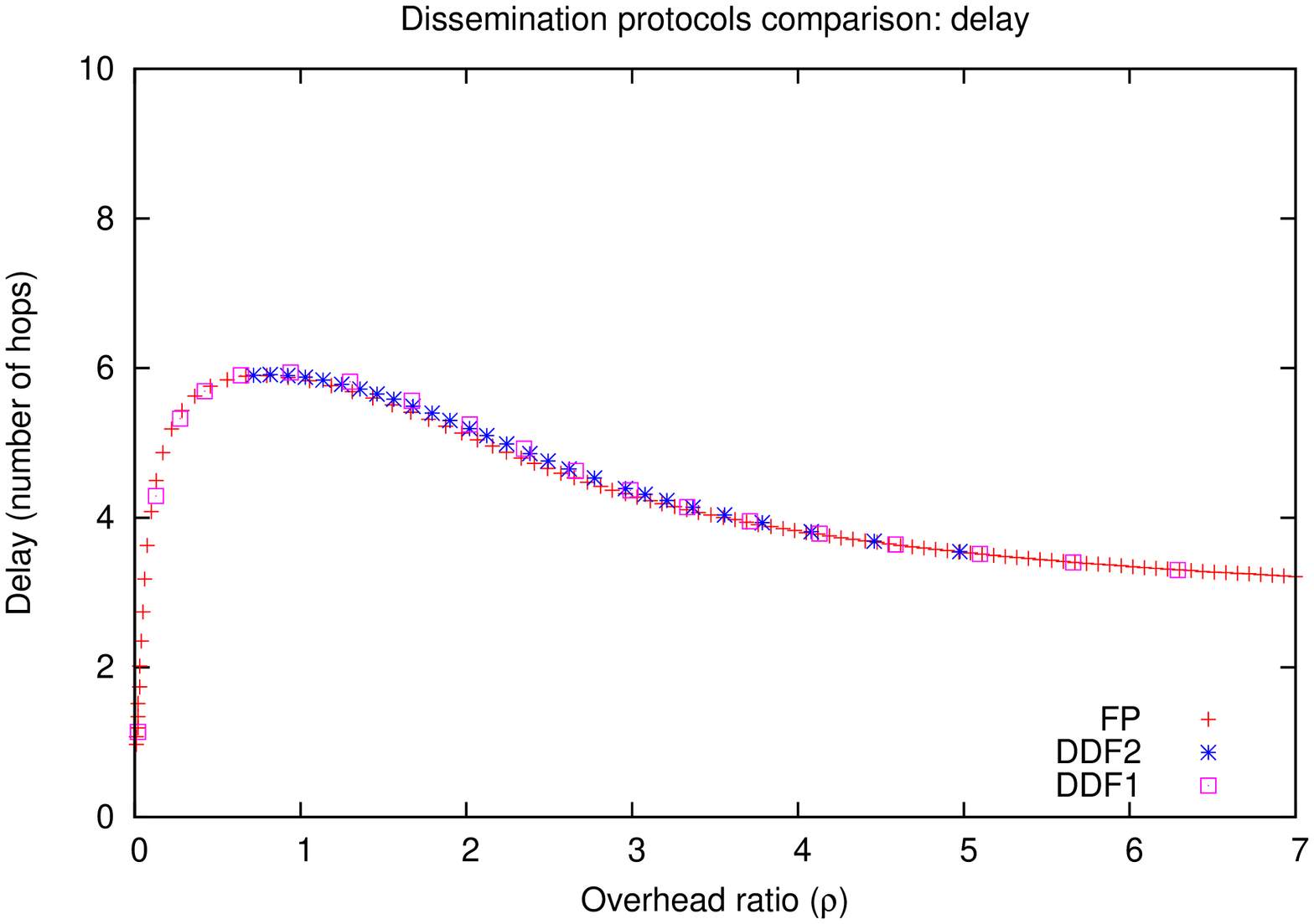}}
\caption{Random graph networks, 500 nodes, 2000 edges, max diameter=6, TTL=8, cache=256.}
\label{fig:random_2000edges-extra}
\end{figure}

In Figures~\ref{fig:random_1000edges-extra},~\ref{fig:random_1500edges-extra} and \ref{fig:random_2000edges-extra}, the FP dissemination is compared with DDF1 and DDF2. In all cases, the degree dependent protocols perform better than FP. 
Table~\ref{tab:random_1000edges} reports the overhead (and delay) that is necessary to the different dissemination protocols for obtaining a given level coverage (100\%, 99\%, 90\%, 75\%) in a random graph with 1000 edges. All protocols have almost the same overhead for a full dissemination. Both DDF1 and DDF2 are better for partial coverages.

\begin{table}[ht]
\begin{center}
\begin{tabular}{ |p{2cm}||p{2cm}|p{2cm}|p{2cm}|p{2cm}|  }
 \hline
 \multicolumn{5}{|c|}{Overhead (and delay) for a given coverage} \\
 \hline
 Algorithm & 100.0\% & 99.0\% & 90.0\% & 75.0\% \\
 \hline
 \hline
 FP   & 3.00 (4.65) & 2.74 (4.86) & 1.80 (6.10) & 1.20 (7.62) \\
 PB   & 3.00 (4.65) & 2.84 (4.76) & 2.03 (5.54) & 1.38 (6.49) \\
 DDF1  & 2.99 (4.66) & 2.24 (5.59) & 1.48 (7.72) & 1.06 (9.11) \\
 DDF2  & 2.99 (4.67) & 2.16 (5.88) & 1.48 (7.74) & 1.07 (8.99) \\
 \hline
\end{tabular}
\end{center}
\caption{Random graph networks, 500 nodes, 1000 edges, max diameter=10, TTL=16, cache=256.}
\label{tab:random_1000edges}
\end{table}

Increasing the number of edges per node, the gap between the algorithms seems to be reduced but DDF2 is slightly better than other protocols (Table~\ref{tab:random_1500edges}).

\begin{table}[ht]
\begin{center}
\begin{tabular}{ |p{2cm}||p{2cm}|p{2cm}| }
 \hline
 \multicolumn{3}{|c|}{Overhead (and delay) for a given coverage} \\
 \hline
 Algorithm & 100.0\% & \\
 \hline
 \hline
 FP   & 5.00 (3.66) & +4.38\% \\
 PB   & 5.00 (4.91) & +4.38\% \\
 DDF1  & 5.00 (3.66) & +4.38\% \\
 DDF2  & 4.79 (3.72) & best \\
 \hline
\end{tabular}
\end{center}
\caption{Random graph networks, 500 nodes, 1500 edges, max diameter=7, TTL=10, cache=256.}
\label{tab:random_1500edges}
\end{table}

With 2000 edges (Table~\ref{tab:random_2000edges}) both DDF1 and DDF2 are better than FP and PB. In this case, the overhead needed by DDF2 to get full coverage is 40.56\% lower than FP and PB. As usual, the overhead reduction is obtained at the cost of a moderate increase in the delay (+10.59\%).
These results suggest that a degree dependent gossip strategy should be used when coverage is the main metric to pursue. In fact, given a certain overhead, degree dependent strategies provide higher coverages, yet at the cost of a slightly higher delay.

Also in the case of DDF protocols, the results shown in the figures are obtained using an indirect method. In this case, it has been varied the $\alpha$ parameter (see Section~\ref{sec:gossip_ddg}) with the goal to obtain full coverage with the lowest possible overhead. This is done, using a sampling of the possible values of $\alpha$ and a fine-grained exploration of the values near the point of interest (that is where there is full coverage). Again, the number of required runs for each evaluation is huge and therefore a high
simulator efficiency is a prerequisite.

\begin{table}[ht]
\begin{center}
\begin{tabular}{ |p{2cm}||p{2cm}|p{2cm}| }
 \hline
 \multicolumn{3}{|c|}{Overhead (and delay) for a given coverage} \\
 \hline
 Algorithm & 100.0\% & \\
 \hline
 \hline
 FP   & 7.00 (3.21) & +40.56\% \\
 PB   & 7.00 (3.21) & +40.56\% \\
 DDF1  & 6.29 (3.30) & +26.30\% \\
 DDF2  & 4.98 (3.55) & best \\
 \hline
\end{tabular}
\end{center}
\caption{Random graph networks, 500 nodes, 2000 edges, max diameter=6, TTL=8, cache=256.}
\label{tab:random_2000edges}
\end{table}

\subsection{Scale-Free Networks}

\begin{figure}[ht]
\centering
\subfloat[\label{fig:scalefree_997edges-coverage_FP-CB}]{\includegraphics[angle=270,width=6.5cm]{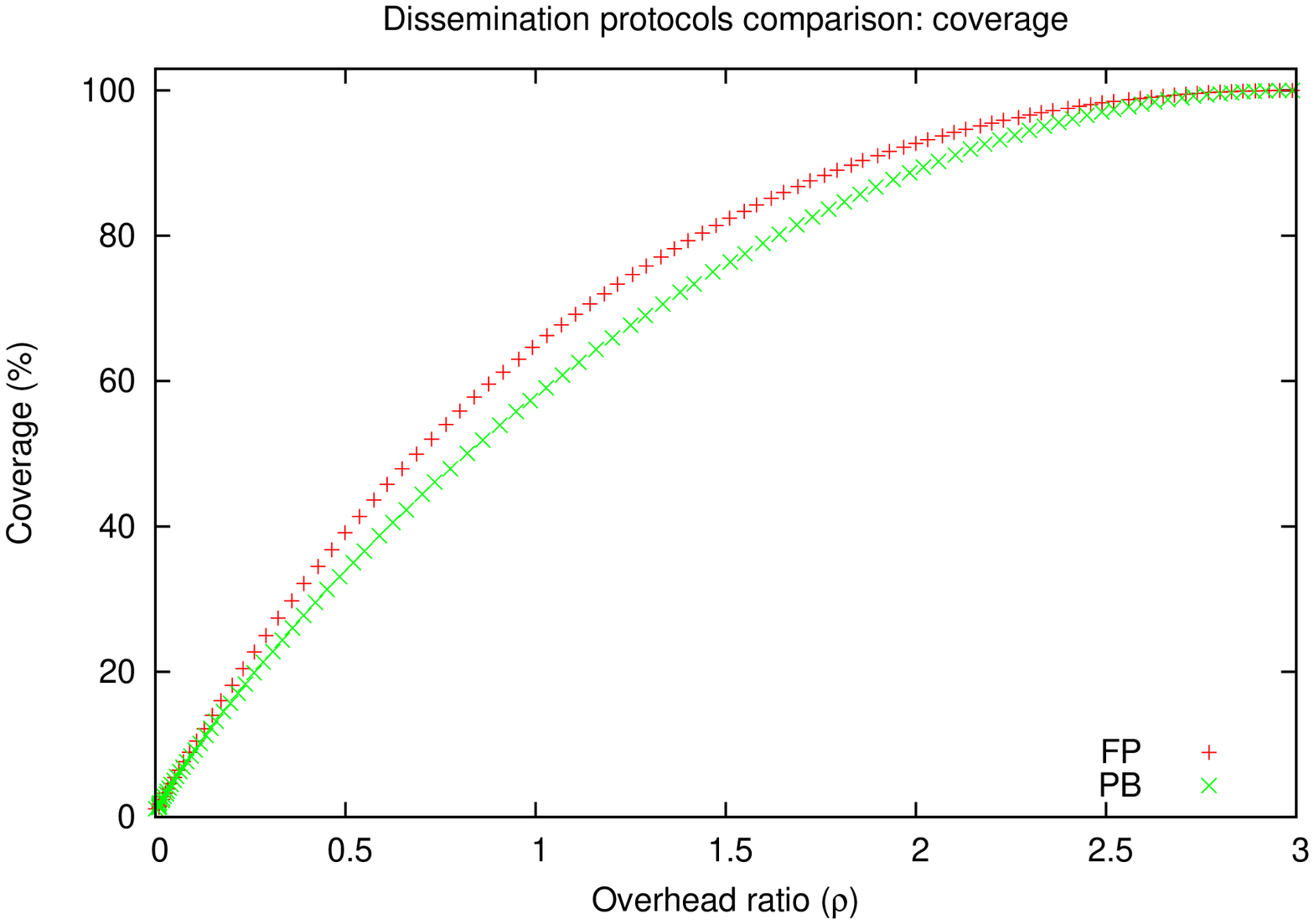}}
\subfloat[\label{fig:scalefree_997edges-delay_FP-CB}]{\includegraphics[angle=270,width=6.5cm]{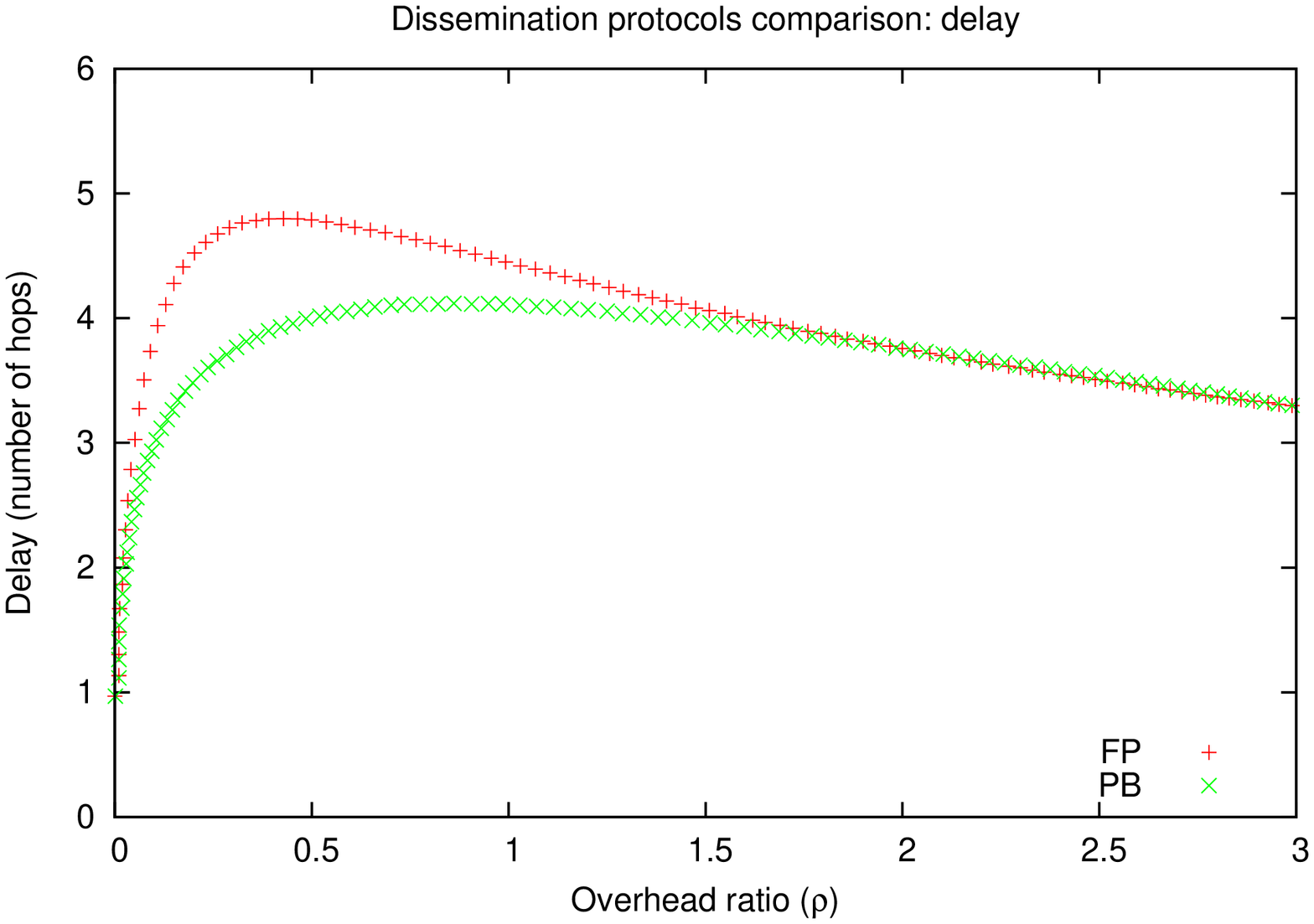}}
\caption{Scale-free networks, 500 nodes, 997 edges, max diameter=7, TTL=10, cache=256.}
\label{fig:scalefree_997edges}
\subfloat[\label{fig:scalefree_1494edges-coverage_FP-CB}]{\includegraphics[angle=270,width=6.5cm]{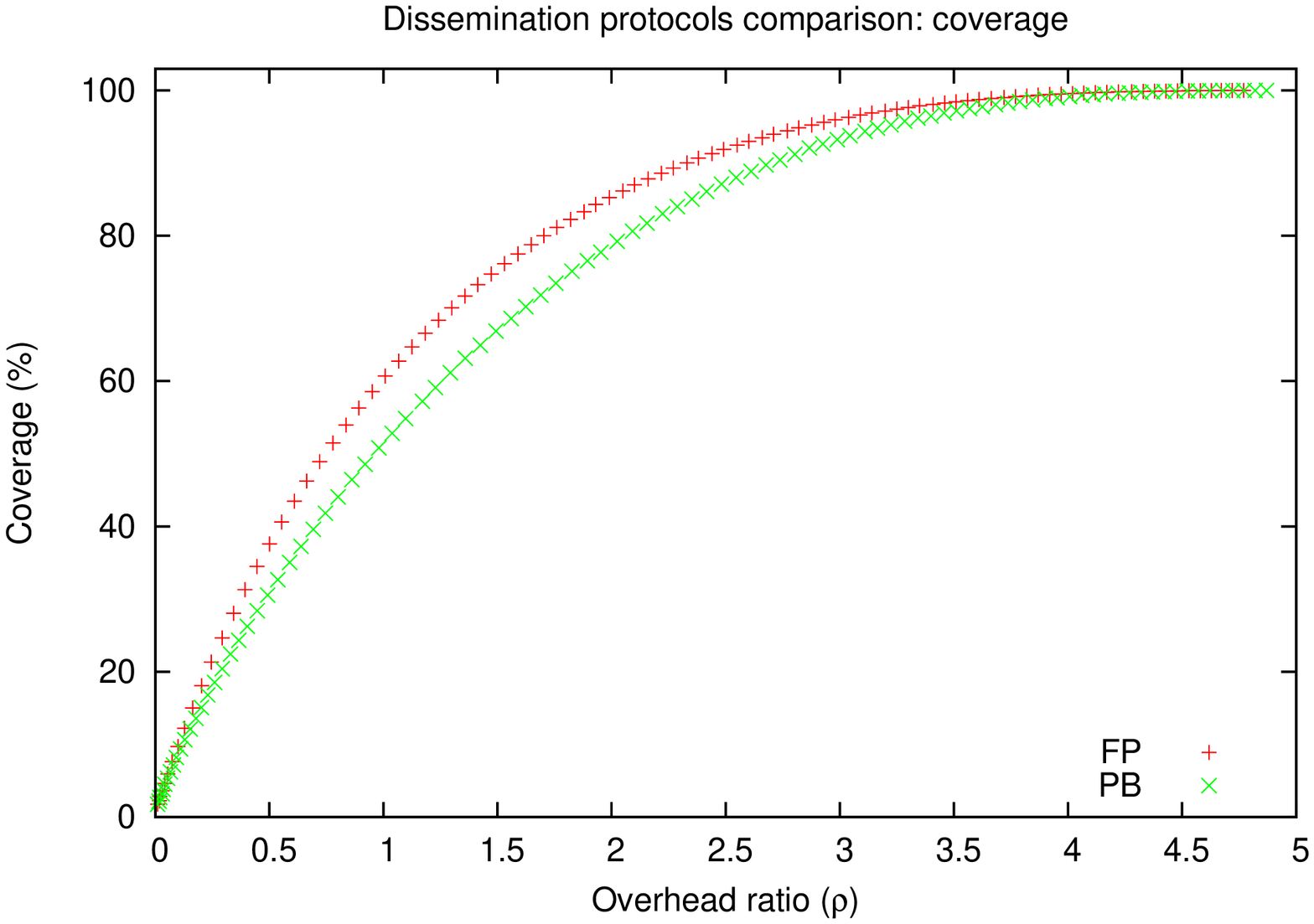}}
\subfloat[\label{fig:scalefree_1494edges-delay_FP-CB}]{\includegraphics[angle=270,width=6.5cm]{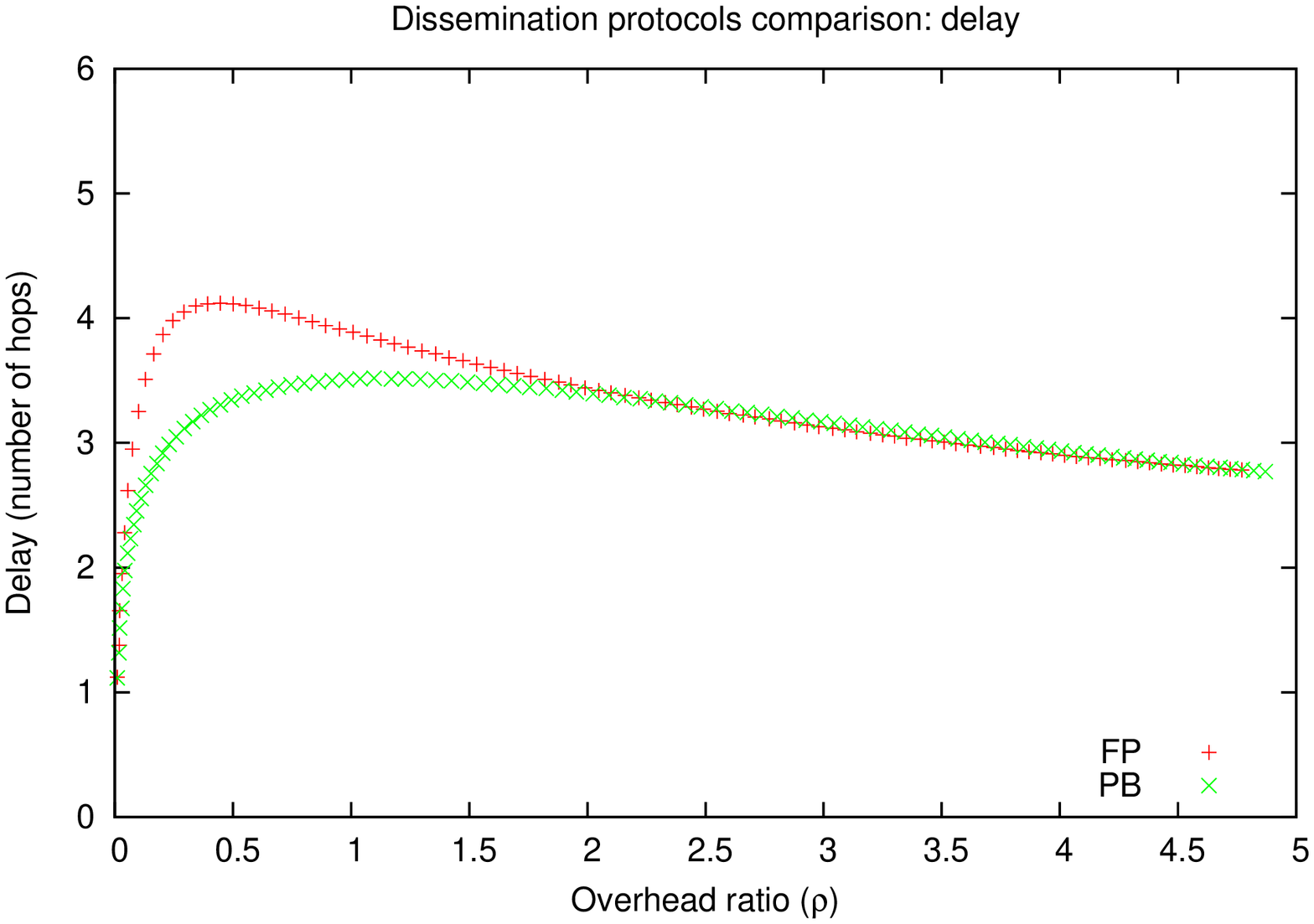}}
\caption{Scale-free networks, 500 nodes, 1494 edges, max diameter=5, TTL=7, cache=256.}
\label{fig:scalefree_1494edges}
\subfloat[\label{fig:scalefree_1990edges-coverage_FP-CB}]{\includegraphics[angle=270,width=6.5cm]{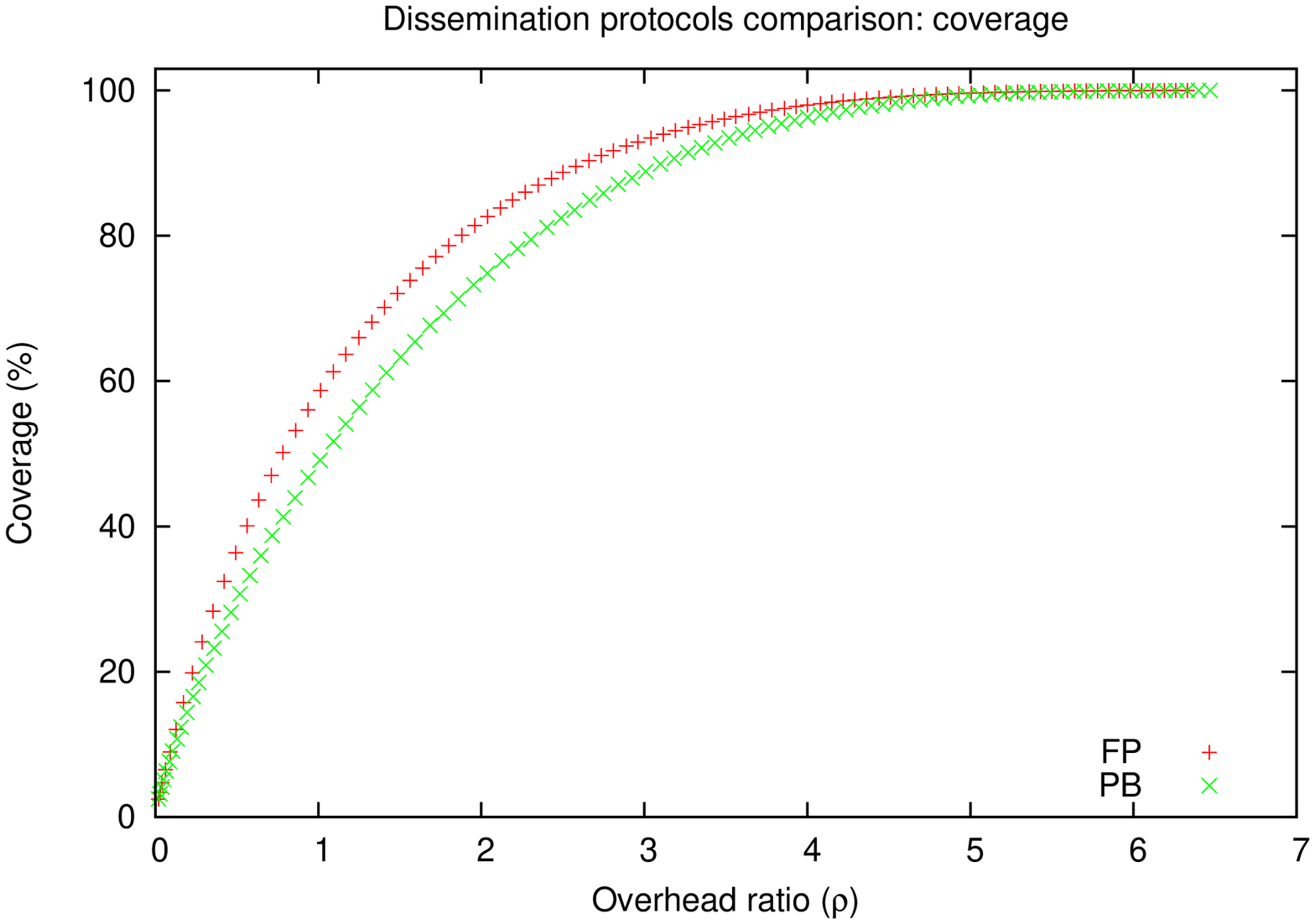}}
\subfloat[\label{fig:scalefree_1990edges-delay_FP-CB}]{\includegraphics[angle=270,width=6.5cm]{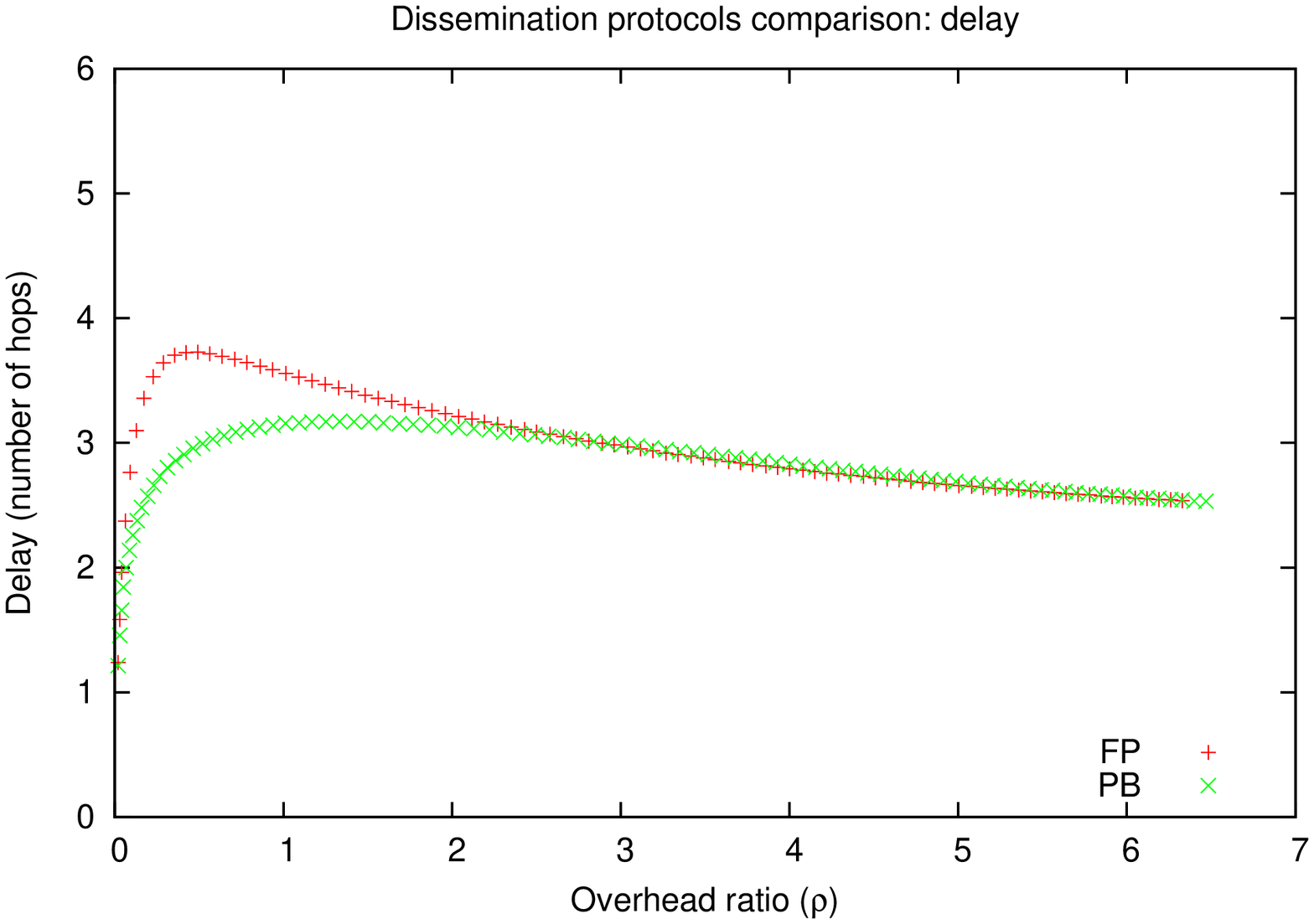}}
\caption{Scale-free networks, 500 nodes, 1990 edges, max diameter=4, TTL=6, cache=256.}
\label{fig:scalefree_1990edges}
\end{figure}

Figures~\ref{fig:scalefree_997edges},~\ref{fig:scalefree_1494edges} and \ref{fig:scalefree_1990edges} show the behavior of the FP and PB protocols on scale-free networks with 997, 1494 and 1990 edges. That is a graph generation with 2, 3 and 4 edges for each node when using the Barab\'asi-Albert model.
Also in this case, FP gets better results than PB for partial coverage and also for full coverage when the edges per node is larger than 2.

\begin{figure}[ht]
\centering
\subfloat[\label{fig:scalefree_997edges-coverage_FP-DDL-DDP}]{\includegraphics[angle=270,width=6.5cm]{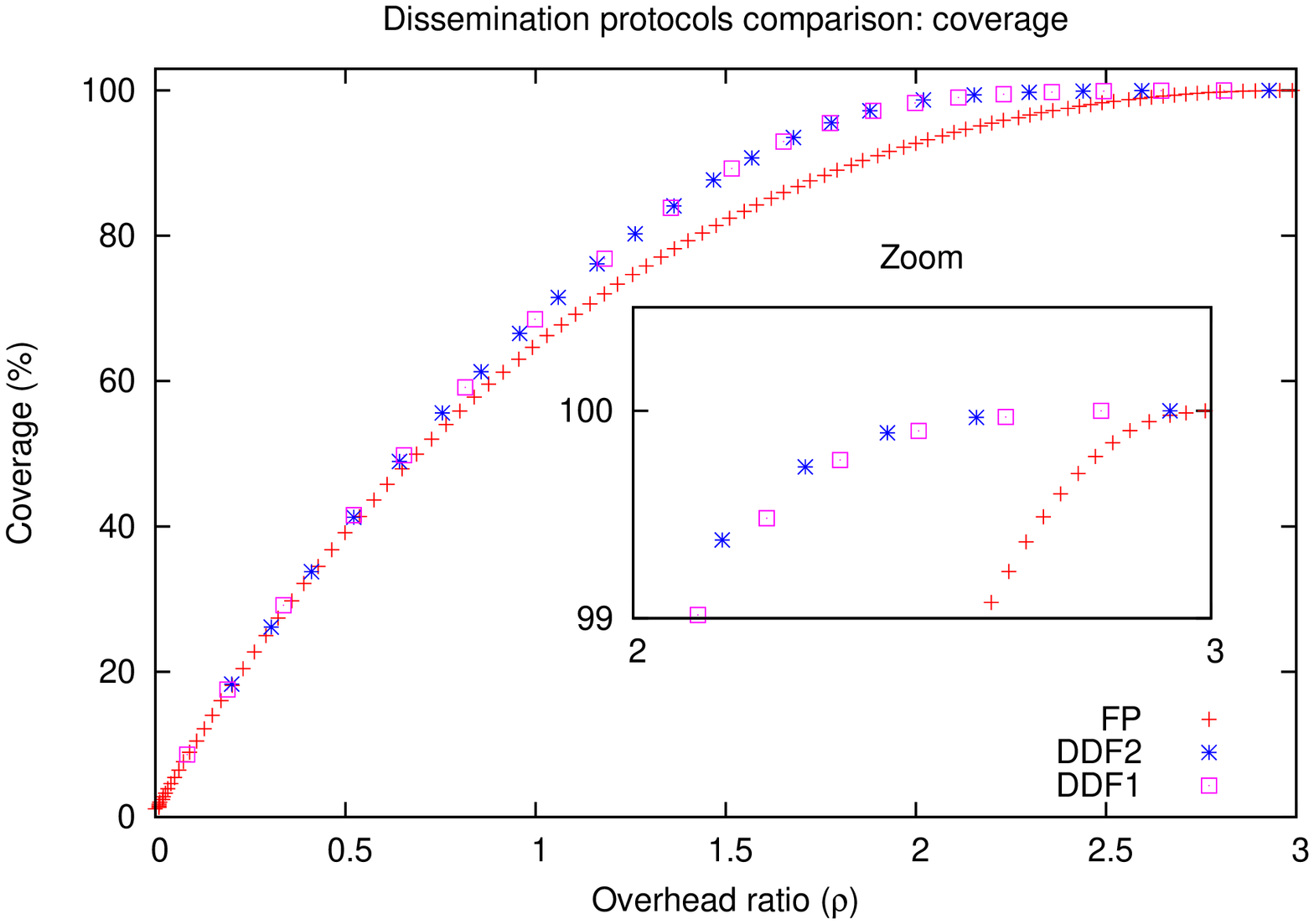}}
\subfloat[\label{fig:scalefree_997edges-delay_FP-DDL-DDP}]{\includegraphics[angle=270,width=6.5cm]{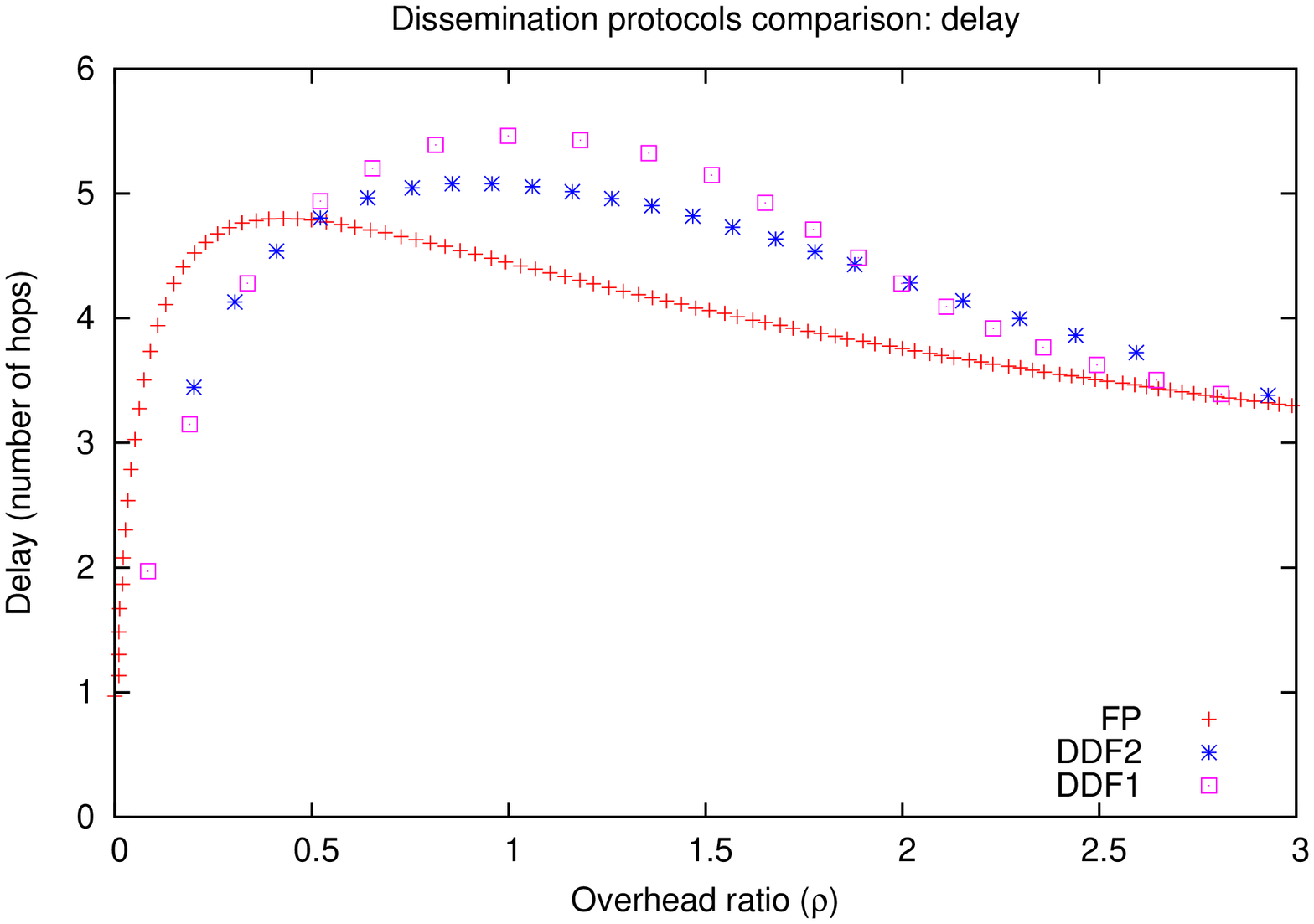}}
\caption{Scale-free networks, 500 nodes, 997 edges, max diameter=7, TTL=10, cache=256.}
\label{fig:scalefree_997edges-extra}
\subfloat[\label{fig:scalefree_1494edges-coverage_FP-DDL-DDP}]{\includegraphics[angle=270,width=6.5cm]{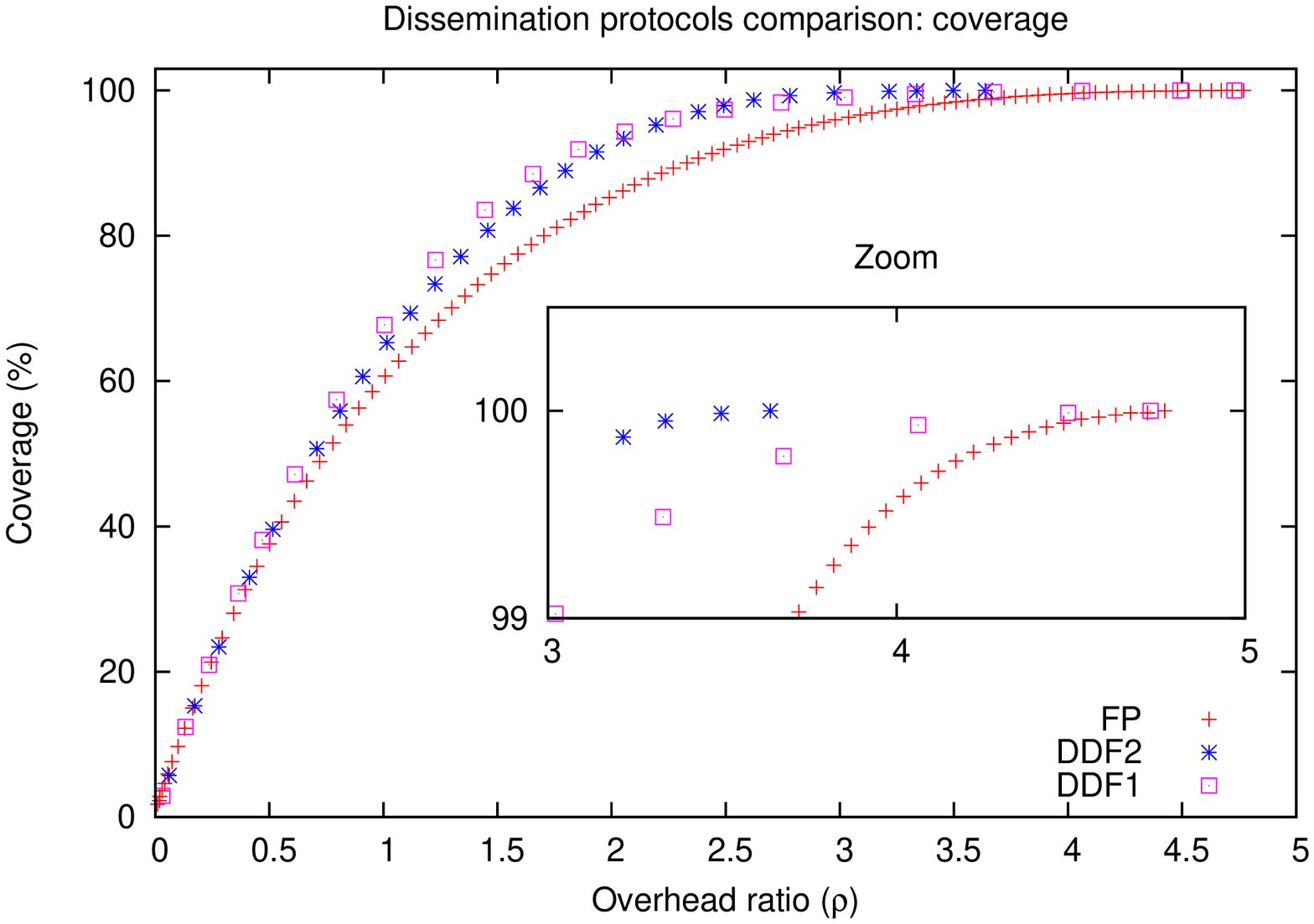}}
\subfloat[\label{fig:scalefree_1494edges-delay_FP-DDL-DDP}]{\includegraphics[angle=270,width=6.5cm]{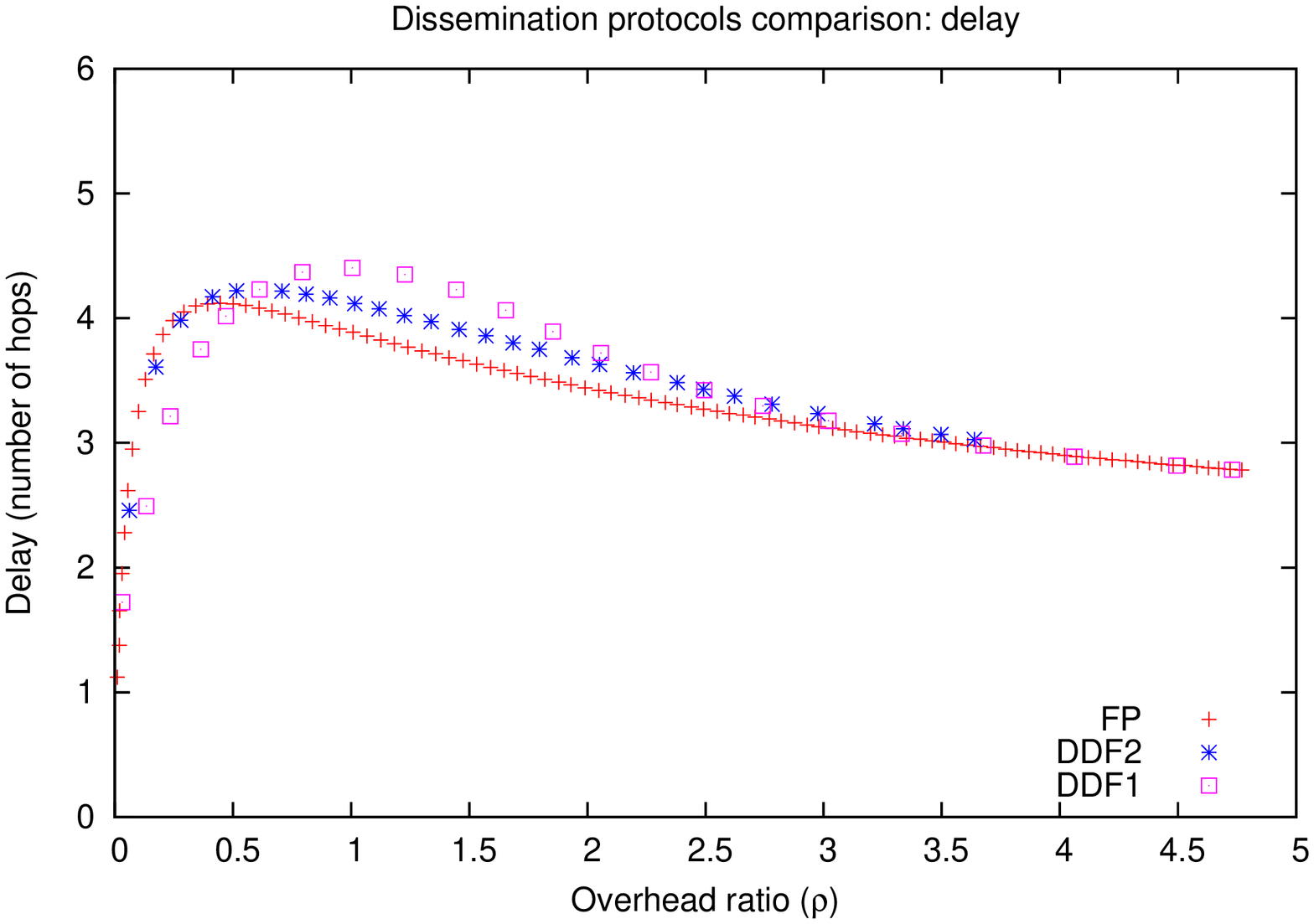}}
\caption{Scale-free networks, 500 nodes, 1494 edges, max diameter=5, TTL=7, cache=256.}
\label{fig:scalefree_1494edges-extra}
\subfloat[\label{fig:scalefree_1990edges-coverage_FP-DDL-DDP}]{\includegraphics[angle=270,width=6.5cm]{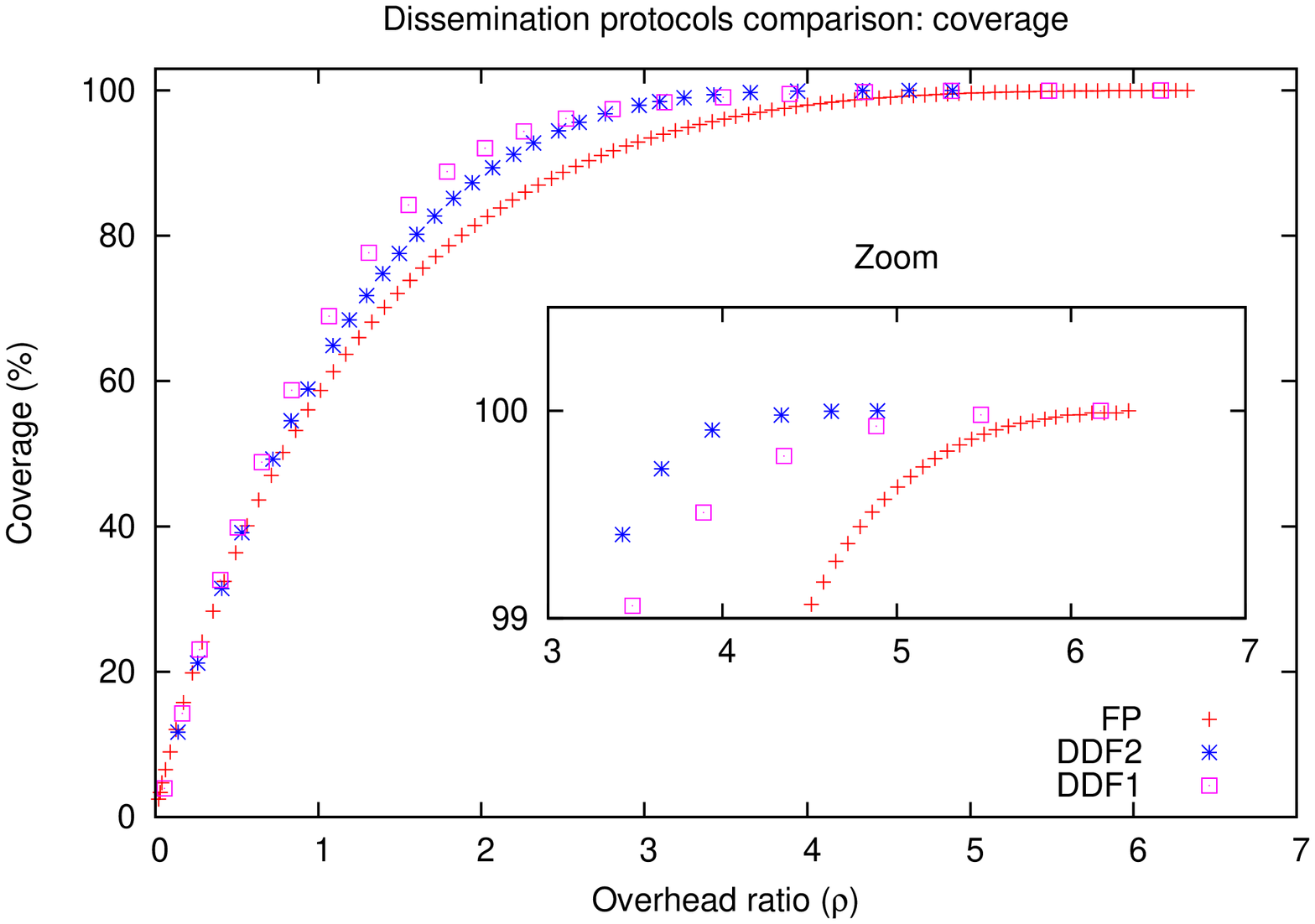}}
\subfloat[\label{fig:scalefree_1990edges-delay_FP-DDL-DDP}]{\includegraphics[angle=270,width=6.5cm]{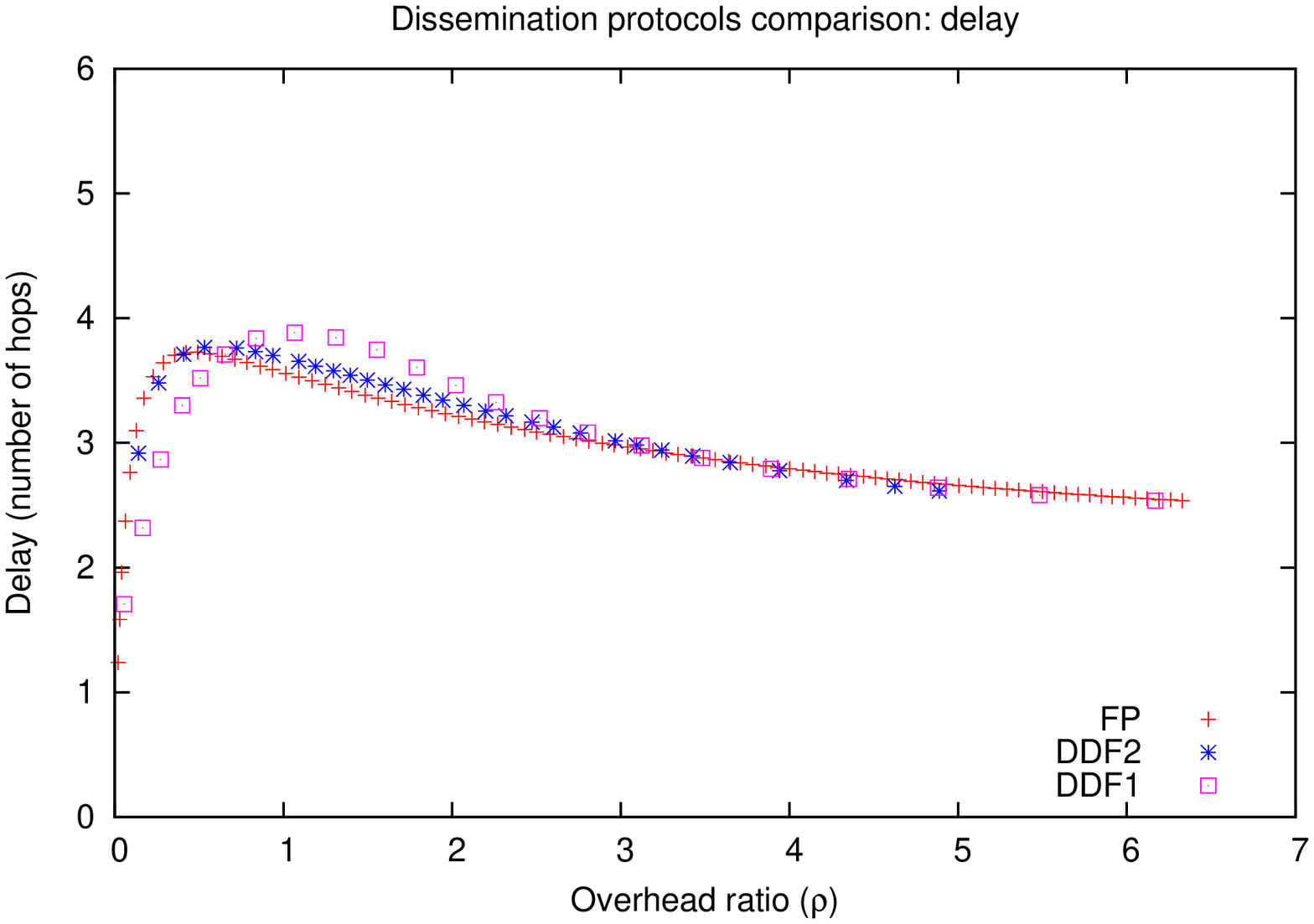}}
\caption{Scale-free networks, 500 nodes, 1990 edges, max diameter=4, TTL=6, cache=256.}
\label{fig:scalefree_1990edges-extra}
\end{figure}

DDF1 performs slightly better than other dissemination algorithms for 2 edges per node (Figure~\ref{fig:scalefree_997edges-extra}). With 3 and 4 edges per node (Figures~\ref{fig:scalefree_1494edges-extra} and \ref{fig:scalefree_1990edges-extra}), DDF2 is largely the best solutions and, in this case, the gain is in the order of 30\% for a full dissemination (Tables~\ref{tab:scalefree_1494edges} and \ref{tab:scalefree_1990edges}). 
Actually, we noticed that when considering mid-range coverage levels, the gain is quite relevant (not shown in the tables for the sake of conciseness). 

\begin{table}[ht]
\begin{center}
\begin{tabular}{ |p{2cm}||p{2cm}|p{2cm}| }
 \hline
 \multicolumn{3}{|c|}{Overhead (and delay) for a given coverage} \\
 \hline
 Algorithm & 100.0\% & \\
 \hline
 \hline
 FP   & 2.99 (3.30) & +6.40\% \\
 PB   & 2.99 (3.30) & +6.40\% \\
 DDF1  & 2.81 (3.39) & best \\
 DDF2  & 2.93 (3.38) & +4.27\% \\
 \hline
\end{tabular}
\end{center}
\caption{Scale-free networks, 500 nodes, 997 edges, max diameter=7, TTL=10, cache=256.}
\label{tab:scalefree_997edges}
\end{table}

\begin{table}[ht]
\begin{center}
\begin{tabular}{ |p{2cm}||p{2cm}|p{2cm}| }
 \hline
 \multicolumn{3}{|c|}{Overhead (and delay) for a given coverage} \\
 \hline
 Algorithm & 100.0\% & \\
 \hline
 \hline
 FP   & 4.77 (2.78) & +31.04\% \\
 PB   & 4.87 (2.77) & +33.79\% \\
 DDF1  & 4.73 (2.78) & +29.94\% \\
 DDF2  & 3.64 (3.03) & best\\
 \hline
\end{tabular}
\end{center}
\caption{Scale-free networks, 500 nodes, 1494 edges, max diameter=5, TTL=7, cache=256.}
\label{tab:scalefree_1494edges}
\end{table}

\begin{table}[ht]
\begin{center}
\begin{tabular}{ |p{2cm}||p{2cm}|p{2cm}| }
 \hline
 \multicolumn{3}{|c|}{Overhead (and delay) for a given coverage} \\
 \hline
 Algorithm & 100.0\% & \\
 \hline
 \hline
 FP   & 6.33 (2.54) & +29.44\%\\
 PB   & 6.47 (2.53) & +32.31\% \\
 DDF1  & 6.17 (2.54) & +26.17\% \\
 DDF2  & 4.89 (2.61) & best \\
 \hline
\end{tabular}
\end{center}
\caption{Scale-free networks, 500 nodes, 1990 edges, max diameter=4, TTL=6, cache=256.}
\label{tab:scalefree_1990edges}
\end{table}

\subsection{Small-World Networks}

\begin{figure}[ht]
\centering
\subfloat[\label{fig:smallworld_1000edges-coverage_FP-CB}]{\includegraphics[angle=270,width=6.5cm]{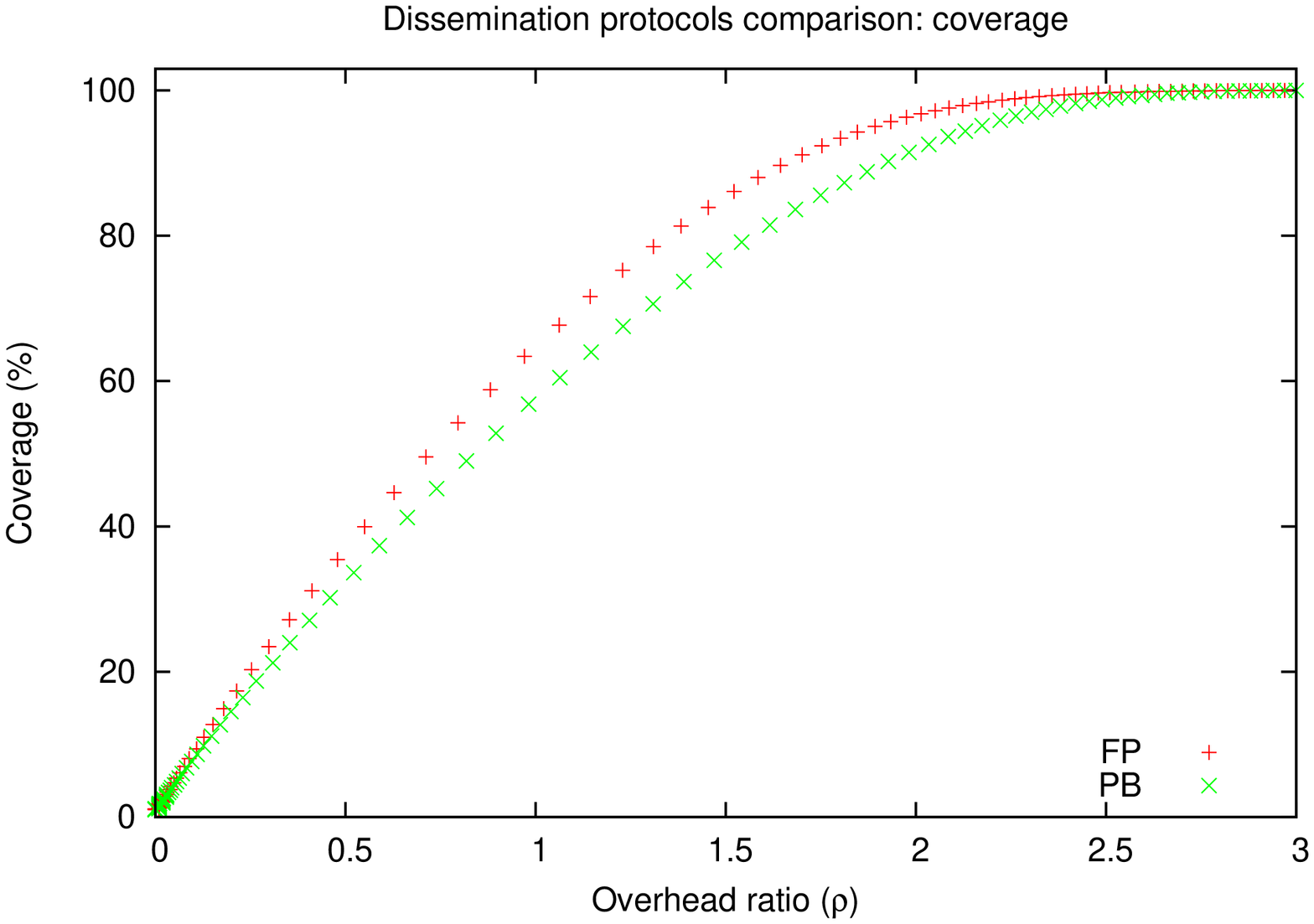}}
\subfloat[\label{fig:smallworld_1000edges-delay_FP-CB}]{\includegraphics[angle=270,width=6.5cm]{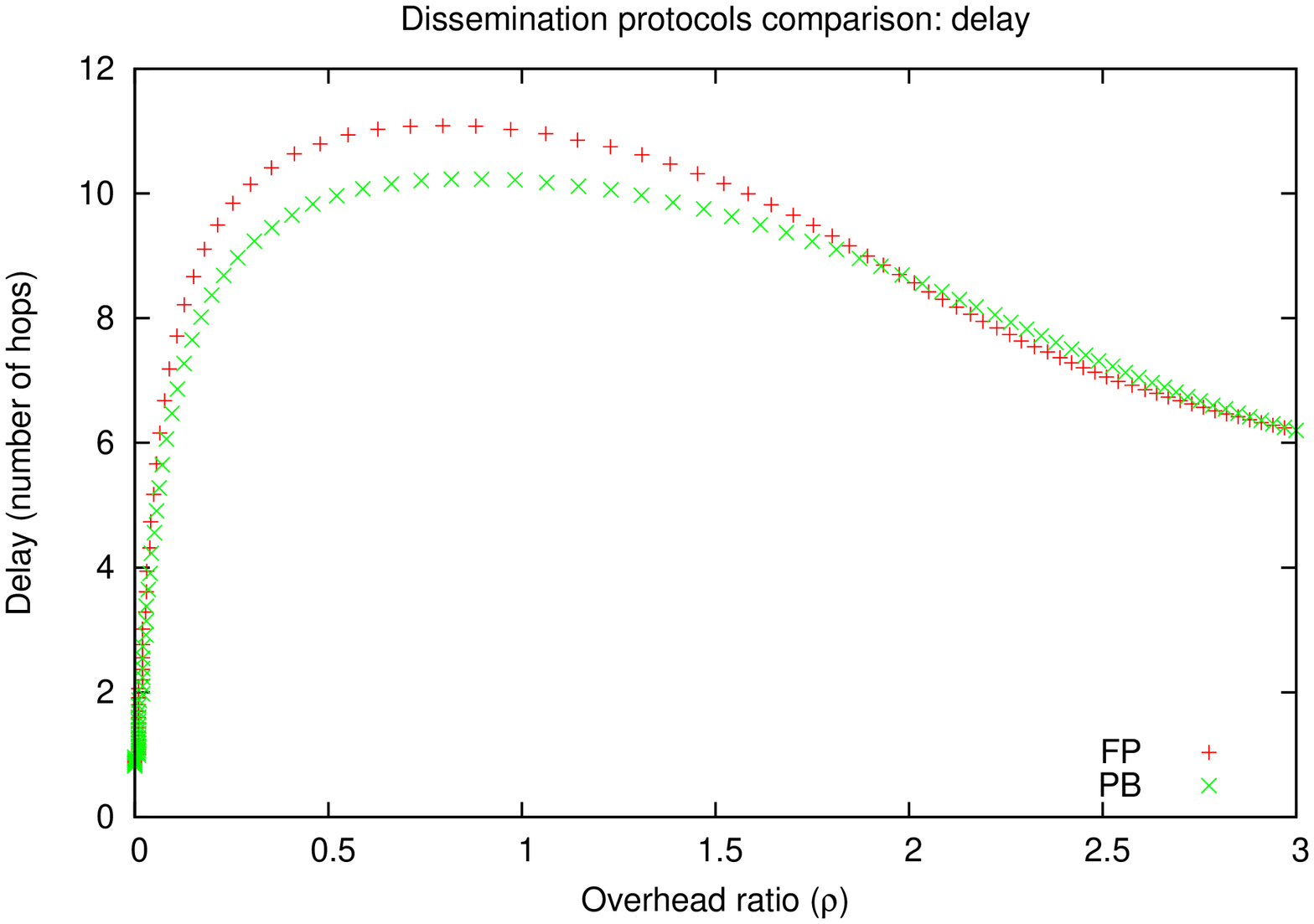}}
\caption{Small-world networks, 500 nodes, 1000 edges, rewiring probability=0.1, max diameter=13, TTL=17, cache=256.}
\label{fig:smallworld_1000edges}
\subfloat[\label{fig:smallworld_1500edges-coverage_FP-CB}]{\includegraphics[angle=270,width=6.5cm]{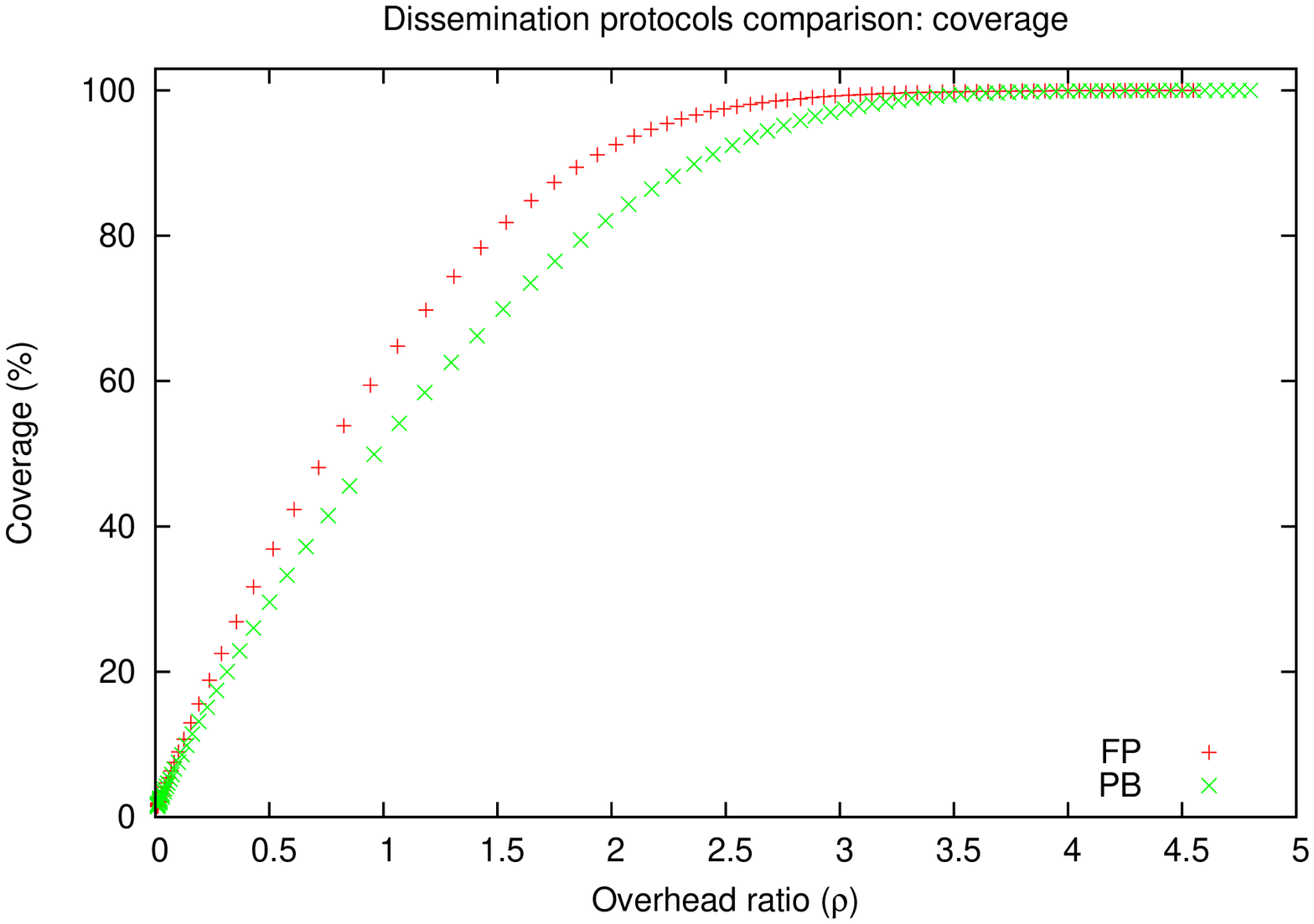}}
\subfloat[\label{fig:smallworld_1500edges-delay_FP-CB}]{\includegraphics[angle=270,width=6.5cm]{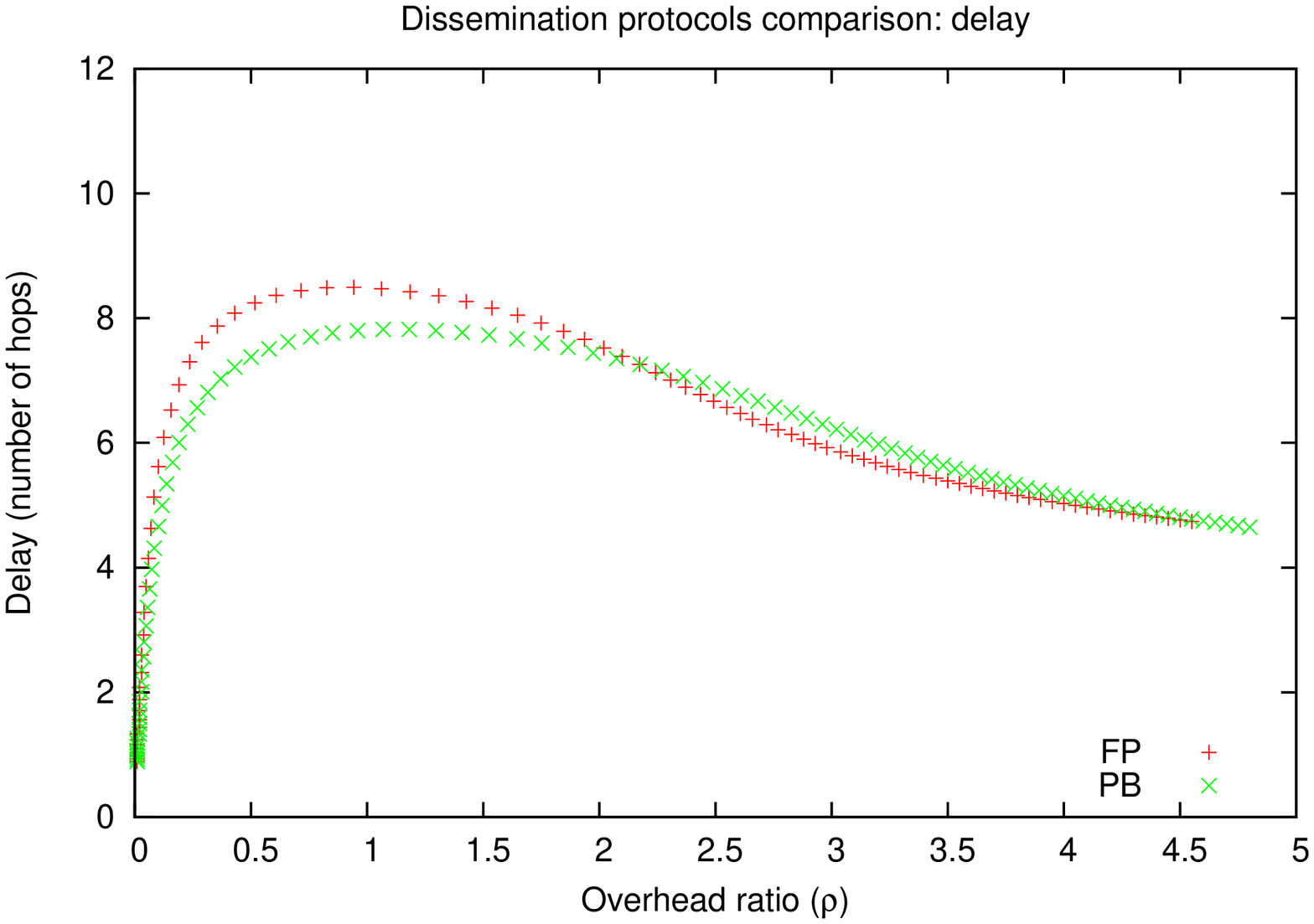}}
\caption{Small-world networks, 500 nodes, 1500 edges, rewiring probability=0.1, max diameter=9, TTL=12, cache=256.}
\label{fig:smallworld_1500edges}
\subfloat[\label{fig:smallworld_2000edges-coverage_FP-CB}]{\includegraphics[angle=270,width=6.5cm]{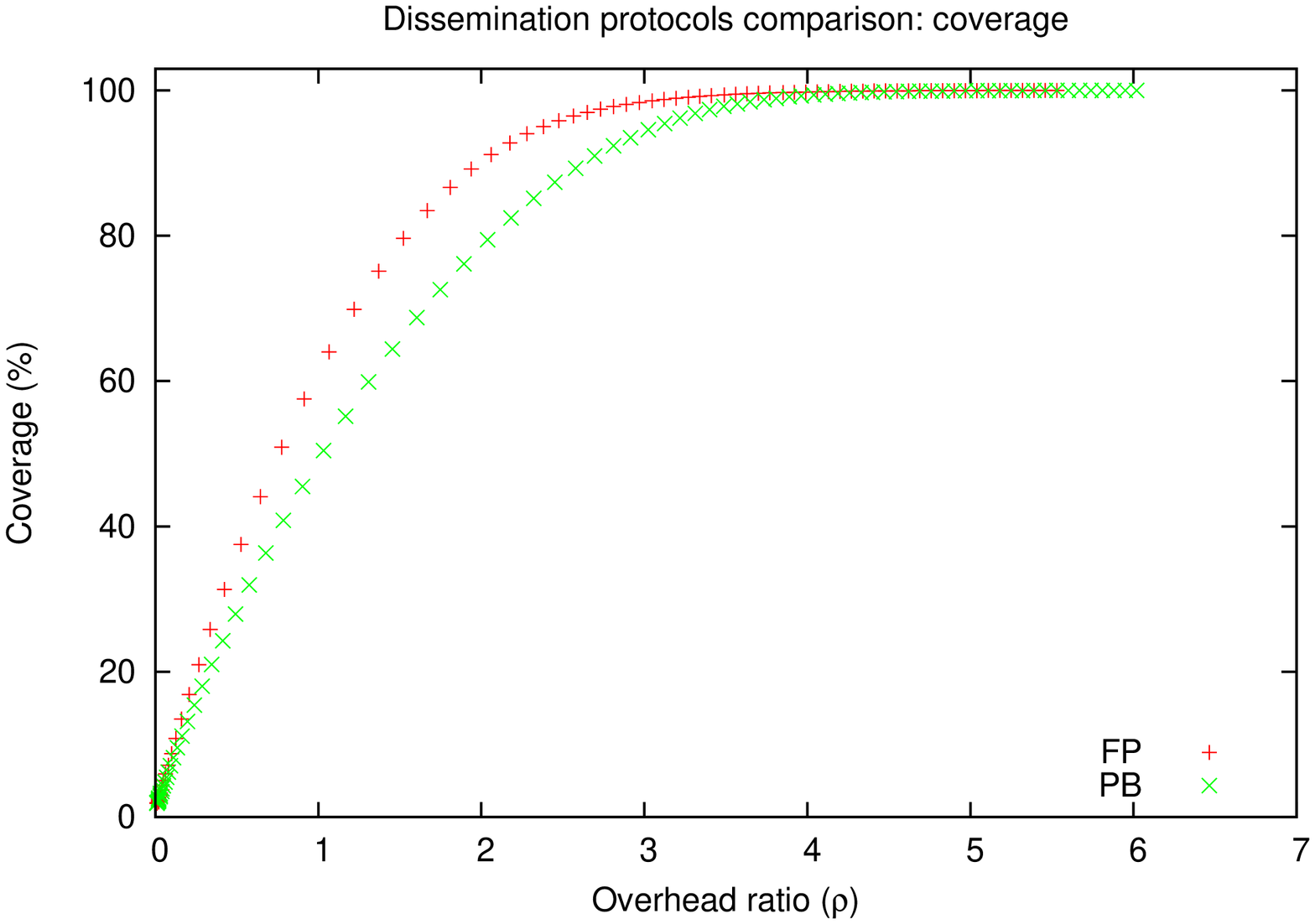}}
\subfloat[\label{fig:smallworld_2000edges-delay_FP-CB}]{\includegraphics[angle=270,width=6.5cm]{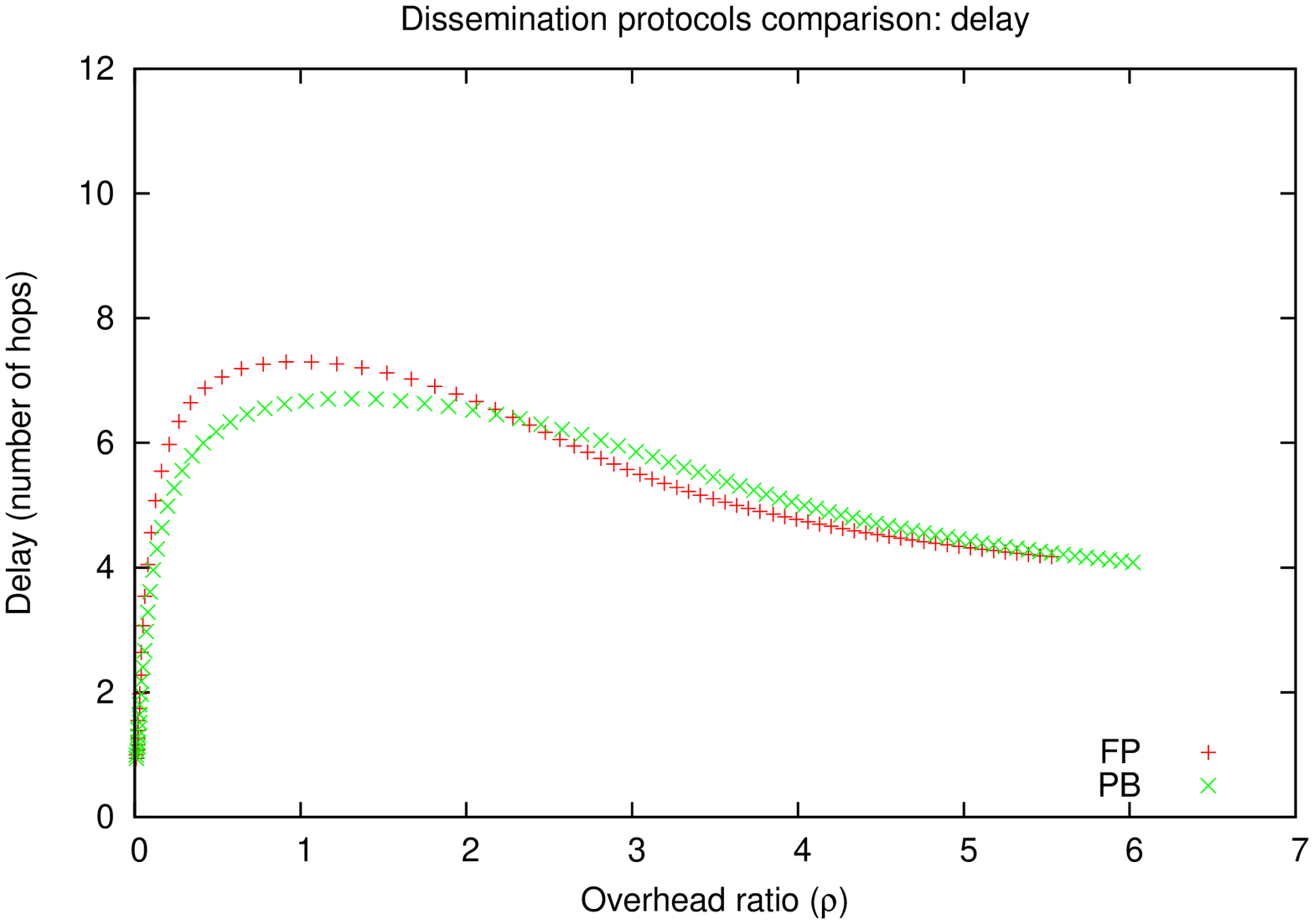}}
\caption{Small-world networks, 500 nodes, 2000 edges, rewiring probability=0.1, max diameter=7, TTL=10, cache=256.}
\label{fig:smallworld_2000edges}
\end{figure}

The dissemination on small-world networks, with a rewiring probability set to $0.1$ (this is a typical value to create a small-world network), further confirms that FP is better than PB for partial and full coverage (Figure~\ref{fig:smallworld_1000edges}). The gap between FP and PB increases with a higher number of edges per node (Figures~\ref{fig:smallworld_1500edges} and \ref{fig:smallworld_2000edges}). 

\begin{figure}[ht]
\centering
\subfloat[\label{fig:smallworld_1000edges-coverage_FP-DDL-DDP}]{\includegraphics[angle=270,width=6.5cm]{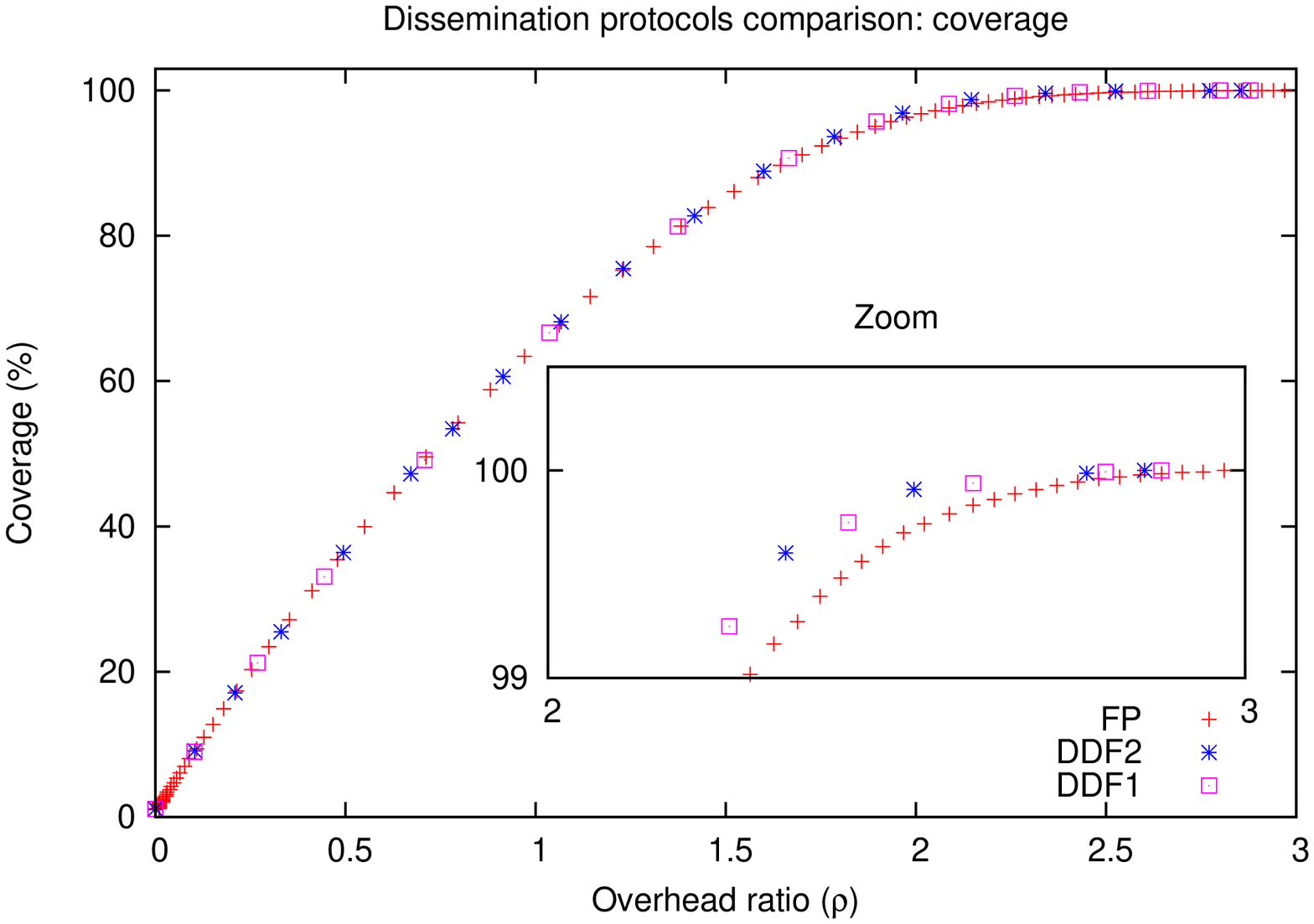}}
\subfloat[\label{fig:smallworld_1000edges-delay_FP-DDL-DDP}]{\includegraphics[angle=270,width=6.5cm]{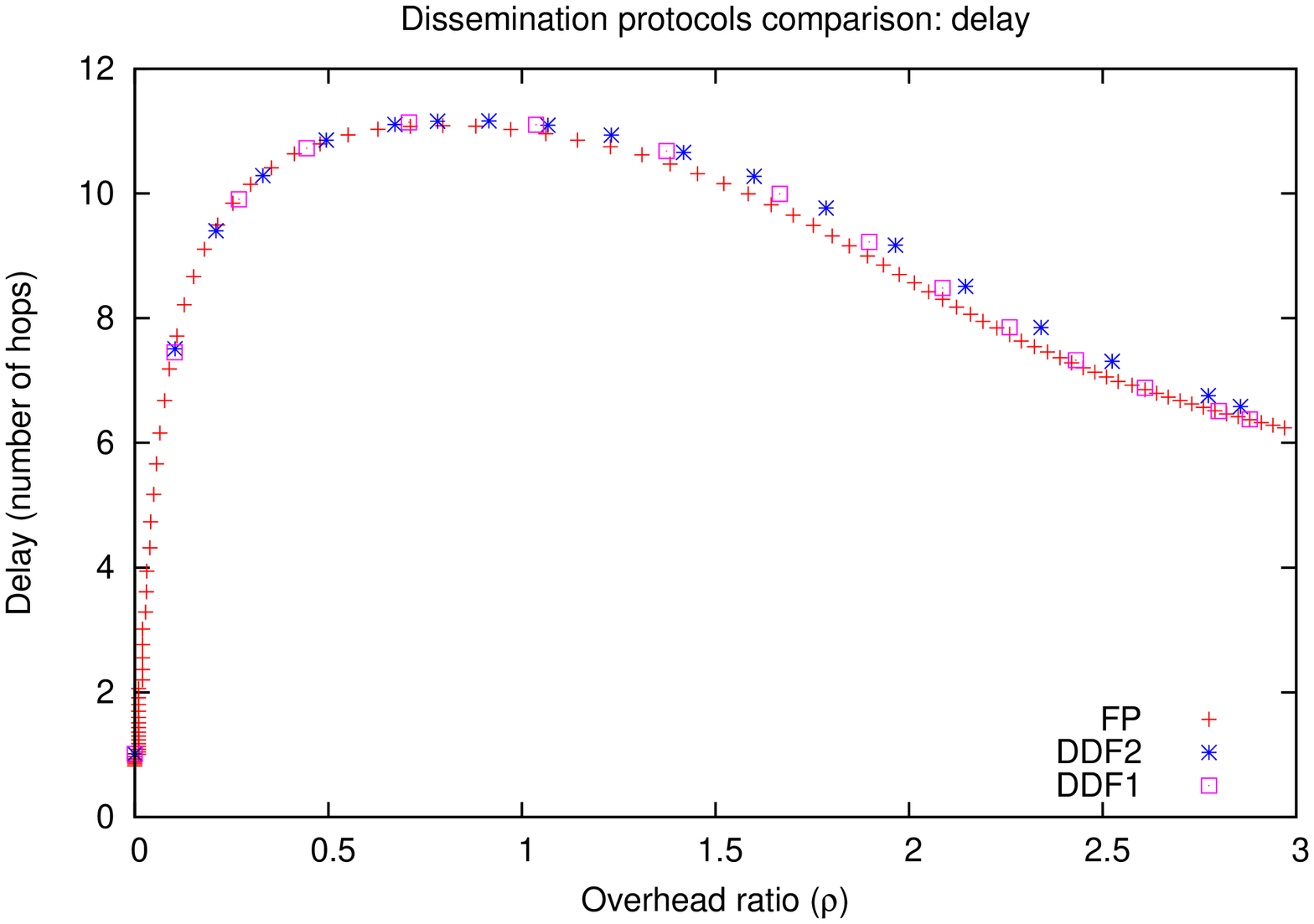}}
\caption{Small-world networks, 500 nodes, 1000 edges, rewiring probability=0.1, max diameter=13, TTL=17, cache=256.}
\label{fig:smallworld_1000edges-extra}
\subfloat[\label{fig:smallworld_1500edges-coverage_FP-DDL-DDP}]{\includegraphics[angle=270,width=6.5cm]{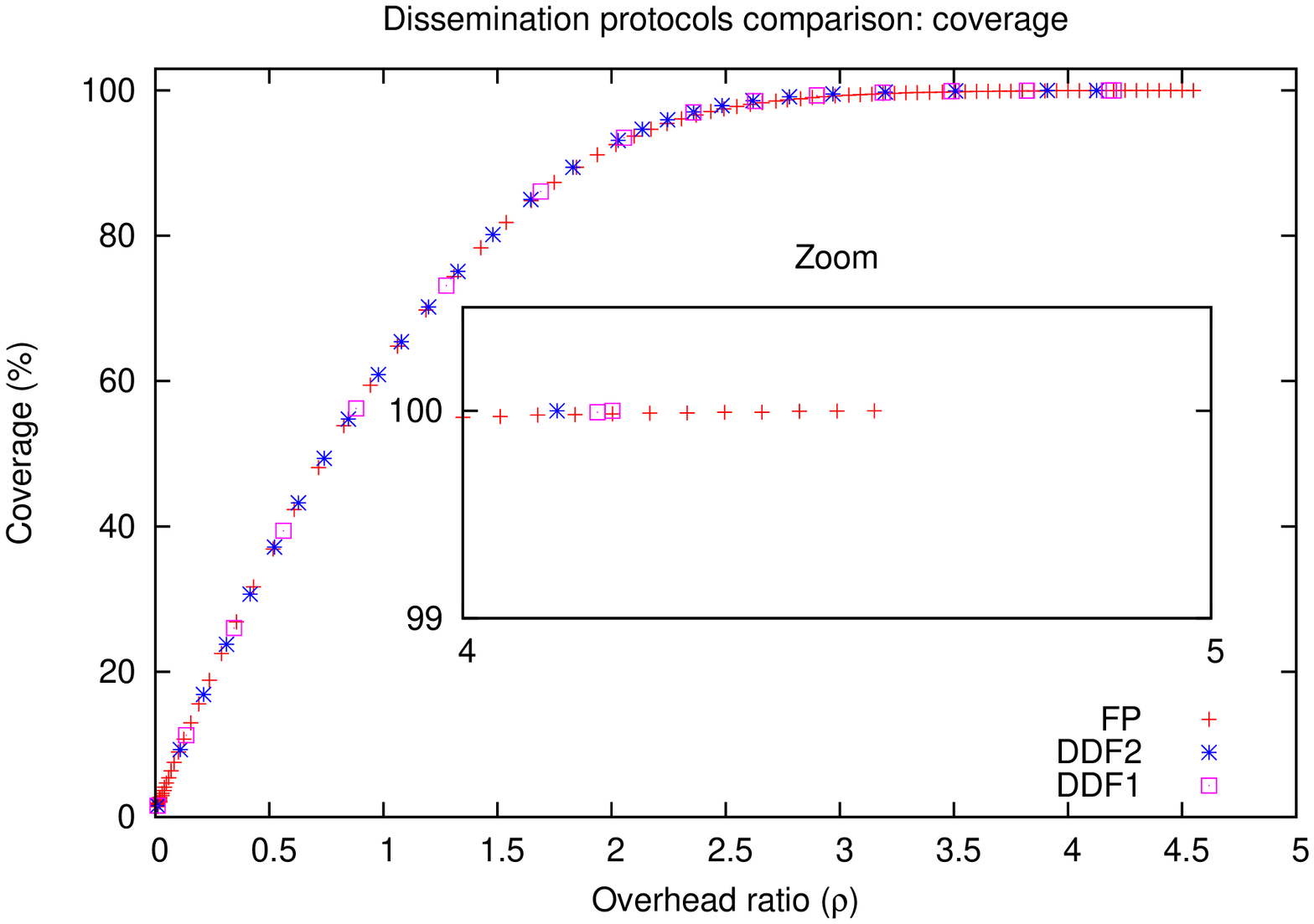}}
\subfloat[\label{fig:smallworld_1500edges-delay_FP-DDL-DDP}]{\includegraphics[angle=270,width=6.5cm]{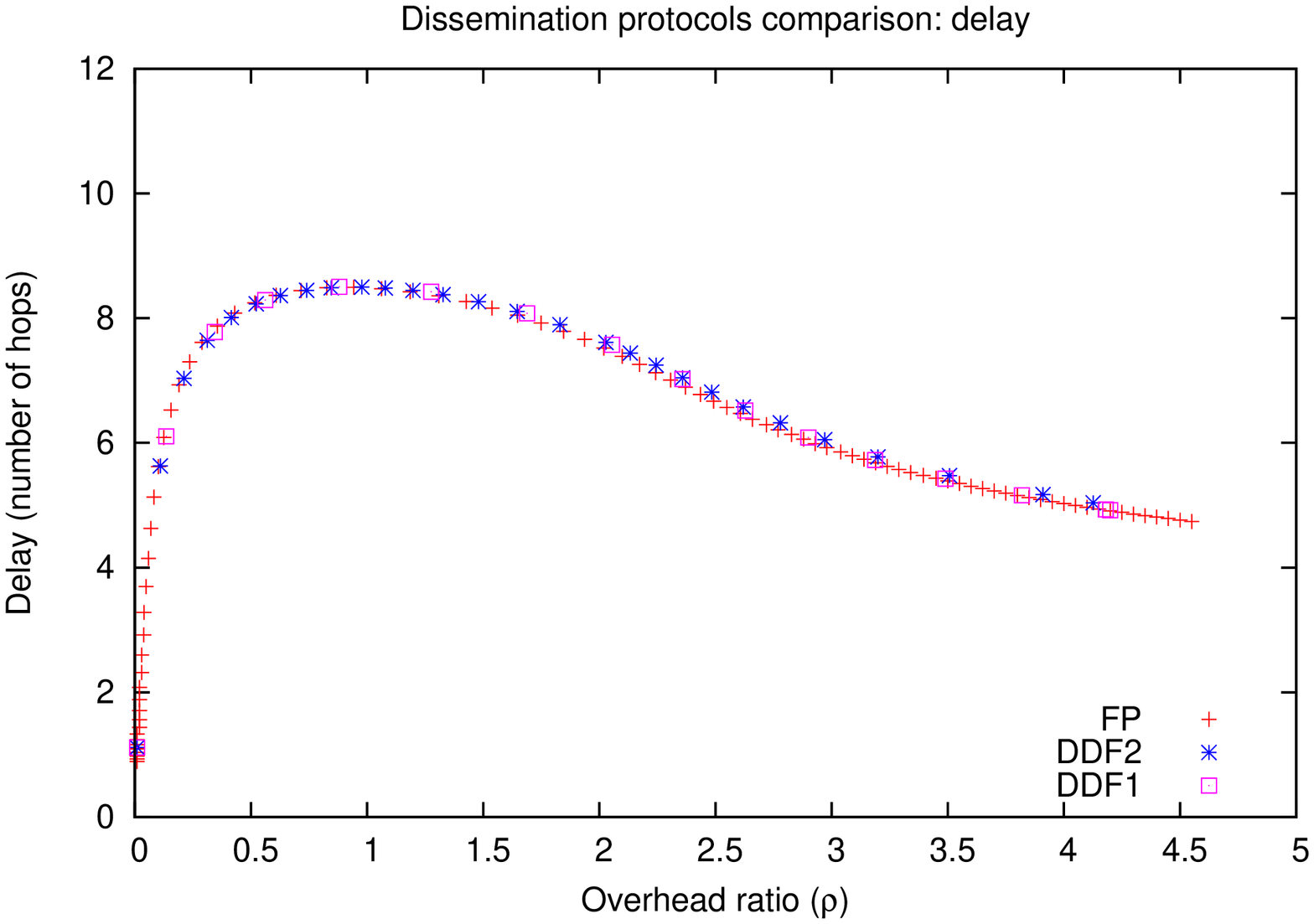}}
\caption{Small-world networks, 500 nodes, 1500 edges, rewiring probability=0.1, max diameter=9, TTL=12, cache=256.}
\label{fig:smallworld_1500edges-extra}
\subfloat[\label{fig:smallworld_2000edges-coverage_FP-DDL-DDP}]{\includegraphics[angle=270,width=6.5cm]{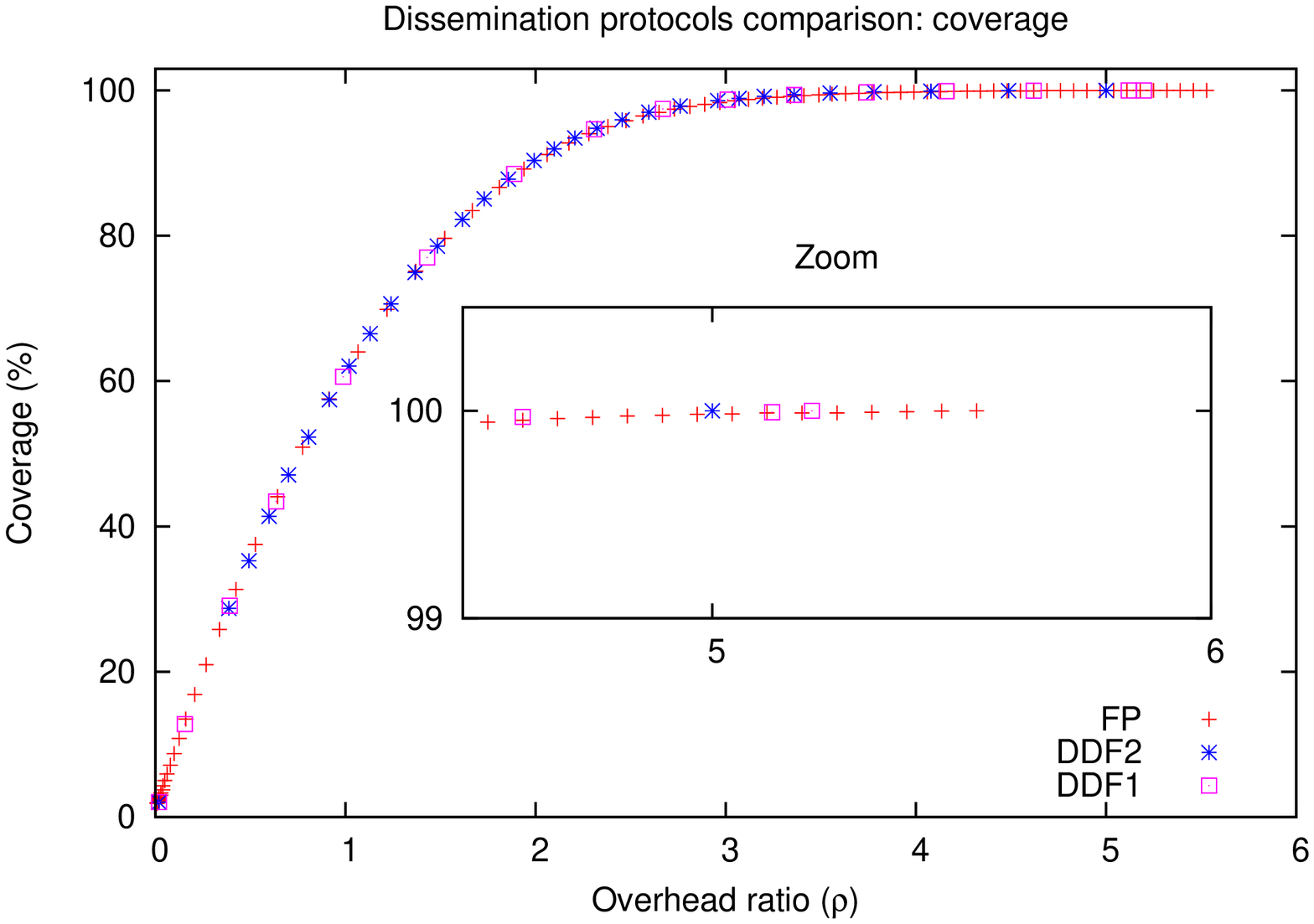}}
\subfloat[\label{fig:smallworld_2000edges-delay_FP-DDL-DDP}]{\includegraphics[angle=270,width=6.5cm]{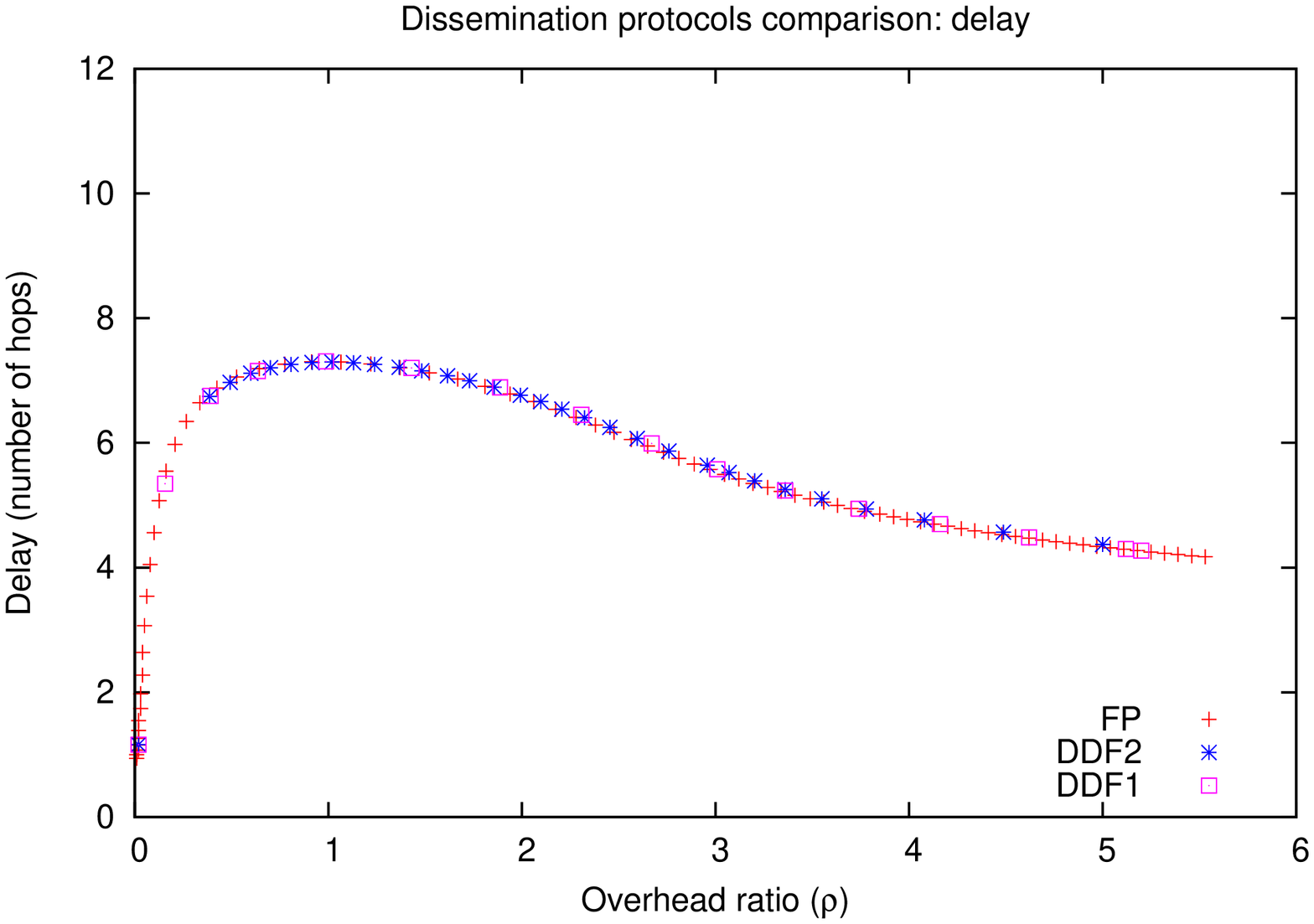}}
\caption{Small-world networks, 500 nodes, 2000 edges, rewiring probability=0.1, max diameter=7, TTL=10, cache=256.}
\label{fig:smallworld_2000edges-extra}
\end{figure}

In this specific scenario, DDF1 and DDF2 are unable to get better results for partial coverage with respect to FP (Figures~\ref{fig:smallworld_1000edges-extra}, \ref{fig:smallworld_1500edges-extra} and \ref{fig:smallworld_2000edges-extra}) but the gain is evident for full coverage (Tables~\ref{tab:smallworld_1000edges}, \ref{tab:smallworld_1500edges} and \ref{tab:smallworld_2000edges}).
When we consider a partial coverage, we are dealing with the need to widely spread a content over a network; when the network is a small-world, then the topology itself provides links to reach different portions of the network. Thus, a simple gossip strategy is quite effective. 
For a full coverage, instead, a dissemination procedure that floods the message when a node has very just one or two links, as degree depending protocols do, avoids that some ``peripheral'' node is left out, during the dissemination.

\begin{table}[ht]
\begin{center}
\begin{tabular}{ |p{2cm}||p{2cm}|p{2cm}| }
 \hline
 \multicolumn{3}{|c|}{Overhead (and delay) for a given coverage} \\
 \hline
 Algorithm & 100.0\% & \\
 \hline
 \hline
 FP   & 2.97 (6.24) & +3.84\% \\
 PB   & 3.00 (6.20) & +4.89\% \\
 DDF1  & 2.88 (6.38) & +0.69\% \\
 DDF2  & 2.86 (6.58) & best \\
 \hline
\end{tabular}
\end{center}
\caption{Small-world networks, 500 nodes, 1000 edges, rewiring probability=0.1, max diameter=13, TTL=17, cache=256.}
\label{tab:smallworld_1000edges}
\end{table}

\begin{table}[ht]
\begin{center}
\begin{tabular}{ |p{2cm}||p{2cm}|p{2cm}| }
 \hline
 \multicolumn{3}{|c|}{Overhead (and delay) for a given coverage} \\
 \hline
 Algorithm & 100.0\% & \\
 \hline
 \hline
 FP   & 4.55 (4.74) & +10.16\% \\
 PB   & 4.75 (4.67) & +15.01\% \\
 DDF1  & 4.20 (4.92) & +1.69\% \\
 DDF2  & 4.13 (5.04) & best \\
 \hline
\end{tabular}
\end{center}
\caption{Small-world networks, 500 nodes, 1500 edges, rewiring probability=0.1, max diameter=9, TTL=12, cache=256.}
\label{tab:smallworld_1500edges}
\end{table}

\begin{table}[ht]
\begin{center}
\begin{tabular}{ |p{2cm}||p{2cm}|p{2cm}| }
 \hline
 \multicolumn{3}{|c|}{Overhead (and delay) for a given coverage} \\
 \hline
 Algorithm & 100.0\% & \\
 \hline
 \hline
 FP   & 5.53 (4.17) & +10.60\% \\
 PB   & 6.02 (4.08) & +20.40\% \\
 DDF1  & 5.20 (4.27) & +4.00\% \\
 DDF2  & 5.00 (4.37) & best \\
 \hline
\end{tabular}
\end{center}
\caption{Small-world networks, 500 nodes, 2000 edges, rewiring probability=0.1, max diameter=7, TTL=10, cache=256.}
\label{tab:smallworld_2000edges}
\end{table}

\subsection{K-regular Networks}

\begin{figure}[ht]
\centering
\subfloat[\label{fig:kregular_1000edges-coverage_FP-CB}]{\includegraphics[angle=270,width=6.5cm]{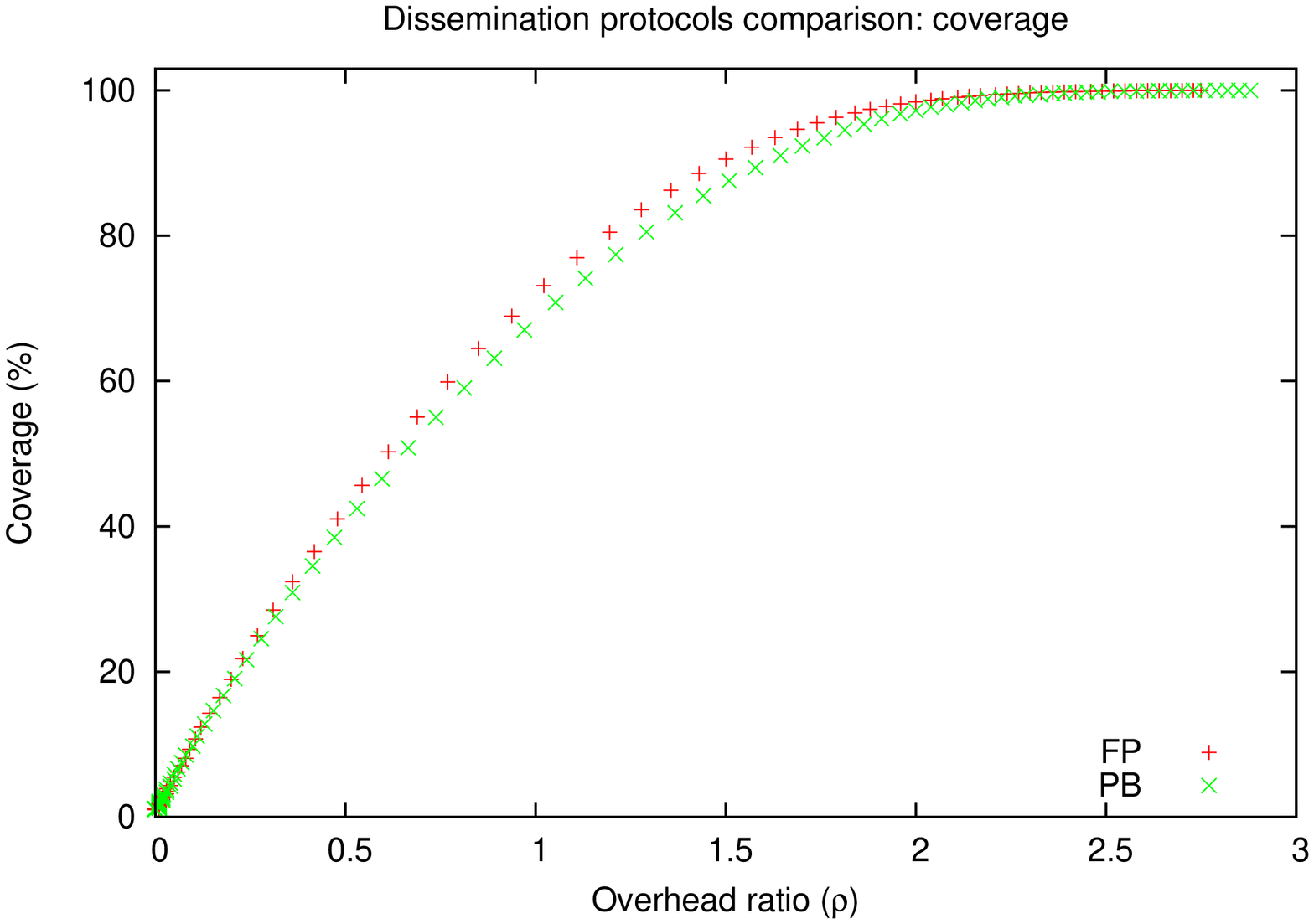}}
\subfloat[\label{fig:kregular_1000edges-delay_FP-CB}]{\includegraphics[angle=270,width=6.5cm]{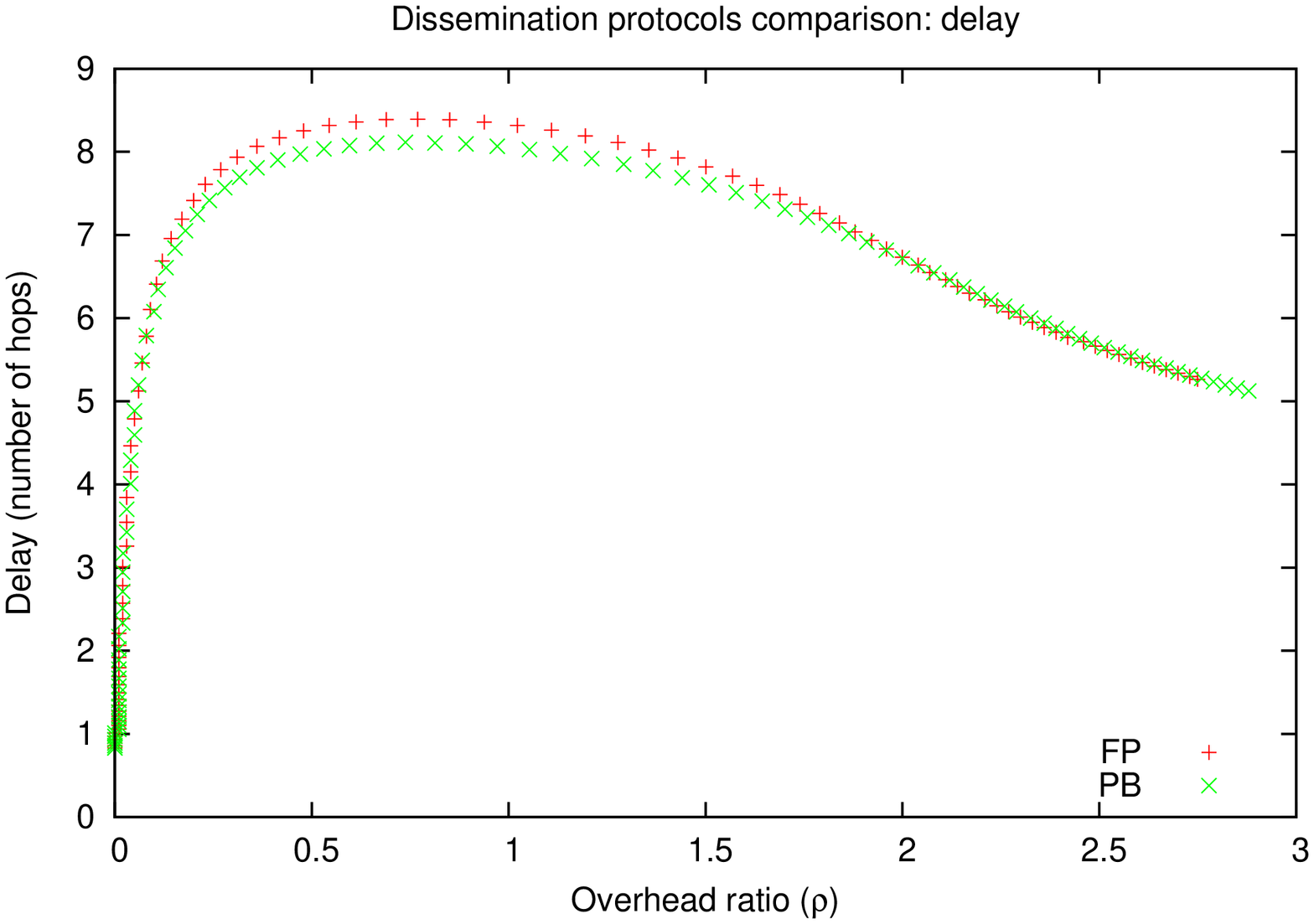}}
\caption{K-regular networks, 500 nodes, 1000 edges, max diameter=8, TTL=11, cache=256.}
\label{fig:kregular_1000edges}
\subfloat[\label{fig:kregular_1500edges-coverage_FP-CB}]{\includegraphics[angle=270,width=6.5cm]{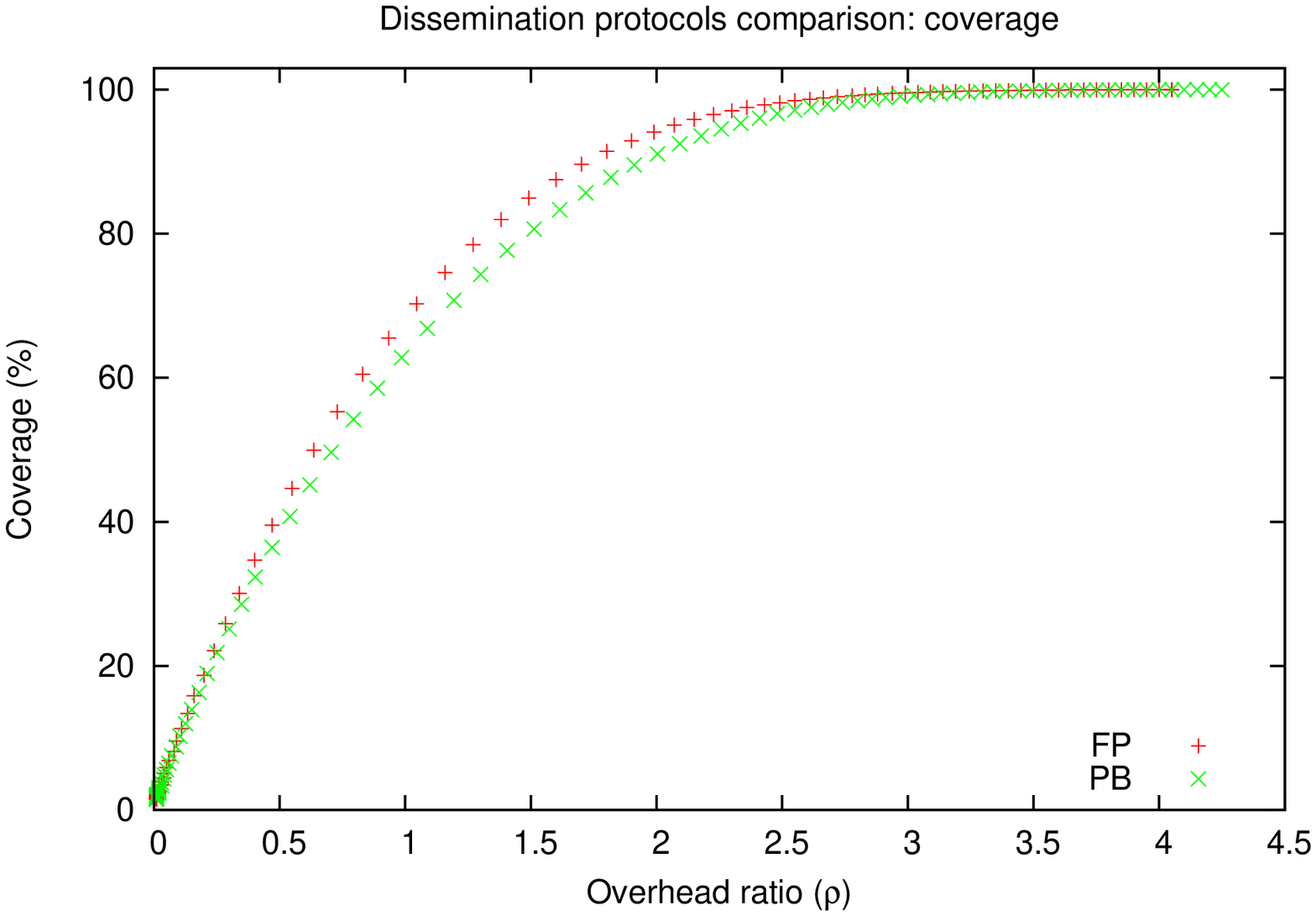}}
\subfloat[\label{fig:kregular_1500edges-delay_FP-CB}]{\includegraphics[angle=270,width=6.5cm]{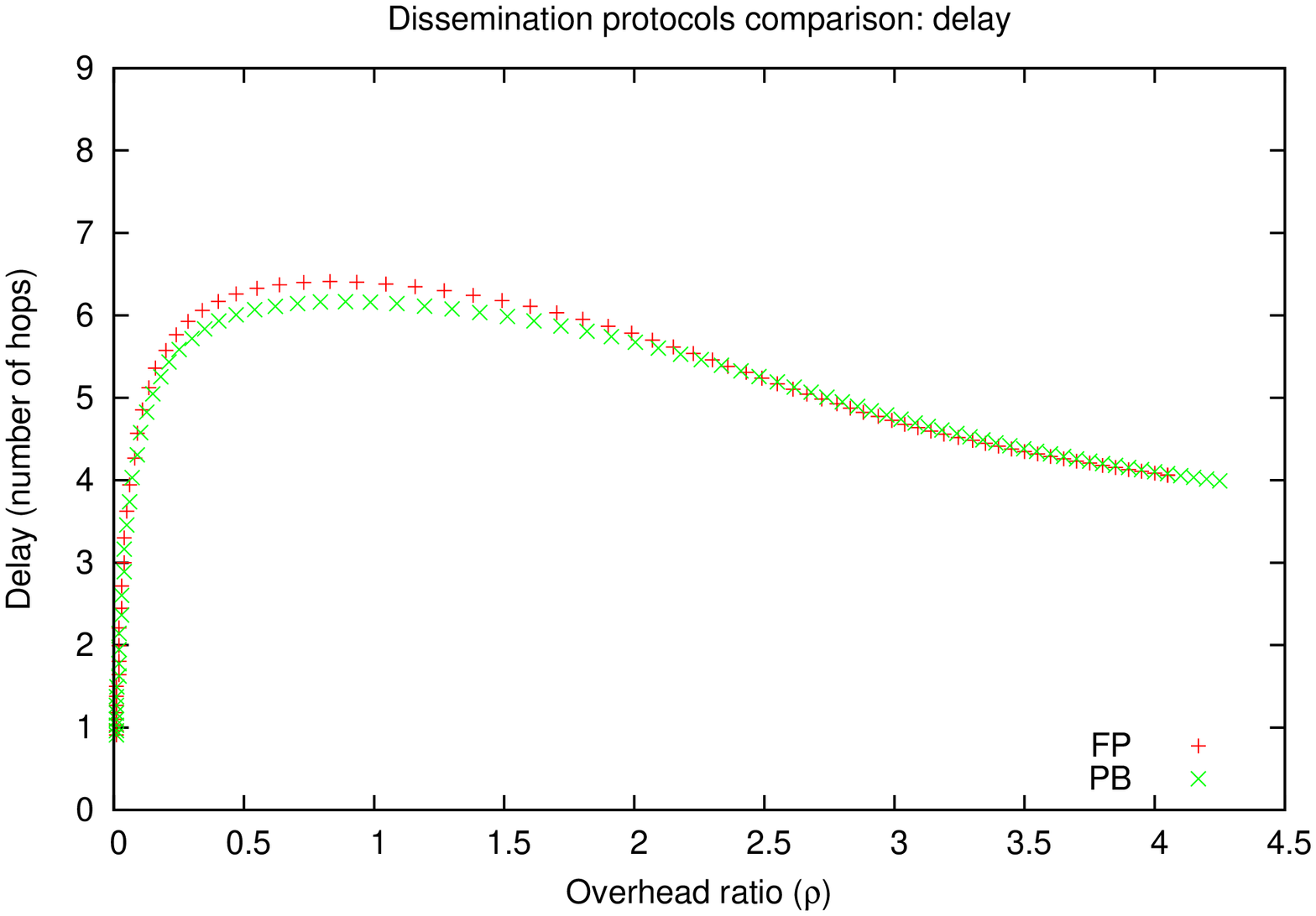}}
\caption{K-regular graphs, 500 nodes, 1500 edges, max diameter=6, TTL=8, cache=256.}
\label{fig:kregular_1500edges}
\subfloat[\label{fig:kregular_2000edges-coverage_FP-CB}]{\includegraphics[angle=270,width=6.5cm]{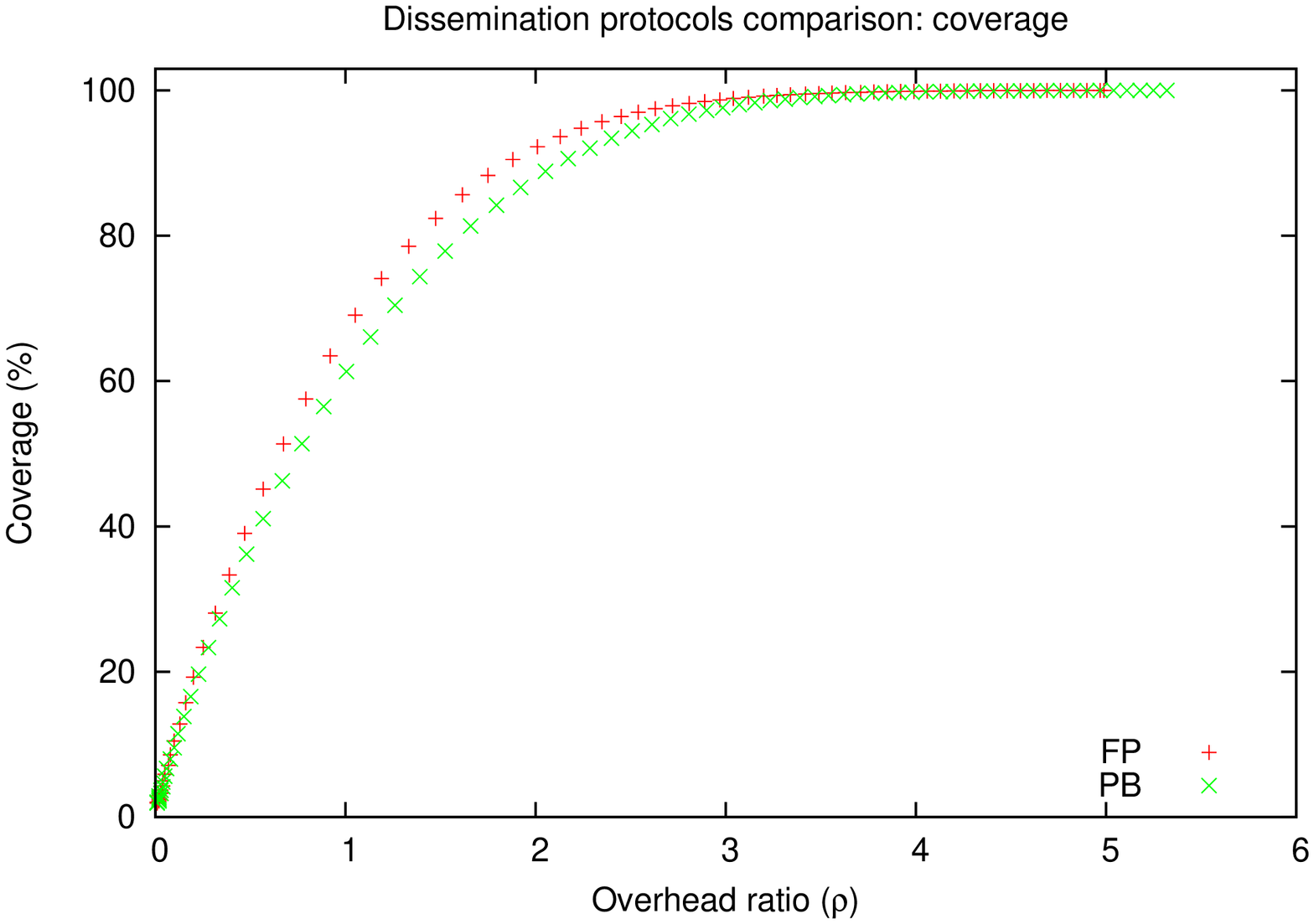}}
\subfloat[\label{fig:kregular_2000edges-delay_FP-CB}]{\includegraphics[angle=270,width=6.5cm]{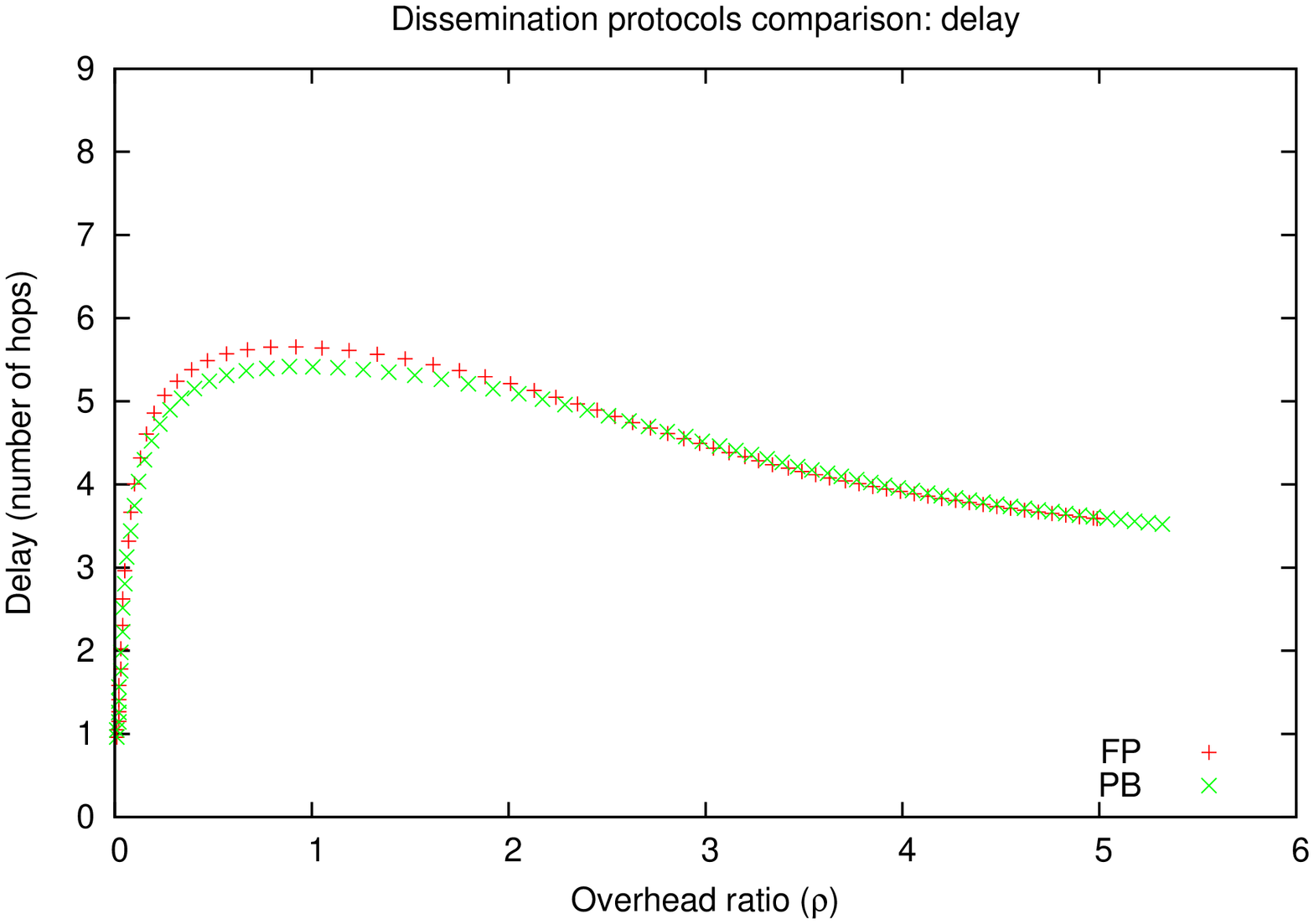}}
\caption{K-regular graphs, 500 nodes, 2000 edges, max diameter=5, TTL=7, cache=256.}
\label{fig:kregular_2000edges}
\end{figure}

As expected, the very regular structure of k-regular networks has an impact on the performance of the dissemination algorithms. More in detail, as usual FP is slightly better than PB (Figures~\ref{fig:kregular_1000edges}, \ref{fig:kregular_1500edges} and \ref{fig:kregular_2000edges}).

\begin{figure}[ht]
\centering
\subfloat[\label{fig:kregular_1000edges-coverage_FP-DDL-DDP}]{\includegraphics[angle=270,width=6.5cm]{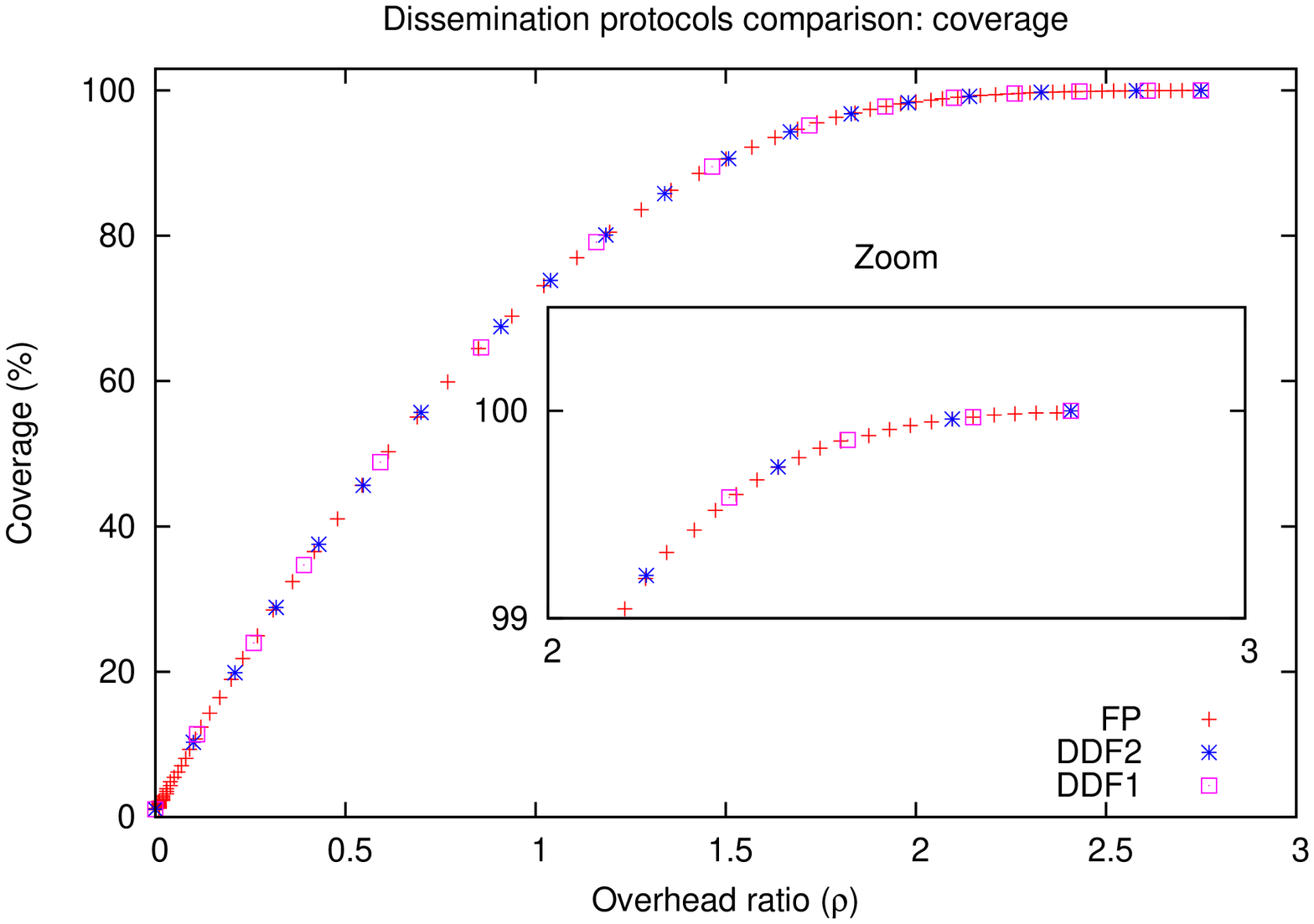}}
\subfloat[\label{fig:kregular_1000edges-delay_FP-DDL-DDP}]{\includegraphics[angle=270,width=6.5cm]{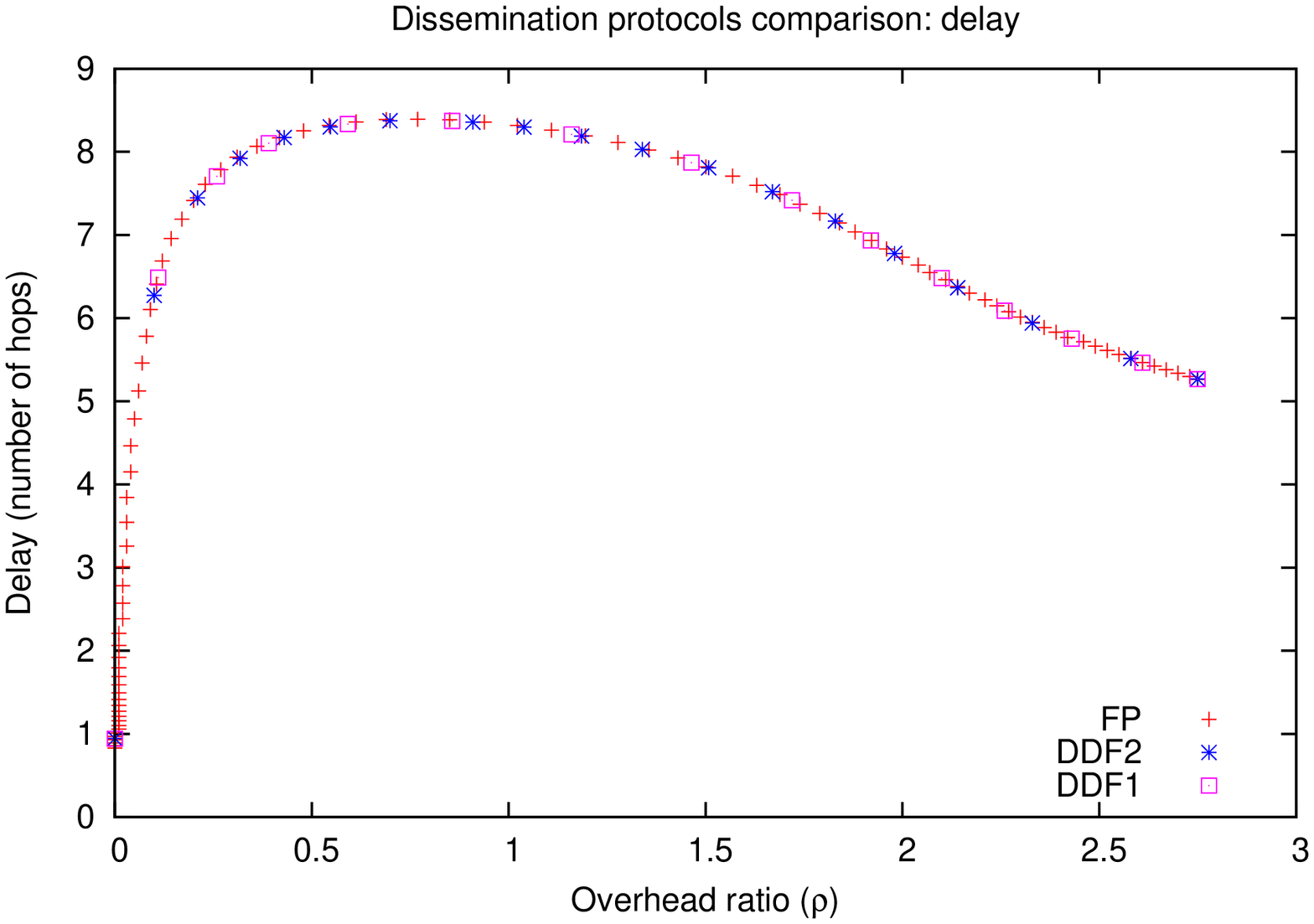}}
\caption{K-regular networks, 500 nodes, 1000 edges, max diameter=8, TTL=11, cache=256.}
\label{fig:kregular_1000edges-extra}
\subfloat[\label{fig:kregular_1500edges-coverage_FP-DDL-DDP}]{\includegraphics[angle=270,width=6.5cm]{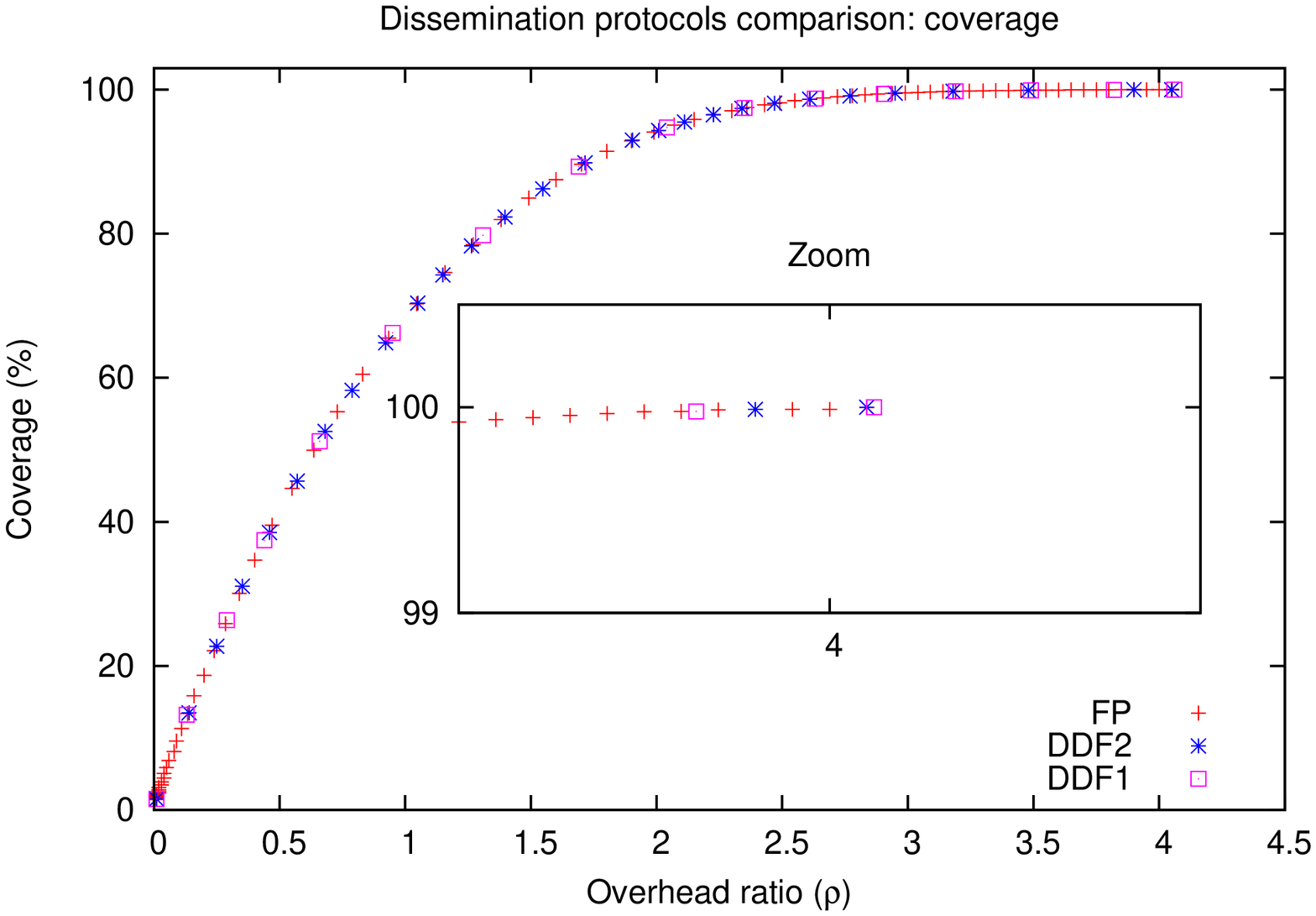}}
\subfloat[\label{fig:kregular_1500edges-delay_FP-DDL-DDP}]{\includegraphics[angle=270,width=6.5cm]{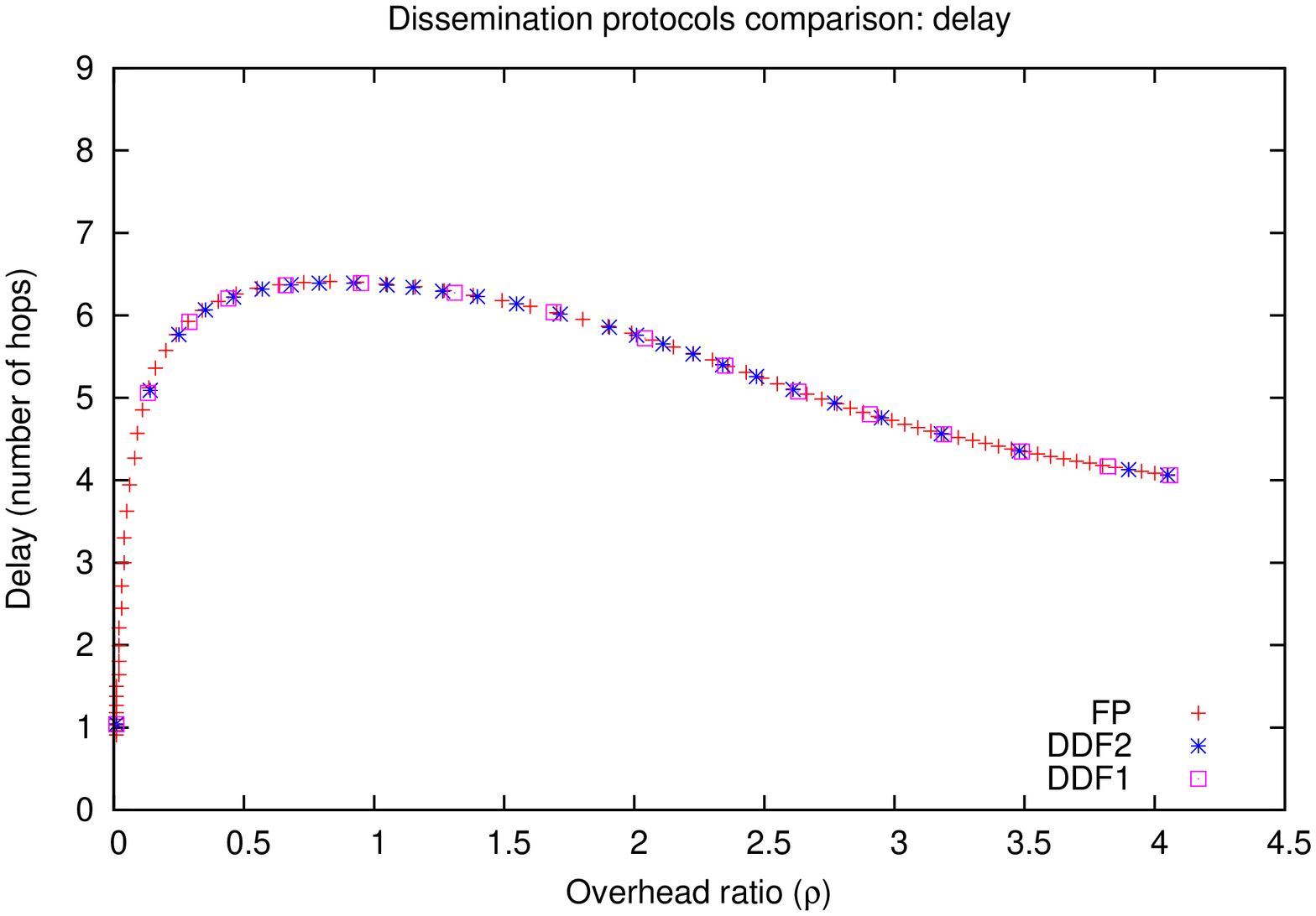}}
\caption{K-regular graphs, 500 nodes, 1500 edges, max diameter=6, TTL=8, cache=256.}
\label{fig:kregular_1500edges-extra}
\subfloat[\label{fig:kregular_2000edges-coverage_FP-DDL-DDP}]{\includegraphics[angle=270,width=6.5cm]{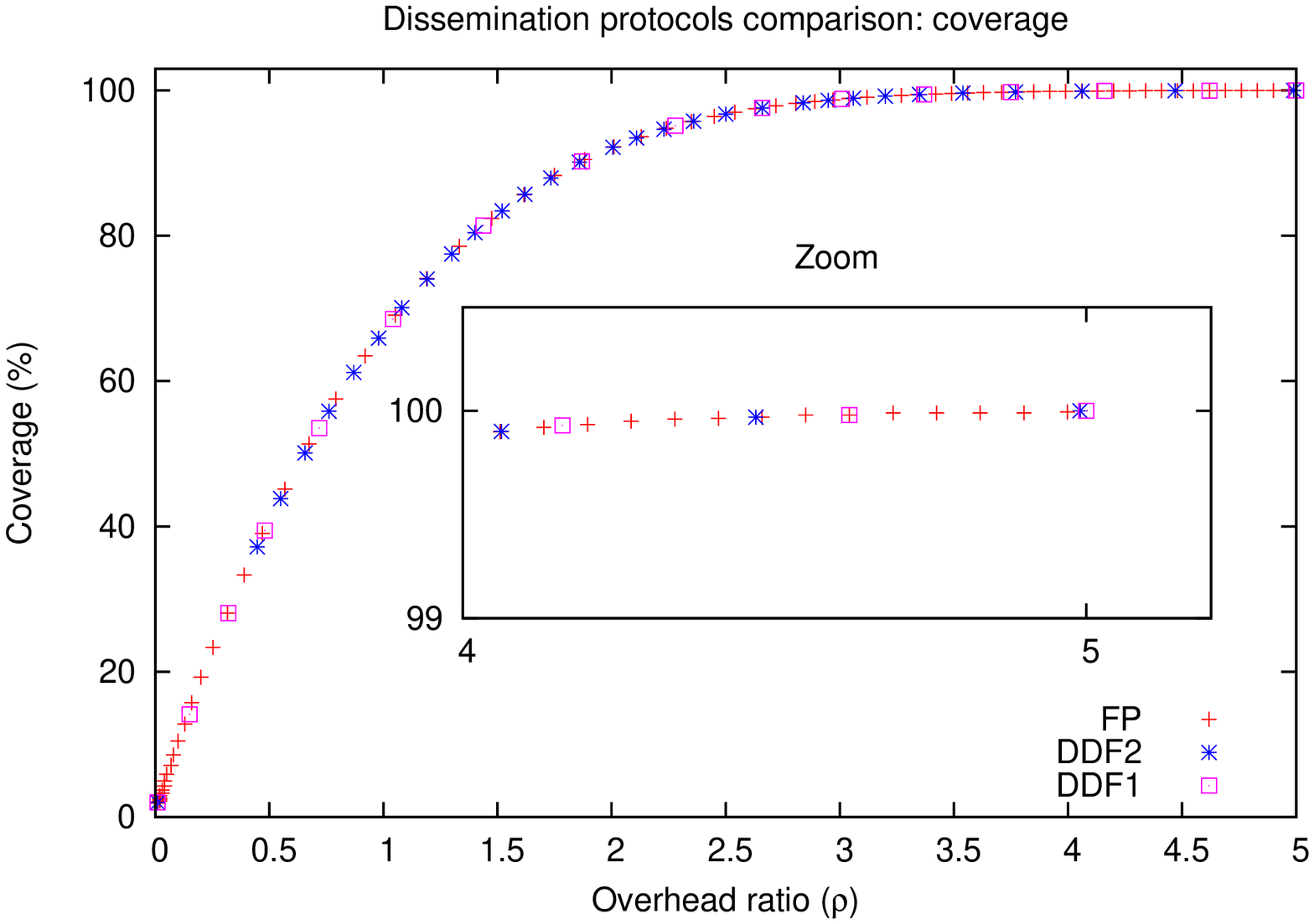}}
\subfloat[\label{fig:kregular_2000edges-delay_FP-DDL-DDP}]{\includegraphics[angle=270,width=6.5cm]{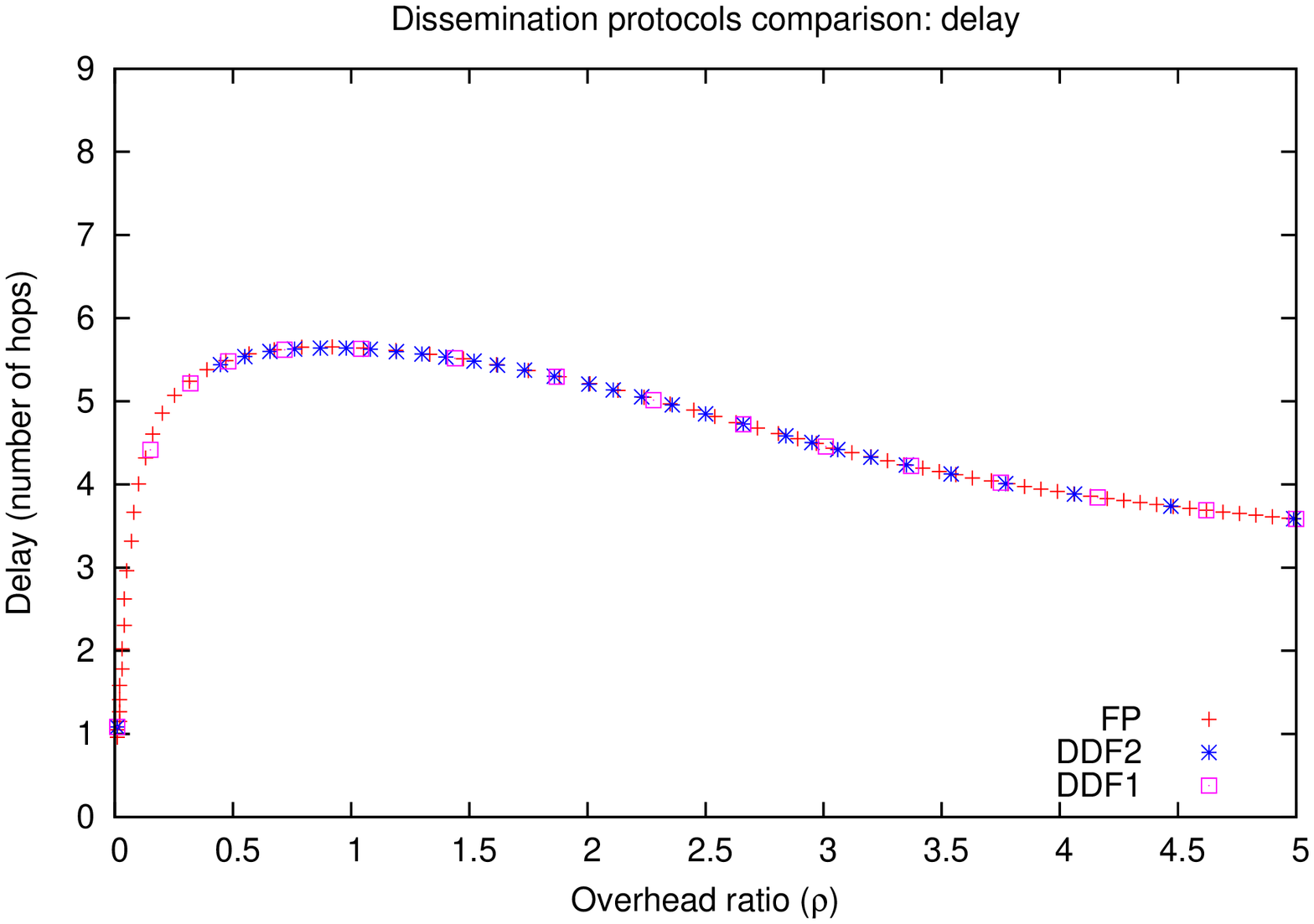}}
\caption{K-regular graphs, 500 nodes, 2000 edges, max diameter=5, TTL=7, cache=256.}
\label{fig:kregular_2000edges-extra}
\end{figure}

The results of DDF1 and DDF2, as expected, are exactly the same of FP (Tables~\ref{tab:kregular_1000edges}, \ref{tab:kregular_1500edges} and \ref{tab:kregular_2000edges} and Figures~\ref{fig:kregular_1000edges-extra}, \ref{fig:kregular_1500edges-extra} and \ref{fig:kregular_2000edges-extra}). 
This is easy to explain given that DDF1 and DDF2 are adaptive variants of FP. In other words, the degree dependent algorithms are unable to tune the dissemination probability in a network in which every node has the same degree by construction. 

\begin{table}[ht]
\begin{center}
\begin{tabular}{ |p{2cm}||p{2cm}|p{2cm}| }
 \hline
 \multicolumn{3}{|c|}{Overhead (and delay) for a given coverage} \\
 \hline
 Algorithm & 100.0\% & \\
 \hline
 \hline
 FP   & 2.75 (5.27) & best \\
 PB   & 2.88 (5.12) & +4.72\% \\
 DDF1  & 2.75 (5.27) & best \\
 DDF2  & 2.75 (5.27) & best \\
 \hline
\end{tabular}
\end{center}
\caption{K-regular networks, 500 nodes, 1000 edges, max diameter=8, TTL=11, cache=256.}
\label{tab:kregular_1000edges}
\end{table}

\begin{table}[ht]
\begin{center}
\begin{tabular}{ |p{2cm}||p{2cm}|p{2cm}| }
 \hline
 \multicolumn{3}{|c|}{Overhead (and delay) for a given coverage} \\
 \hline
 Algorithm & 100.0\% & \\
 \hline
 \hline
 FP   & 4.05 (4.06) & best \\
 PB   & 4.20 (4.01) & +3.70\% \\
 DDF1  & 4.06 (4.06) & +0.24\% \\
 DDF2  & 4.05 (4.06) & best \\
 \hline
\end{tabular}
\end{center}
\caption{K-regular graphs, 500 nodes, 1500 edges, max diameter=6, TTL=8, cache=256.}
\label{tab:kregular_1500edges}
\end{table}

\begin{table}[ht]
\begin{center}
\begin{tabular}{ |p{2cm}||p{2cm}|p{2cm}| }
 \hline
 \multicolumn{3}{|c|}{Overhead (and delay) for a given coverage} \\
 \hline
 Algorithm & 100.0\% & \\
 \hline
 \hline
 FP   & 4.99 (3.59) & best \\
 PB   & 5.25 (3.54) & +5.21\% \\
 DDF1  & 5.00 (3.59) & +0.20\% \\
 DDF2  & 4.99 (3.59) & best \\
 \hline
\end{tabular}
\end{center}
\caption{K-regular graphs, 500 nodes, 2000 edges, max diameter=5, TTL=7, cache=256.}
\label{tab:kregular_2000edges}
\end{table}

\subsection{Free Riding}

Free riding is a common behavior in P2P systems~\cite{hughes2005free}. That is, some nodes benefit from the information provided by the other nodes without offering anything in return to the network. It is clear that the effect of free riding on the protocols performance can be severe~\cite{feldman2004free}. In the case of data dissemination, this means that some nodes (i.e.~free riders) generate new messages to be delivered in the network but they refuse to forward the incoming messages originated by other nodes.

The aim of this section is to investigate the effect of free riding with respect to a specific metric (i.e.~coverage) in presence of different network topologies and dissemination algorithms. For the sake of brevity, only a specific network configuration (i.e.~500 nodes, 3 edges per node) and the best dissemination algorithms resulting from the previous evaluation (i.e.~FP and DDF2) have been tested.

\subsubsection{Random Networks}
In Figure~\ref{fig:random_1500edges-free} is shown the effect of a given percentage of free riders on the coverage that can be obtained for a specific overhead ratio in random networks. Figure~\ref{fig:random_1500edges-coverage_freeFP} demonstrates that even a low percentage of free riders (i.e.~10\%) prevents the FP dissemination to reach full coverage. This means that some messages are discarded by free riders and therefore some nodes (e.g.~leaves) are unable to get such messages from alternative edges. Clearly, this effect on coverage is amplified by a higher percentage of free riders. More specifically, the coverage is reduced by [0.19\%, 0.98\%, 2.43\%] for [10\%, 20\%, 30\%] of free riders. It is worth noting that, even if the configuration with 10\% of free riders is unable to get a full coverage, for some overhead values the coverage is marginally better than without free riders (see the zoom area in Figure~\ref{fig:random_1500edges-coverage_freeFP}). The main reason behind this behavior is cache efficiency. In fact, due to the presence of free riders, a lower number of messages is delivered and the caches implemented in each node are able to operate with a slightly higher efficiency. This leads to less duplicated messages and hence a higher coverage for a given overhead value.

Figure~\ref{fig:random_1500edges-coverage_freeDDF2} shows the effect of free riding on DDF2. In this case, the coverage is reduced by [0.18\%, 1\%, 2.4\%]. This means that, in terms of best coverage, the free riding has the same effect on both dissemination algorithms. In the case of DDF2, a 10\% of free riders reduces the coverage for all the significant overhead values. This means that with DDF2 there is not the cache efficiency improvement seen with FP. This result is in accordance with the characteristics of the degree dependent algorithms.

\subsubsection{Scale-Free Networks}
\begin{figure}[ht]
\centering
\subfloat[\label{fig:random_1500edges-coverage_freeFP}]{\includegraphics[angle=270,width=6.5cm]{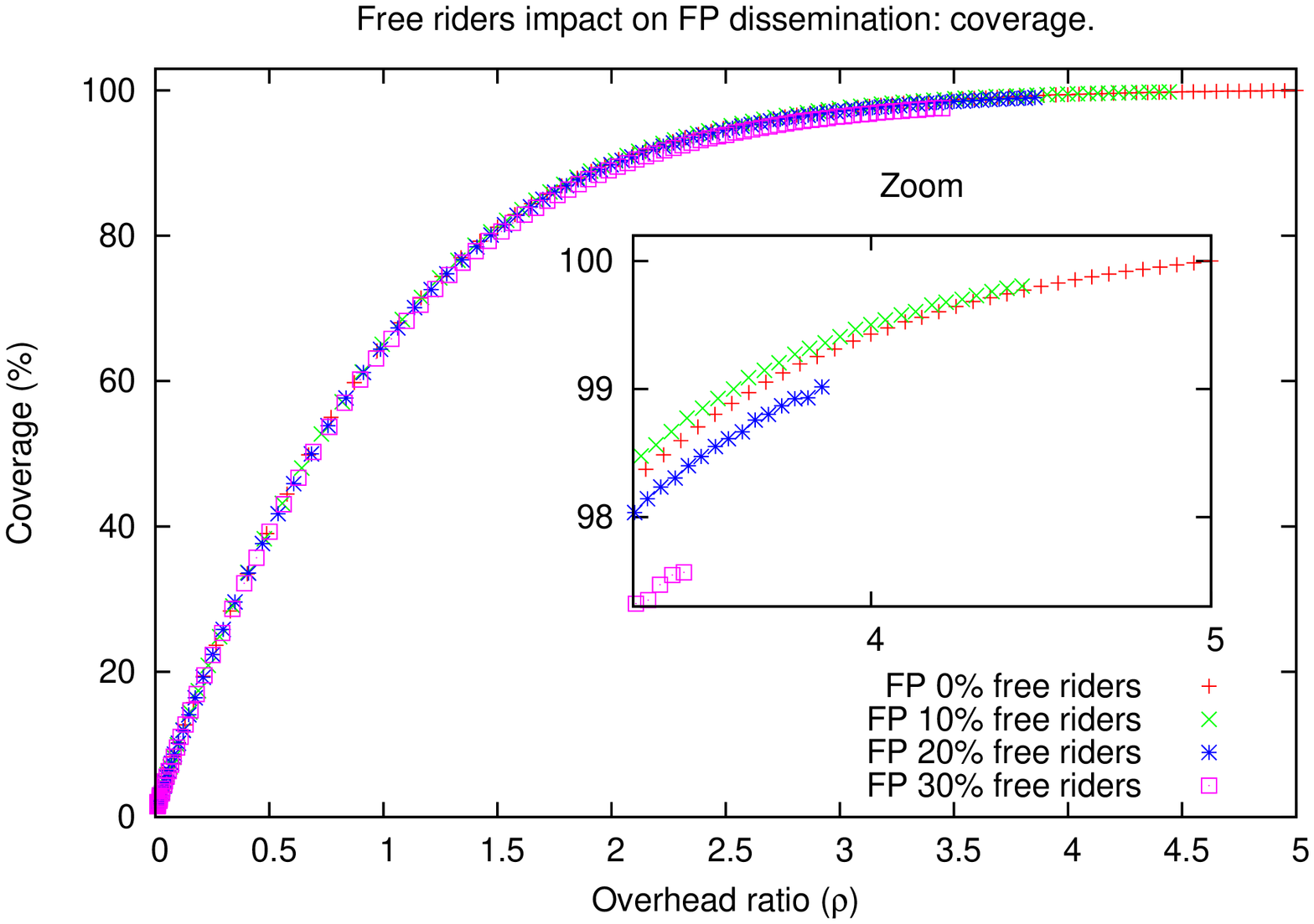}}
\subfloat[\label{fig:random_1500edges-coverage_freeDDF2}]{\includegraphics[angle=270,width=6.5cm]{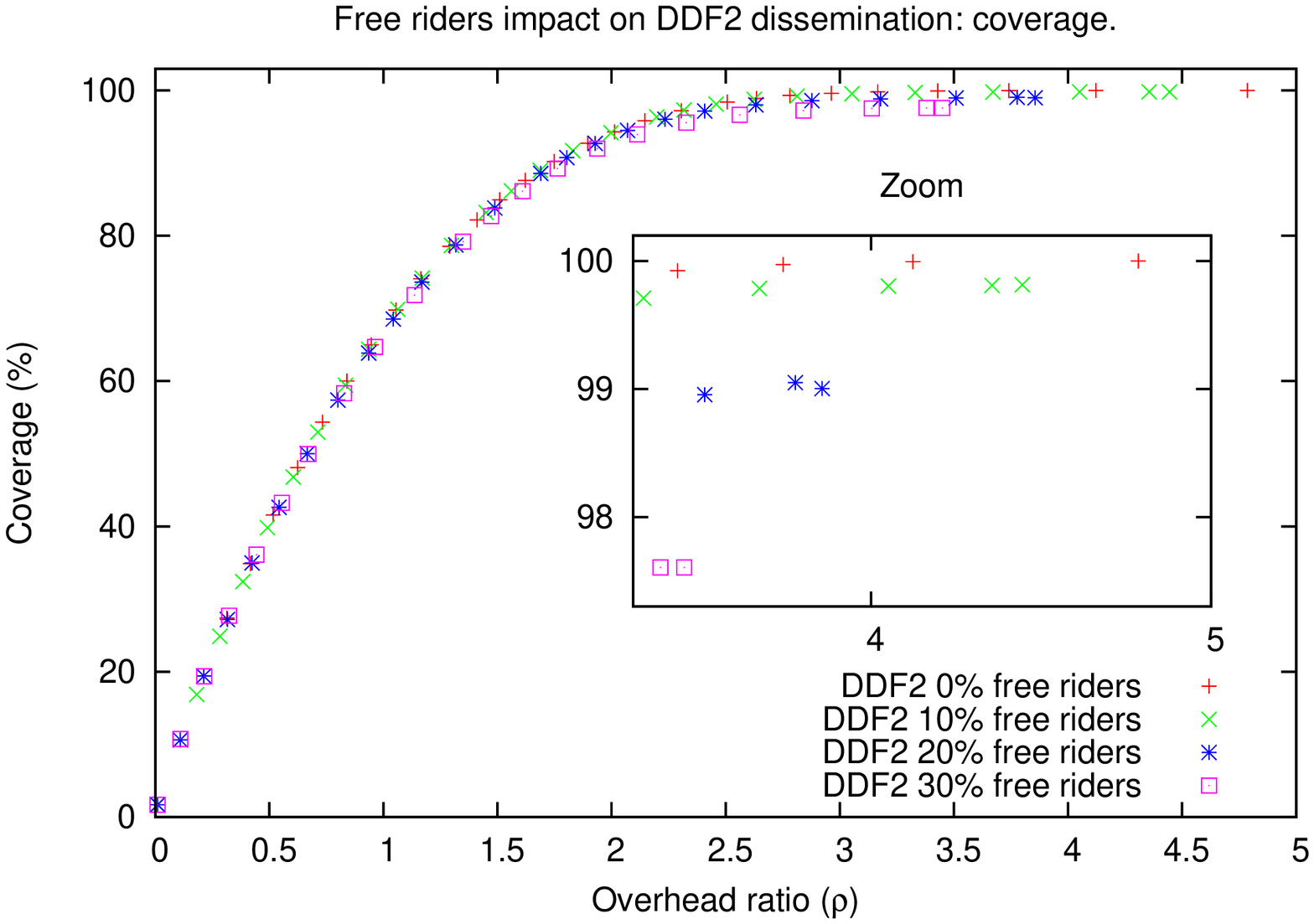}}
\caption{Random networks, 500 nodes, 1500 edges, max diameter=7, TTL=10, cache=256. a) Fixed Probability vs. b) DDF2.}
\label{fig:random_1500edges-free}
\subfloat[\label{fig:scalefree_1494edges-coverage_freeFP}]{\includegraphics[angle=270,width=6.5cm]{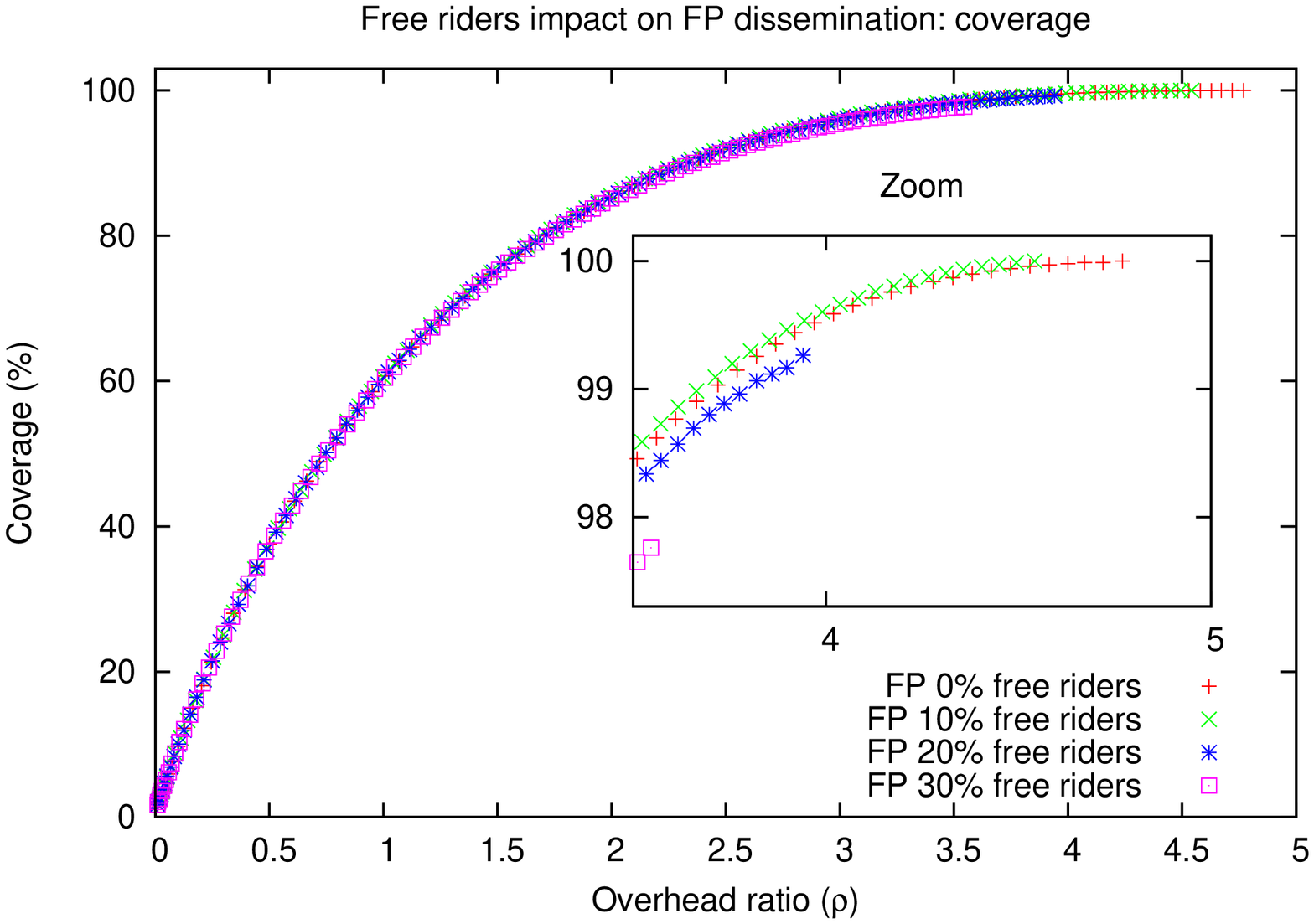}}
\subfloat[\label{fig:scalefree_1494edges-coverage_freeDDF2}]{\includegraphics[angle=270,width=6.5cm]{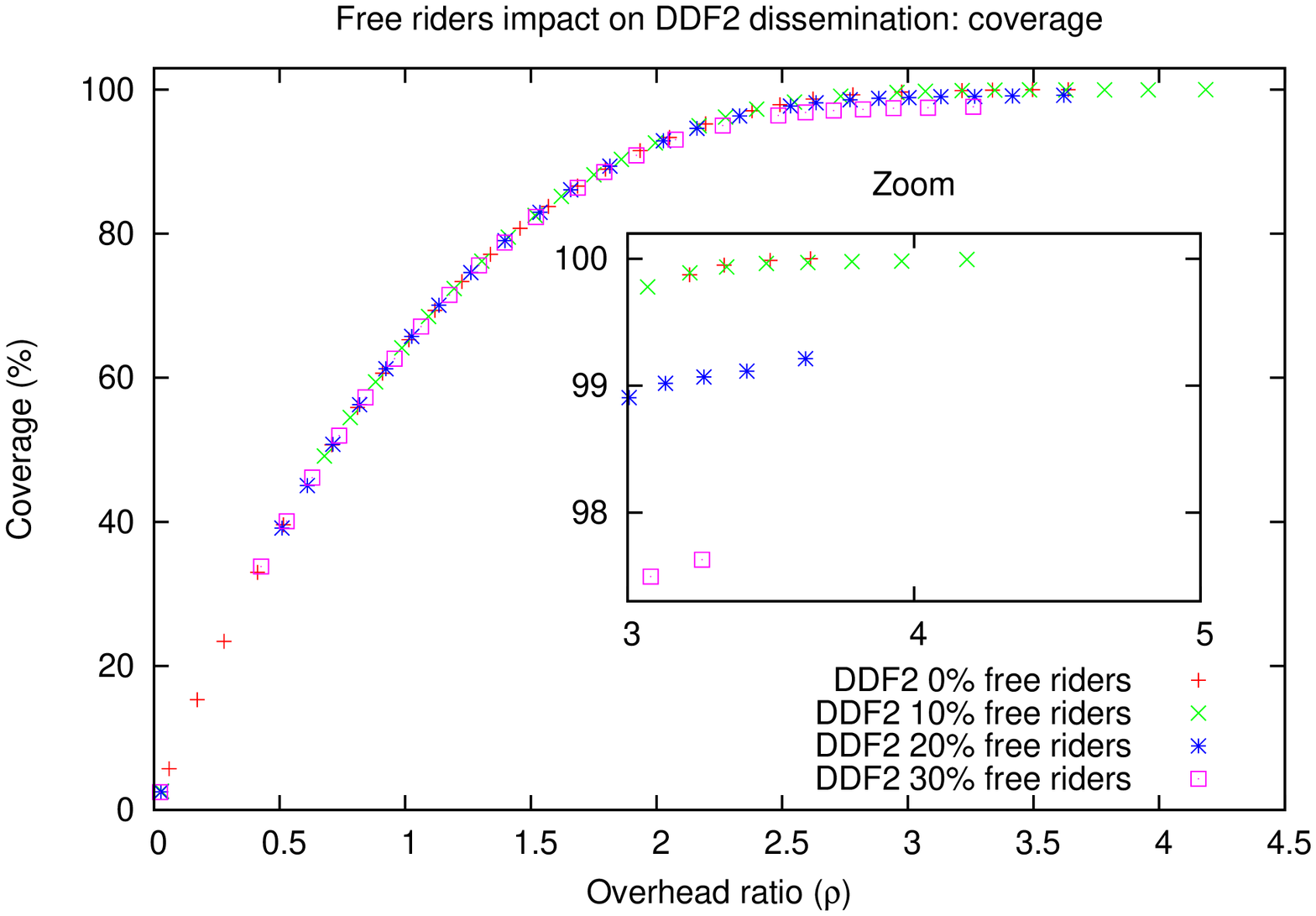}}
\caption{Scale-free networks, 500 nodes, 1494 edges, max diameter=5, TTL=7, cache=256. a) Fixed Probability vs. b) DDF2.}
\label{fig:scalefree_1494edges-free}
\end{figure}

Scale-free networks better deal with free riders than random networks (Figure~\ref{fig:scalefree_1494edges-free}). In fact, with 10\% of free riders both FP and DDF2 are still able to get full dissemination. More specifically, FP obtains a [0\%, 0.74\%, 2.24\%] coverage and DDF2 gets [0\%, 0.79\%, 2.37\%]. It is worth noting that, also in this case, FP and DDF2 obtain results that are comparable. This means that, if we limit our analysis to the FP and DDF2, then the effect of free riding on coverage is more dependent on the network topology than on the dissemination algorithm.

\subsubsection{Small-World Networks}
The effect of free riding on small-world networks is twofold (Figure~\ref{fig:smallworld_1500edges-free}). In fact, a small amount (i.e.~10\%) of free riders is able to prevent full dissemination. On the other hand, for higher levels of free riding (i.e.~20\% and 30\%) the reduction in coverage is more limited than with random and scale-free topologies. Such as in previous cases, FP and DDF2 obtain results that are very close. That is [0.08\%, 0.41\%, 1.13\%] for FP and [0.07\%, 0.40\%, 1.07\%] for DDF2.

\subsubsection{K-regular Networks}
\begin{figure}[ht]
\centering
\subfloat[\label{fig:smallworld_1500edges-coverage_freeFP}]{\includegraphics[angle=270,width=6.5cm]{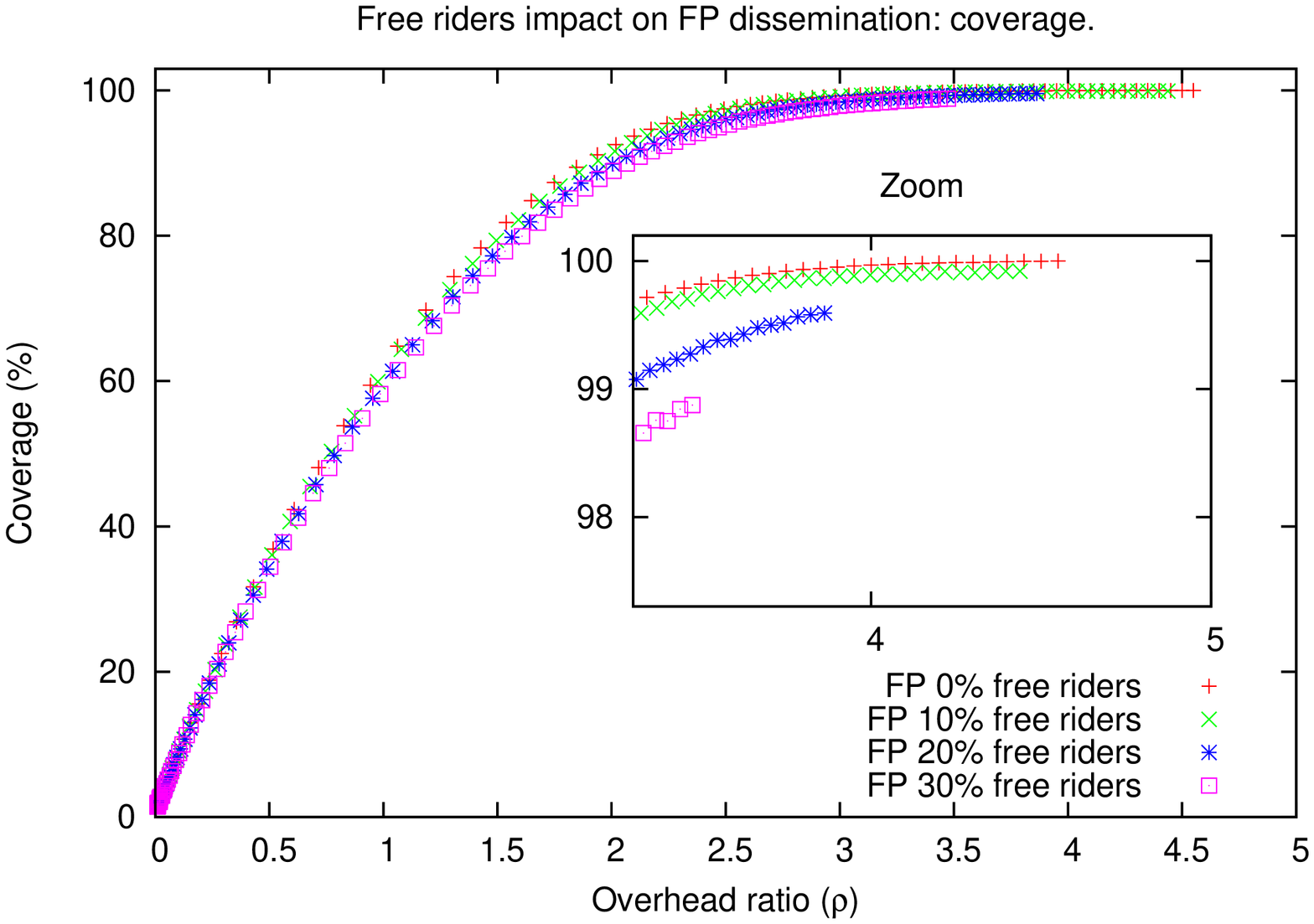}}
\subfloat[\label{fig:smallworld_1500edges-coverage_freeDDF2}]{\includegraphics[angle=270,width=6.5cm]{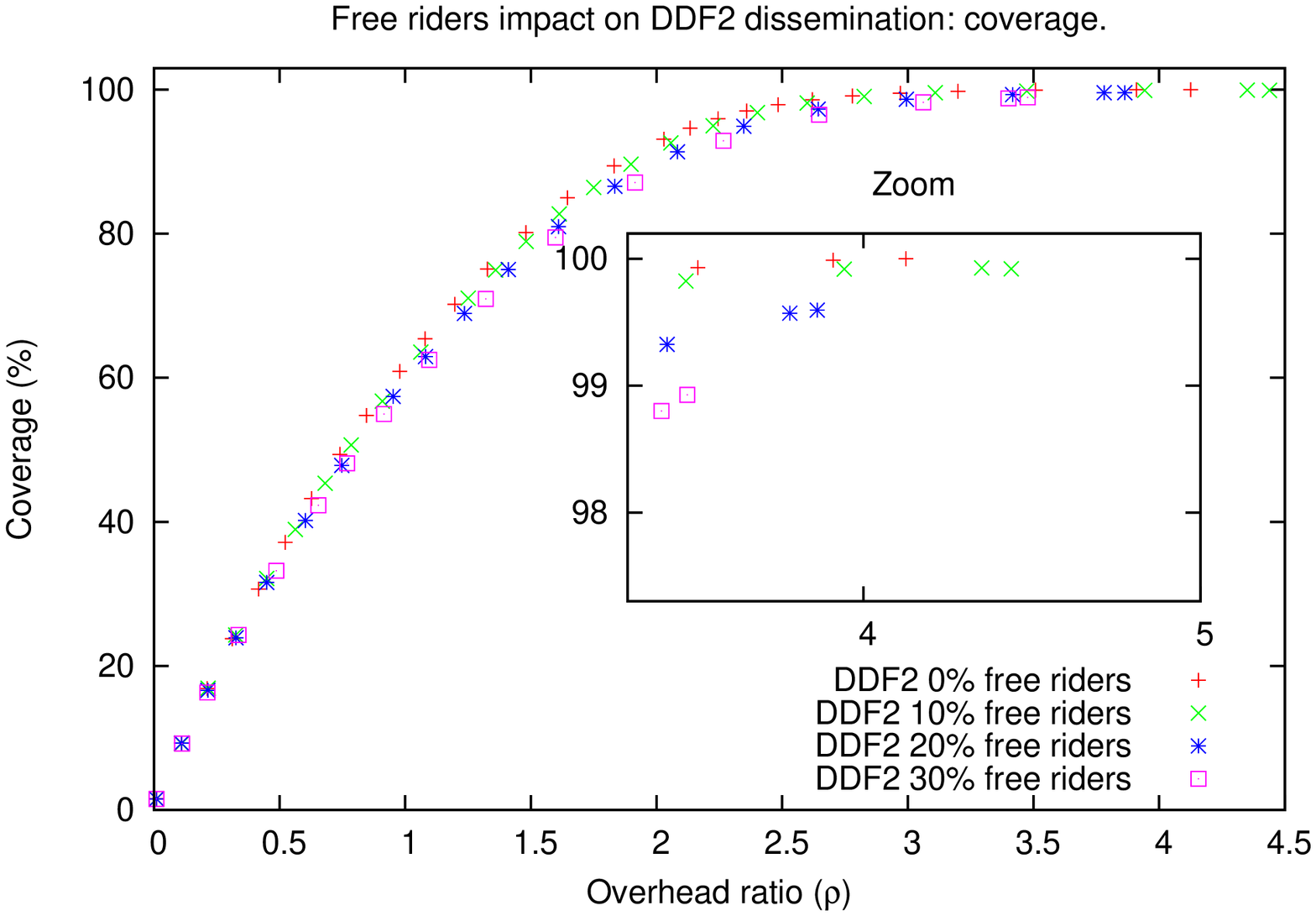}}
\caption{Small-world networks, 500 nodes, 1500 edges, rewiring probability=0.1, max diameter=9, TTL=12, cache=256. a) Fixed Probability vs. b) DDF2.}
\label{fig:smallworld_1500edges-free}
\subfloat[\label{fig:kregular_1500edges-coverage_freeFP}]{\includegraphics[angle=270,width=6.5cm]{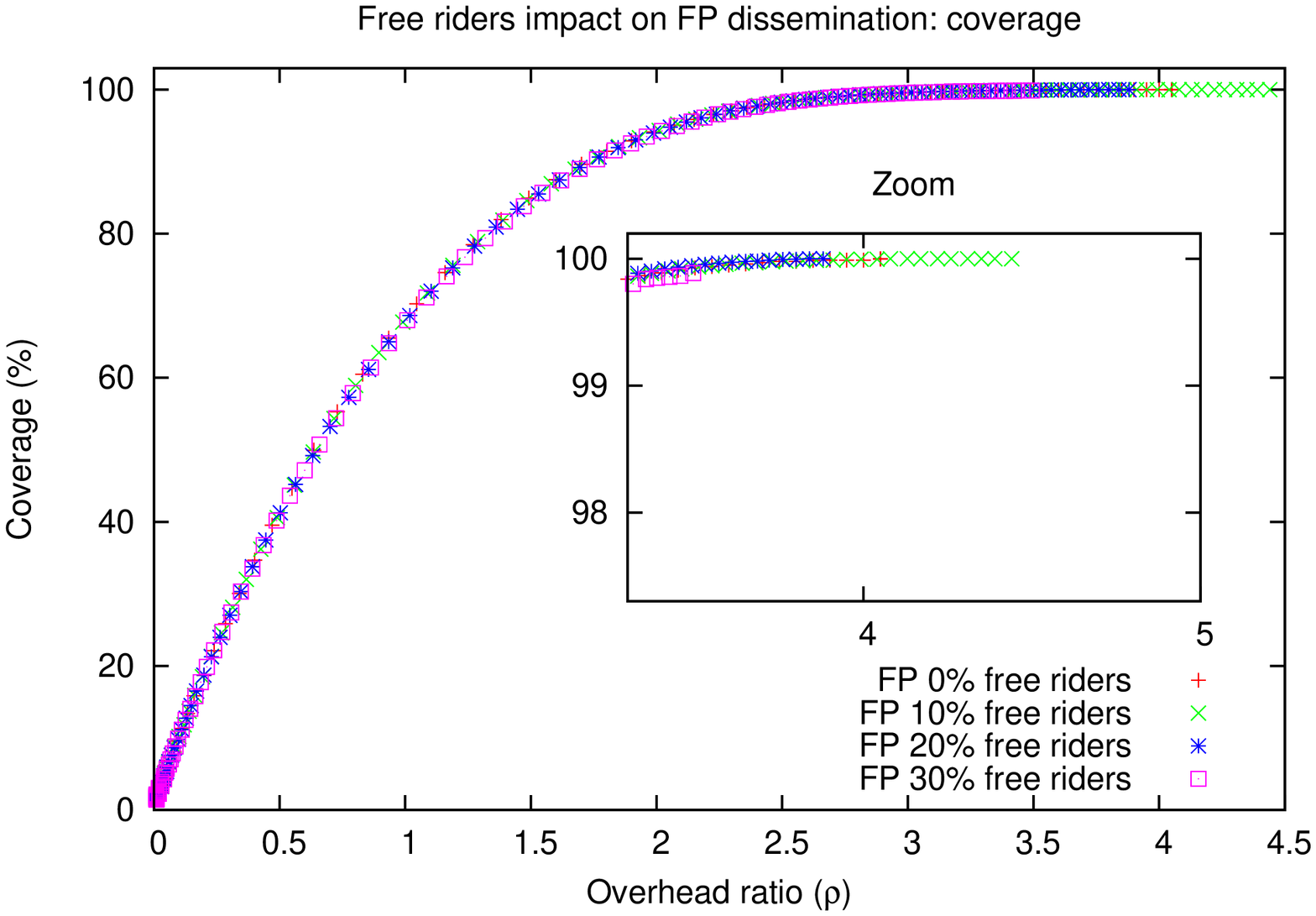}}
\subfloat[\label{fig:kregular_1500edges-coverage_freeDDF2}]{\includegraphics[angle=270,width=6.5cm]{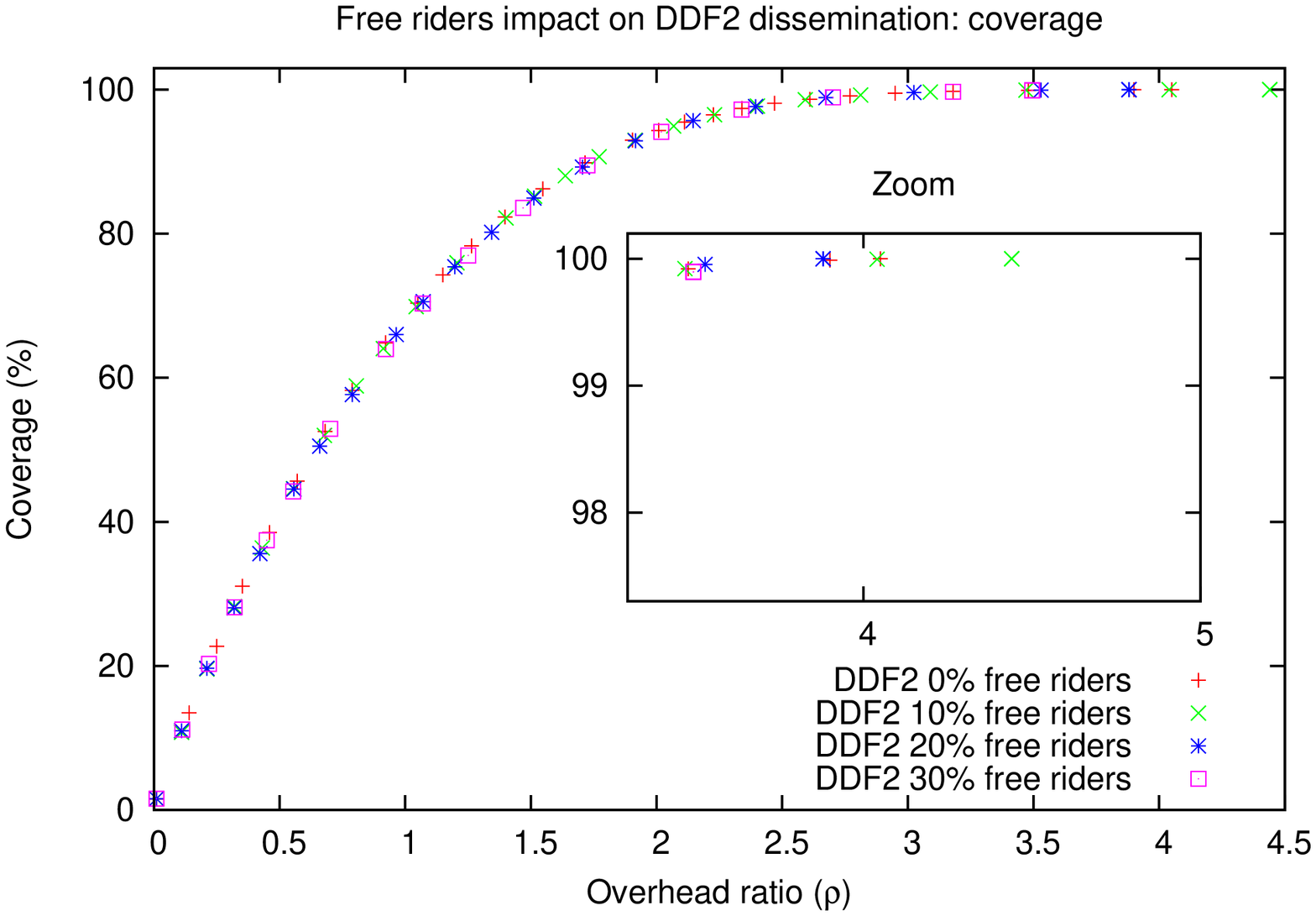}}
\caption{K-regular graphs, 500 nodes, 1500 edges, max diameter=6, TTL=8, cache=256. a) Fixed Probability vs. b) DDF2.}
\label{fig:kregular_1500edges-free}
\end{figure}

K-regular networks are the topology less affected by free riding (i.e.~FP [0\%, 0\%, 0.11\%] and DDF2 [0\%, 0\%, 0.10\%]). As shown in Figure~\ref{fig:kregular_1500edges-free}, for up to 20\% of free riders both the dissemination algorithms are able to get full dissemination. With 30\%, the reduction of coverage is negligible. This good resistance to free riding is given by the uniform structure of k-regular networks. In fact, in the absence of hubs and leaf nodes, the disconnection of nodes from the network (due to free riders) is less likely.

\section{Discussion}\label{sec:discussion}

In this section, the main results from previous performance evaluation are discussed more fully.

\subsection{On the Performance of Gossip}
We already mentioned that the considered schemes allow disseminating messages in an unstructured overlay, whatever its topology, at the cost of some redundancy in the transmission of messages. Degree dependent gossip-based protocols outperform in most cases other standard dissemination schemes, mostly in terms of coverage. In particular, DDF2 performs better than DDF1, in general. However, the dissemination protocol to use might be based on the topology of the P2P unstructured overlay. In fact, results show that the degree dependent gossip protocols work well in random graphs and scale-free networks, where the structure of these topologies is purely based on the degrees setting, and the connections among nodes is arbitrary.
Conversely, in regular networks there is no need to employ a degree dependent gossip protocol, since there are no nodes with degree higher than others; hence, this approach cannot provide important benefits.

Results show that, when dealing with small-world networks, the performance improvements of the degree-dependent gossip scheme is (present but) lower than in random and scale-free topologies. This is probably due by the topology of a small world network, that is typically formed by an amount of local links (i.e., links with nodes that are in turn neighbors by themselves) and by some few ``long distance'' links, that connect a node with others placed in other portions of the network (i.e., apart from that link that connect two nodes, say $x$ and $y$, alternative paths from $x$ to $y$ have much higher distances). Gossip protocols select receivers randomly, without exploiting the presence of such long distance links. Indeed, in these topologies an informed approach might be preferred, that takes into consideration the peculiarity of links in order to fasten the dissemination process. 

There is a trade-off between delay and overhead (i.e.~number of delivered messages over the minimum number of messages needed to obtain a complete coverage). Indeed, delay can be lowered in a gossip protocol if we increase the gossip probability or, if we are able to change the setting parameters of peers, by increasing the node degrees. In fact, this would give to each peer more chances to disseminate a message. But, the higher the amount of relayed messages the higher the overhead.

Finally, it is worth noticing that these results can be used to set a network in order to guarantee certain communication properties. Given an overlay topology, if a certain network coverage should be guaranteed during the dissemination of a message, then a corresponding overhead and delay have to be expected. 

\subsection{On the Gossip Dissemination Algorithms}
In this work, the gossip dissemination algorithms that have been considered are quite simple and all of them are push-based. The rationale of this decision is that, in this paper, we aim to demonstrate that our simulation-based approach permits the analysis of a gossip protocol, determining when and how it is effective, given an underlying network topology. In our view, this is the prerequisite for the study of more complex dissemination strategies. In fact, the conclusions we reached in this paper can be easily extended to other kinds of dissemination protocols, including more complex ones such as pull or push-pull schemes. Thanks to the simulator structure, it is quite easy to implement new dissemination protocols (or to change the behavior of existing ones). This will allow a more comprehensive comparison of dissemination protocols when run on different underlying network topologies.

\subsection{On Churn}
Our analysis assumes that the time required for the dissemination of a message is of an order of magnitude lower than the typical time for perceiving a significant network alteration, in terms of topology. Hence, due to churn, neighbors of a node change in time, but such changes of the network do not alter significantly the topology of an unstructured overlay, during the execution of the gossip procedure. This is a common practice in simulation of P2P systems \cite{gda-simutools-09,Kempe:2004,gda-disio-11-2,Anagnostopoulos:2012:AEI:2365374.2365840,conf/nca/GarbinatoRT07}.

A further motivation for such a simplification is that the probabilistic dissemination of a message in an unstructured network does not require (as the name says by itself) a reconfiguration of the structure of the overlay, if a neighbor node leaves the network. This means that a node in charge of relaying a message decides the set of receiving nodes independently, and the failure of a neighbor node has (in general) no implications on other neighbors. A benefit due to this assumption of considering a overlay static, during the dissemination of a single message, is that it allows more easily measuring metrics of interest such as coverage, delay and overhead ratio. 

\subsection{On Free Riding}
As expected, the presence of free riders in the overlay network has an impact on the evaluated metrics but actually such an impact is quite limited. The previous evaluation shows that the network topology is much important than the dissemination algorithm to limit the effect of free riding on the dissemination coverage. In other words, in the design of P2P overlays, free riding is a parameter that must be considered due to its impact on performance. In practice, thanks to our simulator, it is possible to investigate the effect of free riding also on complex dissemination algorithms. Moreover, setting specific dissemination probabilities for each node, it is even possible to study the effect of ``partial'' and ``transient'' free riders on the dissemination metrics.

\section{Conclusions}\label{sec:conclusions}

In this paper, we presented a study on highly intensive data dissemination protocols over unstructured P2P overlays. We compared four protocols, i.e.~fixed probability, probabilistic broadcast and two degree dependent gossip-based protocols that change the probability of dissemination based on the degree of peers. Our results show that the degree dependent gossip-based protocols outperform in most cases other standard dissemination schemes, mostly in terms of coverage.

We have provided a theoretical model that allows determining the threshold values for having that a disseminated messages reaches a giant component of the network. Moreover, thanks to the features of LUNES, a parallel and distributed simulator we built, we were able to study the impact of cache and TTL on data dissemination, in terms of coverage and delay on disseminations in which the number of delivered messages is in the order of millions. Outcomes confirm that these technical strategies have an important influence on the performance of dissemination schemes. Moreover, dissemination schemes are quite effective to spread information in P2P overlay networks, whatever their topology.

A main result of this study is that, given a set of nodes in a P2P overlay, and the need to design and create an overlay, it is possible to understand how setting the network in order to guarantee certain communication properties. 
In particular, our study shows that given an overlay topology, a certain overhead and delay can be expected to have a particular network coverage, during the dissemination of a message. 
Then, increasing the number of links decreases the average delays, at the cost of higher overheads (i.e.~number of delivered messages over the minimum number of messages needed to obtain a complete coverage). Moreover, the selection of the dissemination protocol should be based on the topology (and vice versa). As mentioned, the degree dependent gossip approach works well in random graphs and scale-free networks. 
Indeed, the structure of these topologies is purely based on the degrees setting, and the connections among nodes is arbitrary.
Conversely, as expected its improvements are negligible in regular networks, since there are no nodes with degree higher than others.
Similarly, in order to take advantage of the peculiarities of a small-world network, an informed approach should be preferred, that for instance, takes into account the distance among nodes. In fact, our results show that in the case of small-world network the degree-dependent scheme improves performances, but less than in random and scale-free topologies.

Finally, a preliminary evaluation of the impact of free riding on the studied metrics has been reported. The main outcome is that, actually, the presence of free riders has an impact on coverage but actually such an impact is quite limited. More specifically, it results that the effect of free riding (on coverage) is more dependent on the network topology than on the dissemination algorithm.

\section*{Acknowledgments} 

We would like to thank Giulio Cirnigliaro for his implementation of the degree dependent dissemination protocols and the anonymous reviewers whose detailed comments (and suggestions) greatly contributed to improve the overall quality of this paper.

\section*{Acronyms}

\begin{tabbing}
\hspace{15mm}\=\kill
LUNES \> Large Unstructured NEtwork Simulator\\
FP \> Fixed Probability (dissemination protocol)\\
PB \> Probabilistic Broadcast (dissemination protocol)\\
DDF1 \> Degree Dependent Function 1 (dissemination protocol)\\
DDF2 \> Degree Dependent Function 2 (dissemination protocol)\\
\end{tabbing}

\bibliographystyle{elsarticle-num}
\bibliography{paper}

\begin{thebibliography}{10}
\expandafter\ifx\csname url\endcsname\relax
  \def\url#1{\texttt{#1}}\fi
\expandafter\ifx\csname urlprefix\endcsname\relax\def\urlprefix{URL }\fi
\expandafter\ifx\csname href\endcsname\relax
  \def\href#1#2{#2} \def\path#1{#1}\fi

\bibitem{gda-simutools-09}
G.~D'Angelo, S.~Ferretti, Simulation of scale-free networks, in: Simutools '09:
  Proceedings of the 2nd International Conference on Simulation Tools and
  Techniques, ICST (Institute for Computer Sciences, Social-Informatics and
  Telecommunications Engineering), ICST, Brussels, Belgium, Belgium, 2009, pp.
  1--10.
\newblock \href
  {http://dx.doi.org/http://dx.doi.org/10.4108/ICST.SIMUTOOLS2009.5672}
  {\path{doi:http://dx.doi.org/10.4108/ICST.SIMUTOOLS2009.5672}}.

\bibitem{KincaidA05}
R.~K. Kincaid, N.~M. Alexandrov, Scale-free networks: A discrete event
  simulation approach, in: International Conference on Computational Science
  (1), 2005.

\bibitem{dobrescu}
R.~Dobrescu, S.~Taralunga, S.~Mocanu, Web traffic simulation with scale-free
  network models, in: AIC'07: Proc. of the 7th Conference on 7th WSEAS
  International Conference on Applied Informatics and Communications, WSEAS,
  2007.

\bibitem{Ferretti:2012}
S.~Ferretti,
  \href{http://doi.acm.org/10.1145/2184356.2184359}{Publish-subscribe systems
  via gossip: A study based on complex networks}, in: Proceedings of the Fourth
  Annual Workshop on Simplifying Complex Networks for Practitioners, SIMPLEX
  '12, ACM, New York, NY, USA, 2012, pp. 7--12.
\newblock \href {http://dx.doi.org/10.1145/2184356.2184359}
  {\path{doi:10.1145/2184356.2184359}}.
\newline\urlprefix\url{http://doi.acm.org/10.1145/2184356.2184359}

\bibitem{Kempe:2004}
D.~Kempe, J.~Kleinberg, A.~Demers,
  \href{http://doi.acm.org/10.1145/1039488.1039491}{Spatial gossip and resource
  location protocols}, J. ACM 51~(6) (2004) 943--967.
\newblock \href {http://dx.doi.org/10.1145/1039488.1039491}
  {\path{doi:10.1145/1039488.1039491}}.
\newline\urlprefix\url{http://doi.acm.org/10.1145/1039488.1039491}

\bibitem{Alistarh:2010:EG:1880999.1881012}
D.~Alistarh, S.~Gilbert, R.~Guerraoui, M.~Zadimoghaddam,
  \href{http://dl.acm.org/citation.cfm?id=1880999.1881012}{How efficient can
  gossip be? (on the cost of resilient information exchange)}, in: Proceedings
  of the 37th International Colloquium Conference on Automata, Languages and
  Programming: Part II, ICALP'10, Springer-Verlag, Berlin, Heidelberg, 2010,
  pp. 115--126.
\newline\urlprefix\url{http://dl.acm.org/citation.cfm?id=1880999.1881012}

\bibitem{Fernandess:2007}
Y.~Fernandess, A.~Fern\'{a}ndez, M.~Monod,
  \href{http://doi.acm.org/10.1145/1317379.1317384}{A generic theoretical
  framework for modeling gossip-based algorithms}, SIGOPS Oper. Syst. Rev.
  41~(5) (2007) 19--27.
\newblock \href {http://dx.doi.org/10.1145/1317379.1317384}
  {\path{doi:10.1145/1317379.1317384}}.
\newline\urlprefix\url{http://doi.acm.org/10.1145/1317379.1317384}

\bibitem{Georgiou:2013}
C.~Georgiou, S.~Gilbert, R.~Guerraoui, D.~R. Kowalski,
  \href{http://doi.acm.org/10.1145/2450142.2450147}{Asynchronous gossip}, J.
  ACM 60~(2) (2013) 11:1--11:42.
\newblock \href {http://dx.doi.org/10.1145/2450142.2450147}
  {\path{doi:10.1145/2450142.2450147}}.
\newline\urlprefix\url{http://doi.acm.org/10.1145/2450142.2450147}

\bibitem{Jelasity:2005}
M.~Jelasity, A.~Montresor, O.~Babaoglu,
  \href{http://doi.acm.org/10.1145/1082469.1082470}{Gossip-based aggregation in
  large dynamic networks}, ACM Trans. Comput. Syst. 23~(3) (2005) 219--252.
\newblock \href {http://dx.doi.org/10.1145/1082469.1082470}
  {\path{doi:10.1145/1082469.1082470}}.
\newline\urlprefix\url{http://doi.acm.org/10.1145/1082469.1082470}

\bibitem{gda-disio-11-2}
G.~D'Angelo, S.~Ferretti, M.~Marzolla, Adaptive event dissemination for
  peer-to-peer multiplayer online games, in: Proceedings of Workshop on
  DIstributed SImulation and Online gaming ({DISIO 2011}), ICST, ICST,
  Brussels, Belgium, Belgium, 2011.

\bibitem{hughes2005free}
D.~Hughes, G.~Coulson, J.~Walkerdine, Free riding on gnutella revisited: the
  bell tolls?, Distributed Systems Online, IEEE 6~(6).

\bibitem{Eugster:2007}
P.~Eugster, P.~Felber, F.~Le~Fessant,
  \href{http://doi.acm.org/10.1145/1317379.1317386}{The "art" of programming
  gossip-based systems}, SIGOPS Oper. Syst. Rev. 41~(5) (2007) 37--42.
\newblock \href {http://dx.doi.org/10.1145/1317379.1317386}
  {\path{doi:10.1145/1317379.1317386}}.
\newline\urlprefix\url{http://doi.acm.org/10.1145/1317379.1317386}

\bibitem{Anagnostopoulos:2012:AEI:2365374.2365840}
C.~Anagnostopoulos, O.~Sekkas, S.~Hadjiefthymiades,
  \href{http://dx.doi.org/10.1016/j.pmcj.2011.06.005}{An adaptive epidemic
  information dissemination model for wireless sensor networks}, Pervasive Mob.
  Comput. 8~(5) (2012) 751--763.
\newblock \href {http://dx.doi.org/10.1016/j.pmcj.2011.06.005}
  {\path{doi:10.1016/j.pmcj.2011.06.005}}.
\newline\urlprefix\url{http://dx.doi.org/10.1016/j.pmcj.2011.06.005}

\bibitem{conf/nca/GarbinatoRT07}
B.~Garbinato, D.~Rochat, M.~Tomassini,
  \href{http://dblp.uni-trier.de/db/conf/nca/nca2007.html#GarbinatoRT07}{Impact
  of scale-free topologies on gossiping in ad hoc networks.}, in: NCA, IEEE
  Computer Society, 2007, pp. 269--272.
\newline\urlprefix\url{http://dblp.uni-trier.de/db/conf/nca/nca2007.html#GarbinatoRT07}

\bibitem{PitreyS13}
C.~Pitrey, F.~Sailhan, Revisiting gossip-style failure detection in wireless
  sensor network, in: H.~L{\"{o}}nn, E.~M. Schiller (Eds.), {SAFECOMP} 2013 -
  Workshop ASCoMS (Architecting Safety in Collaborative Mobile Systems) of the
  32nd International Conference on Computer Safety, Reliability and Security,
  Toulouse, France, 2013, {HAL}, 2013.

\bibitem{Riviere:2007}
E.~Rivi\`{e}re, R.~Baldoni, H.~Li, J.~Pereira,
  \href{http://doi.acm.org/10.1145/1317379.1317387}{Compositional gossip: A
  conceptual architecture for designing gossip-based applications}, SIGOPS
  Oper. Syst. Rev. 41~(5) (2007) 43--50.
\newblock \href {http://dx.doi.org/10.1145/1317379.1317387}
  {\path{doi:10.1145/1317379.1317387}}.
\newline\urlprefix\url{http://doi.acm.org/10.1145/1317379.1317387}

\bibitem{Sarkar:2011}
R.~Sarkar, X.~Zhu, J.~Gao,
  \href{http://doi.acm.org/10.1145/1993042.1993046}{Hierarchical spatial gossip
  for multiresolution representations in sensor networks}, ACM Trans. Sen.
  Netw. 8~(1) (2011) 4:1--4:24.
\newblock \href {http://dx.doi.org/10.1145/1993042.1993046}
  {\path{doi:10.1145/1993042.1993046}}.
\newline\urlprefix\url{http://doi.acm.org/10.1145/1993042.1993046}

\bibitem{Chaintreau:2008}
A.~Chaintreau, P.~Fraigniaud, E.~Lebhar,
  \href{http://doi.acm.org/10.1145/1397735.1397752}{Opportunistic spatial
  gossip over mobile social networks}, in: Proceedings of the First Workshop on
  Online Social Networks, WOSN '08, ACM, New York, NY, USA, 2008, pp. 73--78.
\newblock \href {http://dx.doi.org/10.1145/1397735.1397752}
  {\path{doi:10.1145/1397735.1397752}}.
\newline\urlprefix\url{http://doi.acm.org/10.1145/1397735.1397752}

\bibitem{Ferretti2013481}
S.~Ferretti,
  \href{http://www.sciencedirect.com/science/article/pii/S0140366412004148}{Shaping
  opportunistic networks}, Computer Communications 36~(5) (2013) 481 -- 503.
\newblock \href
  {http://dx.doi.org/http://dx.doi.org/10.1016/j.comcom.2012.12.006}
  {\path{doi:http://dx.doi.org/10.1016/j.comcom.2012.12.006}}.
\newline\urlprefix\url{http://www.sciencedirect.com/science/article/pii/S0140366412004148}

\bibitem{Baldoni:2007}
R.~Baldoni, R.~Beraldi, V.~Quema, L.~Querzoni, S.~Tucci-Piergiovanni,
  \href{http://doi.acm.org/10.1145/1266894.1266898}{Tera: Topic-based event
  routing for peer-to-peer architectures}, in: Proceedings of the 2007
  Inaugural International Conference on Distributed Event-based Systems, DEBS
  '07, ACM, New York, NY, USA, 2007, pp. 2--13.
\newblock \href {http://dx.doi.org/10.1145/1266894.1266898}
  {\path{doi:10.1145/1266894.1266898}}.
\newline\urlprefix\url{http://doi.acm.org/10.1145/1266894.1266898}

\bibitem{VoulgarisRKS06}
S.~Voulgaris, E.~Riviere, A.~Kermarrec, M.~van Steen,
  \href{http://www.iptps.org/papers-2006/Vulgaris-sub06.pdf}{Sub-2-sub:
  Self-organizing content-based publish subscribe for dynamic large scale
  collaborative networks}, in: 5th International workshop on Peer-To-Peer
  Systems, {IPTPS} 2006, Santa Barbara, CA, USA, February 27-28, 2006, 2006.
\newline\urlprefix\url{http://www.iptps.org/papers-2006/Vulgaris-sub06.pdf}

\bibitem{Slavov:2014}
V.~Slavov, P.~Rao, \href{http://dx.doi.org/10.1007/s00778-013-0314-1}{A
  gossip-based approach for internet-scale cardinality estimation of xpath
  queries over distributed semistructured data}, The VLDB Journal 23~(1) (2014)
  51--76.
\newblock \href {http://dx.doi.org/10.1007/s00778-013-0314-1}
  {\path{doi:10.1007/s00778-013-0314-1}}.
\newline\urlprefix\url{http://dx.doi.org/10.1007/s00778-013-0314-1}

\bibitem{Haeupler:2012}
B.~Haeupler, G.~Pandurangan, D.~Peleg, R.~Rajaraman, Z.~Sun,
  \href{http://doi.acm.org/10.1145/2312005.2312031}{Discovery through gossip},
  in: Proceedings of the Twenty-fourth Annual ACM Symposium on Parallelism in
  Algorithms and Architectures, SPAA '12, ACM, New York, NY, USA, 2012, pp.
  140--149.
\newblock \href {http://dx.doi.org/10.1145/2312005.2312031}
  {\path{doi:10.1145/2312005.2312031}}.
\newline\urlprefix\url{http://doi.acm.org/10.1145/2312005.2312031}

\bibitem{Ferretti20131631}
S.~Ferretti,
  \href{http://www.sciencedirect.com/science/article/pii/S0167739X12001367}{Gossiping
  for resource discovering: An analysis based on complex network theory},
  Future Generation Computer Systems 29~(6) (2013) 1631 -- 1644, including
  Special sections: High Performance Computing in the Cloud and Resource
  Discovery Mechanisms for \{P2P\} Systems.
\newblock \href
  {http://dx.doi.org/http://dx.doi.org/10.1016/j.future.2012.06.002}
  {\path{doi:http://dx.doi.org/10.1016/j.future.2012.06.002}}.
\newline\urlprefix\url{http://www.sciencedirect.com/science/article/pii/S0167739X12001367}

\bibitem{khatibi}
E.~Khatibi, S.~Mirtaheri, E.~Khaneghah, M.~Sharifi, Dynamic multilevel
  feedback-based searching strategy in unstructured peer-to-peer systems, in:
  Green Computing and Communications (GreenCom), 2012 IEEE International
  Conference on, 2012, pp. 124--131.
\newblock \href {http://dx.doi.org/10.1109/GreenCom.2012.29}
  {\path{doi:10.1109/GreenCom.2012.29}}.

\bibitem{shah}
B.~Shah, C.~Lee, K.~I. Kim, Fuzzy search controller in unstructured mobile
  peer-to-peer networks, in: Dependable, Autonomic and Secure Computing (DASC),
  2014 IEEE 12th International Conference on, 2014, pp. 173--178.
\newblock \href {http://dx.doi.org/10.1109/DASC.2014.39}
  {\path{doi:10.1109/DASC.2014.39}}.

\bibitem{mbp-2011}
M.~Marzolla, O.~Babaoglu, F.~Panzieri, Server consolidation in clouds through
  gossiping, in: World of Wireless, Mobile and Multimedia Networks (WoWMoM),
  2011 IEEE International Symposium on a, 2011, pp. 1--6.
\newblock \href {http://dx.doi.org/10.1109/WoWMoM.2011.5986483}
  {\path{doi:10.1109/WoWMoM.2011.5986483}}.

\bibitem{Sharifkhani}
F.~Sharifkhani, M.~Pakravan, A review of new advances in resource discovery
  approaches in unstructured p2p networks, in: Advances in Computing,
  Communications and Informatics (ICACCI), 2013 International Conference on,
  2013, pp. 828--833.
\newblock \href {http://dx.doi.org/10.1109/ICACCI.2013.6637283}
  {\path{doi:10.1109/ICACCI.2013.6637283}}.

\bibitem{Asthana:2013}
H.~Asthana, I.~Cox, \href{http://doi.acm.org/10.1145/2505515.2507862}{Retrieval
  of trending keywords in a peer-to-peer micro-blogging osn}, in: Proceedings
  of the 22nd ACM international conference on Conference on information \&\#38;
  knowledge management, CIKM '13, ACM, New York, NY, USA, 2013, pp. 1229--1232.
\newblock \href {http://dx.doi.org/10.1145/2505515.2507862}
  {\path{doi:10.1145/2505515.2507862}}.
\newline\urlprefix\url{http://doi.acm.org/10.1145/2505515.2507862}

\bibitem{Bae:2015}
S.-H. Bae, B.~Howe,
  \href{http://doi.acm.org/10.1145/2807591.2807668}{Gossipmap: A distributed
  community detection algorithm for billion-edge directed graphs}, in:
  Proceedings of the International Conference for High Performance Computing,
  Networking, Storage and Analysis, SC '15, ACM, New York, NY, USA, 2015, pp.
  27:1--27:12.
\newblock \href {http://dx.doi.org/10.1145/2807591.2807668}
  {\path{doi:10.1145/2807591.2807668}}.
\newline\urlprefix\url{http://doi.acm.org/10.1145/2807591.2807668}

\bibitem{amazon}
\href{http://status.aws.amazon.com/s3-20080720.html}{Amazon s3 availability
  event: July 20, 2008} (July 2008) [cited May 2015].
\newline\urlprefix\url{http://status.aws.amazon.com/s3-20080720.html}

\bibitem{DeCandia:2007}
G.~DeCandia, D.~Hastorun, M.~Jampani, G.~Kakulapati, A.~Lakshman, A.~Pilchin,
  S.~Sivasubramanian, P.~Vosshall, W.~Vogels,
  \href{http://doi.acm.org/10.1145/1323293.1294281}{Dynamo: Amazon's highly
  available key-value store}, SIGOPS Oper. Syst. Rev. 41~(6) (2007) 205--220.
\newblock \href {http://dx.doi.org/10.1145/1323293.1294281}
  {\path{doi:10.1145/1323293.1294281}}.
\newline\urlprefix\url{http://doi.acm.org/10.1145/1323293.1294281}

\bibitem{Lakshman:2010}
A.~Lakshman, P.~Malik,
  \href{http://doi.acm.org/10.1145/1773912.1773922}{Cassandra: A decentralized
  structured storage system}, SIGOPS Oper. Syst. Rev. 44~(2) (2010) 35--40.
\newblock \href {http://dx.doi.org/10.1145/1773912.1773922}
  {\path{doi:10.1145/1773912.1773922}}.
\newline\urlprefix\url{http://doi.acm.org/10.1145/1773912.1773922}

\bibitem{vanRenesse:2008}
R.~van Renesse, D.~Dumitriu, V.~Gough, C.~Thomas,
  \href{http://doi.acm.org/10.1145/1529974.1529983}{Efficient reconciliation
  and flow control for anti-entropy protocols}, in: Proceedings of the 2Nd
  Workshop on Large-Scale Distributed Systems and Middleware, LADIS '08, ACM,
  New York, NY, USA, 2008, pp. 6:1--6:7.
\newblock \href {http://dx.doi.org/10.1145/1529974.1529983}
  {\path{doi:10.1145/1529974.1529983}}.
\newline\urlprefix\url{http://doi.acm.org/10.1145/1529974.1529983}

\bibitem{sf_complenet}
S.~Ferretti, \href{http://dx.doi.org/10.1007/978-3-319-05401-8_7}{Searching in
  unstructured overlays using local knowledge and gossip}, in: P.~Contucci,
  R.~Menezes, A.~Omicini, J.~Poncela-Casasnovas (Eds.), Complex Networks V,
  Vol. 549 of Studies in Computational Intelligence, Springer International
  Publishing, 2014, pp. 63--74.
\newblock \href {http://dx.doi.org/10.1007/978-3-319-05401-8_7}
  {\path{doi:10.1007/978-3-319-05401-8_7}}.
\newline\urlprefix\url{http://dx.doi.org/10.1007/978-3-319-05401-8_7}

\bibitem{wkermack27}
W.~O. Kermack, A.~McKendrick, {A Contribution to the Mathematical Theory of
  Epidemics}, Proceedings of the Royal Society of London. Series A, Containing
  Papers of a Mathematical and Physical Character 115~(772) (1927) 700--721.

\bibitem{Newman:2010:NI:1809753}
M.~Newman, Networks: An Introduction, Oxford University Press, Inc., New York,
  NY, USA, 2010.

\bibitem{XgaetaX}
R.~Gaeta, M.~Sereno, Generalized probabilistic flooding in unstructured
  peer-to-peer networks, Parallel and Distributed Systems, IEEE Transactions on
  22~(12) (2011) 2055--2062.
\newblock \href {http://dx.doi.org/10.1109/TPDS.2011.82}
  {\path{doi:10.1109/TPDS.2011.82}}.

\bibitem{gda-mospas-11}
G.~D'Angelo, S.~Ferretti, {LUNES}: Agent-based simulation of {P2P} systems, in:
  Proceedings of the International Workshop on Modeling and Simulation of
  Peer-to-Peer Architectures and Systems (MOSPAS 2011), IEEE, 2011.

\bibitem{EberspacherS05a}
J.~Eberspächer, R.~Schollmeier, First and second generation of peer-to-peer
  systems., in: Peer-to-Peer Systems and Applications, Vol. 3485 of Lecture
  Notes in Computer Science, Springer, 2005, pp. 35--56.

\bibitem{ER}
P.~Erd\"{o}s, A.~R\'{e}nyi,
  \href{http://www.renyi.hu/\~{}p\_erdos/Erdos.html\#1959-11}{{On random
  graphs, I}}, Publicationes Mathematicae (Debrecen) 6 (1959) 290--297.
\newline\urlprefix\url{http://www.renyi.hu/\~{}p\_erdos/Erdos.html\#1959-11}

\bibitem{RevModPhys.74.47}
R.~Albert, A.-L. Barab\'asi,
  \href{http://link.aps.org/doi/10.1103/RevModPhys.74.47}{Statistical mechanics
  of complex networks}, Rev. Mod. Phys. 74 (2002) 47--97.
\newblock \href {http://dx.doi.org/10.1103/RevModPhys.74.47}
  {\path{doi:10.1103/RevModPhys.74.47}}.
\newline\urlprefix\url{http://link.aps.org/doi/10.1103/RevModPhys.74.47}

\bibitem{watts1998cds}
D.~J. Watts, S.~H. Strogatz, {Collective dynamics of'small-world'networks.},
  Nature 393~(6684) (1998) 409--10.

\bibitem{TDGsIMC07}
M.~Iliofotou, P.~Pappu, M.~Faloutsos, M.~Mitzenmacher, S.~Singh, G.~Varghese,
  Network monitoring using traffic dispersion graphs (tdgs), in: Proceedings of
  the 7th ACM SIGCOMM Internet Measurement Conference, ACM, New York, NY, USA,
  2007, pp. 315--320.
\newblock \href {http://dx.doi.org/http://doi.acm.org/10.1145/1298306.1298349}
  {\path{doi:http://doi.acm.org/10.1145/1298306.1298349}}.

\bibitem{Ozkasap:2009}
O.~Ozkasap, M.~Caglar, E.~Cem, E.~Ahi, E.~Iskender,
  \href{http://dx.doi.org/10.1016/j.comnet.2009.03.021}{Stepwise fair-share
  buffering for gossip-based peer-to-peer data dissemination}, Comput. Netw.
  53~(13) (2009) 2259--2274.
\newblock \href {http://dx.doi.org/10.1016/j.comnet.2009.03.021}
  {\path{doi:10.1016/j.comnet.2009.03.021}}.
\newline\urlprefix\url{http://dx.doi.org/10.1016/j.comnet.2009.03.021}

\bibitem{sacha}
J.~Sacha, J.~Dowling, R.~Cunningham, R.~Meier, Discovery of stable peers in a
  self-organising peer-to-peer gradient topology, in: F.~Eliassen, A.~Montresor
  (Eds.), Distributed Applications and Interoperable Systems, Vol. 4025 of
  Lecture Notes in Computer Science, Springer Berlin, Heidelberg, 2006, pp.
  70--83.

\bibitem{Tanta-ngai}
H.~Tanta-ngai, E.~E. Milios, V.~Ke\v{s}elj,
  \href{http://doi.acm.org/10.1145/1651274.1651287}{Self-organizing
  peer-to-peer networks for collaborative document tracking}, in: Proceedings
  of the 1st ACM international workshop on Complex networks meet information \&
  knowledge management, CNIKM '09, ACM, New York, NY, USA, 2009, pp. 59--66.
\newblock \href {http://dx.doi.org/10.1145/1651274.1651287}
  {\path{doi:10.1145/1651274.1651287}}.
\newline\urlprefix\url{http://doi.acm.org/10.1145/1651274.1651287}

\bibitem{verma}
S.~Verma, W.~T. Ooi, Controlling gossip protocol infection pattern using
  adaptive fanout, in: ICDCS '05: Proceedings of the 25th IEEE International
  Conference on Distributed Computing Systems, IEEE Computer Society,
  Washington, DC, USA, 2005, pp. 665--674.
\newblock \href {http://dx.doi.org/http://dx.doi.org/10.1109/ICDCS.2005.20}
  {\path{doi:http://dx.doi.org/10.1109/ICDCS.2005.20}}.

\bibitem{newmanHandbook}
M.~E.~J. Newman, \href{http://dx.doi.org/10.1002/3527602755.ch2}{Random graphs
  as models of networks}, Wiley-VCH Verlag GmbH and Co. KGaA, 2005, pp. 35--68.
\newblock \href {http://dx.doi.org/10.1002/3527602755.ch2}
  {\path{doi:10.1002/3527602755.ch2}}.
\newline\urlprefix\url{http://dx.doi.org/10.1002/3527602755.ch2}

\bibitem{DBLP:conf/bioadit/OkuyamaTK06}
T.~Okuyama, T.~Tsuchiya, T.~Kikuno, Improving the robustness of epidemic
  communication in scale-free networks, in: Proceedings of Biologically
  Inspired Approaches to Advanced Information Technology, Second International
  Workshop, BioADIT 2006, Vol. 3853 of Lecture Notes in Computer Science,
  Springer, 2006, pp. 294--305.

\bibitem{Portmann20031159}
M.~Portmann, A.~Seneviratne,
  \href{http://www.sciencedirect.com/science/article/B6TYP-478YXBY-3/2/45607964354b5739028117010ebf1585}{Cost-effective
  broadcast for fully decentralized peer-to-peer networks}, Computer
  Communications 26~(11) (2003) 1159--1167, ubiquitous Computing.
\newblock \href {http://dx.doi.org/DOI: 10.1016/S0140-3664(02)00250-5}
  {\path{doi:DOI: 10.1016/S0140-3664(02)00250-5}}.
\newline\urlprefix\url{http://www.sciencedirect.com/science/article/B6TYP-478YXBY-3/2/45607964354b5739028117010ebf1585}

\bibitem{Wilf_1994}
H.~S. Wilf, {Generatingfunctionology}, 2nd Edition, Academic Press, Inc., 1994.

\bibitem{PhysRevLett.85.4626}
R.~Cohen, K.~Erez, D.~ben Avraham, S.~Havlin,
  \href{http://link.aps.org/doi/10.1103/PhysRevLett.85.4626}{Resilience of the
  internet to random breakdowns}, Phys. Rev. Lett. 85 (2000) 4626--4628.
\newblock \href {http://dx.doi.org/10.1103/PhysRevLett.85.4626}
  {\path{doi:10.1103/PhysRevLett.85.4626}}.
\newline\urlprefix\url{http://link.aps.org/doi/10.1103/PhysRevLett.85.4626}

\bibitem{p2p09-peersim}
A.~Montresor, M.~Jelasity, {PeerSim}: A scalable {P2P} simulator, in: Proc. of
  the 9th Int. Conference on Peer-to-Peer (P2P'09), Seattle, WA, 2009, pp.
  99--100.

\bibitem{igraph}
igraph – the network analysis package, http://http://igraph.org (2015).

\bibitem{Boyd:2006}
S.~Boyd, A.~Ghosh, B.~Prabhakar, D.~Shah,
  \href{http://dx.doi.org/10.1109/TIT.2006.874516}{Randomized gossip
  algorithms}, IEEE/ACM Trans. Netw. 14~(SI) (2006) 2508--2530.
\newblock \href {http://dx.doi.org/10.1109/TIT.2006.874516}
  {\path{doi:10.1109/TIT.2006.874516}}.
\newline\urlprefix\url{http://dx.doi.org/10.1109/TIT.2006.874516}

\bibitem{gda-simpat-2014}
G.~D’Angelo, M.~Marzolla,
  \href{http://www.sciencedirect.com/science/article/pii/S1569190X14001014}{New
  trends in parallel and distributed simulation: From many-cores to cloud
  computing}, Simulation Modelling Practice and Theory (SIMPAT)\href
  {http://dx.doi.org/http://dx.doi.org/10.1016/j.simpat.2014.06.007}
  {\path{doi:http://dx.doi.org/10.1016/j.simpat.2014.06.007}}.
\newline\urlprefix\url{http://www.sciencedirect.com/science/article/pii/S1569190X14001014}

\bibitem{pads}
{Parallel And Distributed Simulation (PADS) Research Group},
  \url{http://pads.cs.unibo.it} (2015).

\bibitem{feldman2004free}
M.~Feldman, C.~Papadimitriou, J.~Chuang, I.~Stoica, Free-riding and
  whitewashing in peer-to-peer systems, in: Proceedings of the ACM SIGCOMM
  workshop on Practice and theory of incentives in networked systems, ACM,
  2004, pp. 228--236.

\end{thebibliography}

\end{document}